\documentclass[10pt,a4paper]{book}

\usepackage{float}
\usepackage{psfrag}
\usepackage{amssymb}
\usepackage{amsmath}
\usepackage{graphics}
\usepackage{amsthm}
\usepackage{hyperref}
\usepackage{breakurl}
\usepackage[all]{xy}
\usepackage{color}
\usepackage{xcolor}
\usepackage{mathtools}
\usepackage{listings}
\usepackage{rotating}
\usepackage{fancyhdr}
\usepackage{epigraph}
\usepackage{array}
\usepackage{cancel}
\usepackage{subfigure}
\usepackage{wasysym}
\usepackage[font=small,format=plain,labelfont=bf,up,textfont=up]{caption}
\usepackage{mathrsfs}
\usepackage{makeidx}
\makeindex
\newcommand{\symin}[3]{\index{#1}\index{#2@${#3}$|see{#1}}}

\newcommand{\unit}[1]{\,\mathrm{#1}}

\CompileMatrices

\DeclareCaptionFont{white}{\color{white}}
\DeclareCaptionFormat{listing}{\colorbox{gray}{\parbox{\textwidth}{#1#2#3}}}
\title{Automated Synthesis of Controllers for Search and Rescue from Temporal Logic Specifications}
\date{\today}
\author{Clemens Wiltsche}

\newcommand{\alabel}[1]{({#1})\label{#1}}
\newcommand{\aref}[1]{\hyperref[#1]{(#1)}}
\newcommand{\fig}[1]{\hyperref[fig:#1]{Figure~\ref{fig:#1}}}
\newcommand{\tab}[1]{\hyperref[tab:#1]{Table~\ref{tab:#1}}}
\renewcommand{\sec}[1]{\hyperref[sec:#1]{Section~\ref{sec:#1}}}

\theoremstyle{plain}

\newcounter{definition}
\newenvironment{definition}[1][\relax]{\begin{list}%
{}{\leftmargin 0pt\rightmargin 0pt\labelsep 3pt\parsep 0pt%
\setlength{\listparindent}{\parindent}
}
   \item \textbf{Definition\refstepcounter{definition}\ \thedefinition \MakeUppercase{#1}.\ }}{\hspace*{\fill}$\Box$\end{list}}

\newcommand{\e}{\epsilon}
\newcommand{\A}{\mathscr{A}}
\newcommand{\C}{\mathcal{C}}
\newcommand{\CC}{\mathfrak{C}}
\newcommand{\E}{\mathcal{E}}
\renewcommand{\L}{\mathcal{L}}
\newcommand{\M}{\mathcal{M}}
\renewcommand{\P}{\mathcal{P}}
\renewcommand{\S}{\mathcal{S}}
\newcommand{\T}{\mathcal{T}}
\renewcommand{\v}{\varphi}
\newcommand{\V}{\mathcal{V}}
\newcommand{\X}{\mathcal{X}}
\newcommand{\Y}{\mathcal{Y}}
\newcommand{\seq}[1]{\L_{#1}}

\let\next\undefined
\DeclareMathOperator{\next}{\mathop\bigcirc}
\DeclareMathOperator{\strongnext}{\mathop\odot}
\DeclareMathOperator{\eventually}{\mathop\lozenge}
\DeclareMathOperator{\pasteventually}{\mathop{\blacklozenge}}
\DeclareMathOperator{\always}{\mathop\square}
\DeclareMathOperator{\pastalways}{\mathop{\blacksquare}}
\DeclareMathOperator{\until}{\mathbin\mathcal{U}}
\DeclareMathOperator{\weakuntil}{\mathbin\mathrm{U}}
\DeclareMathOperator{\defines}{\mathrel\triangleq}
\DeclareMathOperator{\Coloneqq}{\mathrel::\joinrel=}
\DeclareMathOperator{\derives}{\vdash}
\let\models\undefined
\DeclareMathOperator{\models}{\vDash}
\let\nmodels\undefined
\DeclareMathOperator{\nmodels}{\nvDash}
\DeclareMathOperator{\trans}{\mathrel\rightarrow}
\DeclareMathOperator{\qdot}{\mathrel.}
\DeclareMathOperator{\comp}{\mid}

\newcommand{\ass}[2]{\v^e_{{#1},{#2}}}
\newcommand{\guar}[2]{\v^s_{{#1},{#2}}}
\newcommand{\vstate}[1]{\S_v{#1}}
\newcommand{\estate}[1]{\S_e{#1}}
\newcommand{\cmove}{g}
\newcommand{\cseq}{\Gamma}

\newcommand{\namefont}{\mathit}

\newcommand{\stay}[1]{\namefont{stay}(#1)}
\newcommand{\rais}[1]{\namefont{raise}(#1)}
\newcommand{\clear}[1]{\namefont{clear}(#1)}

\newcommand{\DfA}{\namefont{DEQf\_ACK}}
\newcommand{\Df}{\namefont{DEQf}}
\newcommand{\FIFOf}{\namefont{FIFOf}}
\newcommand{\FIFOr}{\namefont{FIFOr}}
\newcommand{\AMM}{\namefont{AMM}}
\newcommand{\cellID}{\namefont{cellID}}
\newcommand{\flag}{\namefont{flag}}
\newcommand{\store}{\namefont{store}}

\newcommand{\Adj}[1]{\namefont{Adj}({#1})}

\newcommand{\Cin}[1]{C_{\mathrm{in}{#1}}}

\newcommand{\cnt}{\namefont{cnt}}
\newcommand{\mcnt}{\namefont{mcnt}}
\newcommand{\cst}{\namefont{cst}}
\newcommand{\ackin}{\namefont{ackin}}
\newcommand{\ackout}{\namefont{ackout}}
\newcommand{\Cout}[1]{C_{\mathrm{out}{#1}}}

\newcommand{\ecl}{\mathrm{ecl}}
\newcommand{\eco}{\mathrm{eco}}

\newcommand{\vcl}{\mathrm{vcl}}
\newcommand{\vco}{\mathrm{vco}}
\newcommand{\vcr}{\mathrm{vcr}}
\newcommand{\vpc}{\mathrm{vpc}}

\newcommand{\sv}{\mathrm{\mathbf{sv}}}
\newcommand{\ccl}{\mathrm{cl}}
\newcommand{\cpc}{\mathrm{pc}}
\newcommand{\ccr}{\mathrm{cr}}
\newcommand{\csv}{\sv}

\newcommand{\mgu}{\mathrm{gu}}
\newcommand{\mpc}{\mathrm{sp}}
\newcommand{\mcs}{\mathrm{cs}}
\newcommand{\mcl}{\mathrm{sl}}
\newcommand{\Mcs}{M = \mcs}
\newcommand{\Mgu}{M = \mgu}
\newcommand{\Mcl}{M = \mcl}
\newcommand{\Mpc}{M = \mpc}
\newcommand{\Mncs}{M \neq \mcs}

\newcommand{\Mncl}{M \neq \mcl}
\newcommand{\Mnpc}{M \neq \mpc}

\newcommand{\Cops}{\mathscr{C}}
\newcommand{\Robbers}{\mathscr{R}}
\newcommand{\freelist}{\mathfrak{R}}
\newcommand{\guardlist}{\mathfrak{G}}

\newcommand{\pref}[1]{\namefont{pref}({#1})}
\newcommand{\badpref}[1]{\namefont{badpref}({#1})}
\newcommand{\nextval}[1]{\namefont{nextval}({#1})}

\begin{document}

\pagestyle{empty}

{\color{white}.}\\[1cm]

\begin{center}
{\LARGE Automated Synthesis of Controllers for Search}\\[0.2cm]
{\LARGE and Rescue from Temporal Logic Specifications}\\

\vspace{1cm}

{\large Clemens Wiltsche}\\[0.2cm]
{\today}

\vspace{2cm}

{\large \sc Master Thesis}\\[0.2cm]
{\large \sc Institute for Automation}\\[0.2cm]
{\large \sc ETH Zurich}

\end{center}

\vspace{9cm}

{\large
\begin{minipage}[t]{0.6\textwidth}
Supervising Professor:\newline
Prof.\ John Lygeros
\end{minipage}
\begin{minipage}[t]{0.4\textwidth}
\begin{flushright}
Supervisors:\\
Dr.\ Federico A. Ramponi\\
Sean Summers
\end{flushright}
\end{minipage}
}

\cleardoublepage

\pagestyle{fancy}
\pagenumbering{roman}
\setcounter{page}{1}

\chapter*{Abstract}

In this thesis, the synthesis of correct-by-construction controllers for robots assisting in Search and Rescue (SAR) is considered. In recent years, the development of robots assisting in disaster mitigation in urban environments has been actively encouraged, since robots can be deployed in dangerous and hazardous areas where human SAR operations would not be possible.\\

In order to meet the reliability requirements in SAR, the specifications of the robots are stated in Linear Temporal Logic and synthesized into finite state machines that can be executed as controllers. The resulting controllers are purely discrete and maintain an ongoing interaction with their environment by changing their internal state according to the inputs they receive from sensors or other robots.\\

Since SAR robots have to cooperate in order to complete the required tasks, the synthesis of controllers that together achieve a common goal is considered. This distributed synthesis problem is provably undecidable, hence it cannot be solved in full generality, but a set of design principles is introduced in order to develop specialized synthesizable specifications. In particular, communication and cooperation are resolved by introducing a verified standardized communication protocol and preempting negotiations between robots.\\

The robots move on a graph on which we consider the search for stationary and moving targets. Searching for moving targets is cast into a game of cops and robbers, and specifications implementing a winning strategy are developed so that the number of robots required is minimized.\\

The viability of the methods is demonstrated by synthesizing controllers for robots performing search and rescue for stationary targets and searching for moving targets. It is shown that the controllers are guaranteed to achieve the common goal of finding and rescuing the targets.\\

\cleardoublepage
\tableofcontents

\chapter*{List of Abbreviations}
 \addcontentsline{toc}{chapter}{List of Abbreviations}
 
\begin{tabbing}
 \hspace*{1.6cm}  \= \kill
 A/G \> Assumption-Guarantee \\[0.5ex]
 BNF \> Backus-Naur Form \\[0.5ex]
 CTL \> Computation Tree Logic \\[0.5ex]
 FDS \> Fair Discrete System \\[0.5ex]
 FIFO \> First in first out \\[0.5ex]
 FSA \> Finite State Automaton \\[0.5ex]
 GCS \> Graph Clearing Sequence \\[0.5ex]
 GR[1] \> Generalized Reactivity[1] \\[0.5ex]
 JTLV \> Java-TLV \\[0.5ex]
 LTL \> Linear Temporal Logic \\[0.5ex]
 SAR \> Search And Rescue \\[0.5ex]
 TL \> Temporal Logic \\[0.5ex]
 TLA \> Temporal Logic of Actions \\[0.5ex]
 TLV \> Temporal Verification System \\[0.5ex]
 USAR \> Urban Search And Rescue \\[0.5ex]
 WiSAR \> Wilderness Search and Rescue \\
\end{tabbing}



\cleardoublepage

\pagenumbering{arabic}
\setcounter{page}{1}
\chapter{Introduction}

Robot-assisted Search and Rescue (SAR) has received increased public attention after the 1995 Hanshin-Awaji Earthquake in Kobe, Japan, and the attacks in the same year on the Murrah Federal Building in Oklahoma City, USA~\cite{RobotsRescue}: The RoboCup-Rescue project, started in 1998 at several Japanese universities is one of the first reactions of robotics researchers to the aforementioned earthquake, encouraging the development of robots assisting in disaster mitigation~\cite{RoboCup}.\\

The US Federal Emergency Management Agency gives the following definition:

\begin{quote}
Urban search-and-rescue (USAR)\index{USAR} involves the location, rescue (extrication), and initial medical stabilization of victims trapped in confined spaces.\footnote{\url{http://www.fema.gov/emergency/usr/}, accessed 19 Oct 2011}
\end{quote}

A recent survey by Goodrich et al.\ \cite{WiSAR} extends the application of SAR operations from confined urban spaces:

\begin{quote}
Wilderness search and rescue (WiSAR)\index{WiSAR} operations include finding and providing assistance to humans who are lost or injured in mountains, deserts, lakes, rivers, or other remote settings.
\end{quote}

Thus, SAR entails searching for people that are in distress or in need of medical support and providing relief to their situation. In an abstract sense, SAR includes the searching of any kind of problematic object and alleviating the situation. This requires teams of trained personnel, a well-managed organizational structure and management of equipment logistics.\\

\subsubsection{Motivating Robot-Assisted SAR}

Consider the scenario of a burning building, or the aftermath of a devastating hurricane. Usually there is only limited information available about whether there were people in the affected area at the time of the incident and what the imminent dangers to the rescue personnel are. Deploying SAR teams exposes personnel to the dangers of the site, and rescue efforts might be slow and costly.\\

Supporting the SAR efforts by robots has several advantages. Robots can be deployed in hazardous areas where human operations would not be possible. A major issue during disasters is the physical exhaustion and lack of sleep of SAR personnel, resulting in an increased susceptibility to misjudgements. Robots do not suffer from this kind of fatigue, even though their power supply needs regular recharging. Also, the precision of the sensors and speed of robots can by far outmatch those of humans. This is usually partially alleviated by canine-supported or mounted SAR teams, with the obvious limitations in hazardous or physically demanding environments.\\

As SAR robots obtain higher levels of autonomy, it can also be expected that the number of human operators per robot can be decreased. For each SAR robot that was employed during the operations after the September 11 attacks on the World Trade Center in 2001, at least two or more operators were needed, even though the robots had only very basic reconnaissance capabilities. Current examples of autonomous vehicles with greater operational capabilities are deployed in military applications, and require even a larger number of operators, e.g.\ $180$ for the General Atomics Predator (but here the degree of autonomy is partially limited due to legal constraints).\footnote{The Economist, Drones and the Man, Jul 30th 2011, \url{http://www.economist.com/node/21524876}, accessed 28 Oct 2011.}\\

Automated synthesis for a particular class of specifications can be done efficiently in polynomial time. This could even allow feasible on-the-fly synthesis of controllers if the configuration changes during a field operation, e.g.\ by attaching a new sensor or by changing the means of locomotion~\cite{Marsupials}. In this case, it is merely required to add the appropriate rules to the specification, as highlighted in the discussion of incremental modification at the beginning of \sec{specs}. However, the main limitation of synthesis approach is the large state space size of the resulting controllers. Hand-coded controllers are less prone to this problem (see the comparisons in Bloem et al.\ \cite{HWPSL}), but cannot be scaled to arbitrary size and complexity.\\

\subsubsection{Contributions}

In this thesis, the development of controllers for autonomous SAR robots is investigated. In order to act autonomously, a robot must be enabled to make decisions about the actions it performs, based on its observations about the environment and the other robots.\\

We consider purely discrete controllers  for SAR robots that maintain an ongoing interaction with their environment by changing their internal state according to the inputs they receive. The specification of such controllers consists of a set of rules or guarantees such as ``each location in the search area is visited repeatedly,'' while allowing assumptions about the environment to be made, such as  ``targets cannot move faster than $3\unit{mph}$.'' Such statements are be expressed in a formal high-level specification language which closely matches how humans would naturally express such properties.\\

Instead of merely verifying properties, we are concerned with synthesizing\index{Synthesis} controllers that are correct by construction. Using such an approach shifts the validation effort to the specification. However, expressing the specifications in a high-level language can greatly simplify this process.\\

SAR is typically performed using several agents that have to coordinate and communicate. We therefore need to consider the synthesis of controllers that \emph{together} achieve a common goal. This is called compositional (or distributed) synthesis\index{Synthesis!Compositional Synthesis}, and is in general undecidable~\cite{HardSynth}. The compositional synthesis problem therefore cannot be solved in its full generality, but we introduce a set of design principles that help to develop appropriate specifications that \emph{can} be synthesized. Communication and coordination are resolved by introducing a standardized control protocol and preempting negotiations\index{Negotiation} between robots.\\

The viability of our methods is demonstrated by synthesizing controllers for robots performing search and rescue for stationary targets and searching for moving targets. A four-phase handshake protocol is used to communicate commands between the agents\index{Communication!Protocol}. When interpreting the controllers in real-time, our main motivation is not to attain temporal efficiency, but to develop controllers that are proven to be reliable. We show that the controllers are guaranteed to achieve the common goal of finding and rescuing the targets and provide simulation results demonstrating the validity of the controllers.\\

\section{Robot-Assisted SAR Scenarios}

In this section, we consider two paradigms for robot-assisted SAR and outline the scenarios. The stationary target search in USAR and the moving target search in WiSAR guide our development of the specifications that are synthesized to be executed as controllers on SAR robots.\\

\subsection{USAR} \label{sec:usar} \index{USAR}

Robots assisting in USAR can be deployed for a variety of tasks, like inspecting damaged or collapsed buildings for imminent danger to human rescuers. Robots can also be used to monitor the temperature and air quality in a burning building or provide comfort to victims by establishing a two-way audio connection to rescue personnel. Casper and Murphy~\cite{WTCRescue} provide a survey of such tasks related to the September 11, 2001 attacks on the World Trade Center.\\

We study controllers for USAR to develop an understanding of the tasks of searching and rescuing when a central coordinating authority exists. In USAR we can assume that injured or entrapped persons or objects are not able to move. It is therefore sufficient to develop methods for stationary target search, which considerably simplifies the controllers.\\

\subsection{WiSAR} \label{sec:wisar} \index{WiSAR}

While USAR is mostly concerned with small to medium scale sites such as burning or collapsed buildings, WiSAR entails searching over large regions in often rugged remote areas~\cite{WiSAR}. Robot-assisted WiSAR often combines several types of Unmanned Air- and Ground Vehicles in order to find a target that might be moving in order to get to safety itself.\\

The search strategies required for WiSAR are fundamentally different from those sufficient for USAR. This is mainly because the targets are assumed to be able to move. In this case, the number of robots required for reliable search depends on the topology of the environment. Also, the search radius has to increase with time, since a target might try to move away from the location at which it is initially expected to be, trying to reach safety itself. In this thesis we concentrate on developing a reliable distributed search strategy for a constant search radius.\\

\section{Challenges in Robot-Assisted SAR}

There are several challenges in developing and deploying SAR robots. The most critical ones are summarized below.\\

\subsubsection{Unknown Environment Topology} \index{Topology}

In a typical SAR scenario, little is known about the topology of the environment in which the robots are operating. In USAR, even if building plans are available to a robot, passages and corridors might be blocked, voids might be opened and in a collapsed building the entire structure might be changed. Also, in a WiSAR application, the terrain might be unknown, and in particular, accurate elevation data is not always available. This is exacerbated by the need to expand the search radius as time progresses.\\

Searching an unknown environment can be accomplished by first obtaining an internal representation of the topology in which the robot is operating. The main drawback of this approach is that time is wasted on building this internal map that could be dedicated to rescuing. An alternative is to learn the map of the environment concurrently while searching~\cite{ProbLoc, ProbPurs}.  These approaches obtain a probabilistic representation of the environment and are therefore not applicable to the controller framework considered in this thesis.\\

\subsubsection{Changing Environment Topology}

Even if the robot has an accurate and reliable representation of the environment topology at some point in time, in a real-world scenario this knowledge has to be updated continually. An example of an application in which this capability is necessary is USAR in a burning building in which hallways and staircases might collapse. Another example is WiSAR after natural disasters such as landslides, floods or earthquakes in which the landscape keeps changing. Thus SAR robots need a way of detecting the environment topology and adapting their movements through the SAR site accordingly.\\

This adds another layer of complexity to the task of synthesizing controllers that are correct by construction, since it has to be taken into account that the ``knowledge'' a robot has at each point in time might be changing. Synthesis guaranteeing such epistemic (knowledge-based) properties has been considered before~\cite{EpSynth} but not in the input-output driven way that is required for controllers.\\

Due to the problems with unknown and changing environment topologies, in this thesis we assume a constant environment and that an accurate and reliable representation of the environment topology is known at specification-time, i.e.\ when the specification is written for the SAR robots.\\

\subsubsection{Cooperation and Communication} \index{Communication}\index{Cooperation}

Often a SAR task requires the cooperation of many agents. For example, finding a moving target can be done reliably by a single robot only in a very limited number of cases, see \sec{movtsearch}. In order to cooperate, agents must be able to share relevant information about their own actions and adapt their strategies accordingly.\\

Communication between robots is governed by protocols which must be encoded in the specification. However, this introduces circularity: A robot has to guarantee to satisfy its part of the protocol's contract only if the robot with which it communicates satisfies its own part. A na\"ive formalization of this principle would allow the robots to exhibit \emph{any} behavior in the case that no robot satisfies its guarantees initially. Resolving these circularities requires to impose restrictions on the specifications pertaining to communication, giving rise to assumption/guarantee specifications~\cite{AGSpec} and the corresponding composition rules, see \sec{distrsped}.\\

\subsubsection{Realizability and Synthesizability} \index{Synthesizability}\index{Realizability}

Synthesis of correct-by-construction controllers has the obvious advantage that the specification is guaranteed to be satisfied. However, there exist specifications that cannot be realized (or satisfied) by any controller. Thus, if the specification is unrealizable, no controller can be generated from it, independently of the synthesis method used.\\

Not all specifications that are realizable can also be synthesized into a controller, depending on the synthesis method used. When choosing the synthesis method, there is a tradeoff between the generality of the specifications supported and the efficiency of the synthesis and of the resulting controllers.\\

The realizability of some specifications depends on the topology of the environment. For example, the property ``each location in the search area is visited repeatedly'' requires that the topology can be represented by a strongly connected graph, see \sec{maprepresentation}. Deciding realizability and synthesizing controllers for a particular topology can be handled reasonably well by existing methods. However, controllers that provide guarantees for all graphs classified by some user-specified properties (such as connectedness or Hilbertness) pose a much harder problem and is not treated here.\\

\subsubsection{Integration and Human-Robot Interaction}

One of the main challenges in developing robots assisting SAR is that they have to be integrated in existing organizational structures. This includes technical aspects such as the tasks that a robot is capable to perform, and the improvements that this yields over not deploying the robots alongside human SAR personnel. However, legal considerations such as liability in the case of failure or lack of trust of team leaders in the robots' capabilities may prevent SAR robots from being deployed even when there are no technical difficulties or limitations.\\

Moreover, the interaction of humans with the robots must be taken into account. One major concern is the need for trained operators that must dispatch, control and maintain the robots. Also, automatic detection of victims is not yet reliable enough for completely autonomous SAR, so human operators are required to interpret the sensor data~\cite{WTCRescue}. While those aspects mainly concern the search for the victims, automated on-site rescue and medical treatment poses a variety of new challenges.\\

\section{Previous Work}

The previously mentioned RoboCup-Rescue project initiated a surge of research in robot-assisted SAR. However, even as early as 1984, computer assistance to SAR has been considered by Belardo et al.\ \cite{SRComp}. Early attempts to robot-assisted SAR include supportive technologies for robot deployment~\cite{AIUSAR}, coordinating robots to perform rescue-like tasks ~\cite{RetrTasks, BoxPushing} and combining this with cooperative search~\cite{CooperativeSearch}.\\

The RoboCup-Rescue project and other similar projects simulate USAR scenarios in which several robot-developing teams can test their devices and hold competitions. With the advancements from such early research projects, several SAR robots were deployed during the World Trade Center rescue response in 2001~\cite{WTCRescue}. While robot deployment time was limited, this unfortunate incident provided valuable information for further development for SAR robots.\footnote{CNN, J.D. Sutter, How 9/11 Inspired a New Era of Robotics, 07 Sept 2011, \url{http://articles.cnn.com/2011-09-07/tech/911.robots.disaster.response_1_packbot-robots-disasters?_s=PM:TECH}, accessed 30 Nov 2011}\\








Recent research into robots assisting in tasks connected with SAR has demonstrated several successful implementations. Several types of robots such as tracked, legged and flying vehicles, snakes and even shape-shifters have been developed that allow access into damaged buildings, hazardous areas and confined spaces~\cite{Legged, Snakes, Flying, Marsupials}. Some of these robots have already been deployed in actual SAR tasks~\cite{WTCRescue, Cologne, Mines}.\\

Instead of focusing on the physical properties of the SAR robots, this thesis is predominantly concerned with the control of teams of robots. This has been considered previously, e.g.\ for building inspection and searching~\cite{SensTeam, ProbPurs}. However, often these teams need to perform complex tasks that can be described quite straightforwardly in a rule-based way, but are hard to implement manually into a discrete controller that obeys all the rules at the same time. A promising approach is to automatically synthesize correct-by-construction controllers from high-level descriptions~\cite{KressFainekos}.\\

The synthesis\index{Synthesis} of correct-by-construction controllers has been considered since the 1960's for example by Church, B\"uchi, Landweber, and Rabin. Initially, the results were developed for very general properties, leading to a pessimistic outlook on the computational complexity of synthesis. Also, only recently have methods been developed that are directly applicable for controller synthesis~\cite{Pnueli}. This seminal work by Piterman et al.\ enabled efficient synthesis of controllers from temporal logic specifications and forms the basis of synthesis tools developed at the California Institute of Technology and the University of Pennsylvania~\cite{KressFainekos, TuLiP}. The respective research groups augment the discrete controllers resulting from the temporal logic specifications with continuous controllers. Other approach to synthesizing such hybrid controllers that are not based on the work by Piterman et al.\ were developed by Kloetzer and Belta~\cite{LTLPlanning} and Loizou and Kyriakopoulos~\cite{FirstSpec}. An overview of current techniques is given in the work by Belta et al., which goes beyond the techniques covered in this thesis~\cite{SymbolicPlanning}.\\

Several successful examples and implementations of controllers that have been synthesized from high level specifications have been developed~\cite{LTLMop, PTZ, VehicleManagement}. Toy examples of robots moving in buildings and completing tasks include a conceptual implementation of USAR~\cite{Waldo}. Controller synthesis has also been suggested to be used for vehicles competing in the DARPA Grand Challenge but for such a large-scale application several hurdles have to be overcome first~\cite{DARPA}. Still, modelling and synthesizing controllers from high-level specifications for symbolic control and planning is becoming increasingly powerful~\cite{SymbolicPlanning}, and several software front-ends for synthesis of discrete and hybrid systems such as TuLiP~\cite{TuLiP}, LTLCon~\cite{LTLCon}, LTLMop~\cite{LTLMop} and PESSOA~\cite{PESSOA} have been developed since.\\

\chapter{Theoretical Background}

We consider the synthesis of controllers for SAR robots that are guaranteed to satisfy their specifications by construction. Besides the synthesis procedure, a formal description language of the resulting controllers as well as a formal specification language have to be introduced.\\

In \sec{reactive} we introduce an abstract model of discrete reactive controllers and how to reason about their properties. The synthesis procedure of a single correct-by-construction controller from its specification is outlined in \sec{synth}. Since several agents cooperate in an SAR scenario, compositional synthesis is introduced in \sec{distrsynth}.\\

\section{Reactive Systems} \label{sec:reactive}

A {\bf controller}\index{Controller} must maintain a continuous interaction with its {\bf environment}\index{Environment} and the {\bf system}\index{System} it controls. The system can be controlled by changing set points of actuators, but the environment may only be sensed and not influenced by the controller. The system keeps a state that is changed according to the behavior of its controller and its environment. The controller maintains its knowledge about the system's state and thus the controller is simply considered to be part of the system.\footnote{In traditional control terminology the system is considered to be fully observable.}\\

A traditional computer program can be regarded as relation between a set of initial states and a set of final states, acting as state transformers. However, this paradigm is not appropriate for describing the program of a controller, since it has to refer to the controller's ongoing behavior. A controller \emph{reacts} to changes in its inputs and its program is therefore called a {\bf reactive system}\index{Reactive System}~\cite{CurrTrend}. The behavior of a reactive system can be described by a (possibly infinite) sequence of states of the system. This sequence can be influenced by the environment via events or shared variables, which can be modelled by incorporating appropriate atomic propositions into the specification language.\\

We consider a controller as a reactive system in discrete time. Since in this thesis the synthesized controllers are simulated as if they were running directly on digital microprocessors, this temporal abstraction is sensible. Moreover, we are mainly concerned with specifying and synthesizing controllers that decide on a robot's high-level moves, similar to the moves in a game of chess, rather than the robot's continuous trajectories.\\

\subsection{Fair Discrete Systems} \label{sec:fsa}

A reactive system can be represented as a Finite State Automaton\index{Finite State Automaton} (FSA) that captures how the system state evolves over time. Piterman et al.\ suggest the use of a Fair Discrete System (FDS)\footnote{Strictly, we introduce what is called \emph{fairness-free} FDS's. Fairness requires that certain transitions are taken infinitely many times~\cite{CurrTrend}.} as an abstract computational model for reactive systems~\cite{Pnueli}. We make the inputs, the outputs and the state space explicit, as it will later simplify the description of systems consisting of several FDS's.

\begin{definition}
  A {\bf Fair Discrete System}\symin{Fair Discrete System}{M}{\M} $\M$ is a tuple $(V, \X, \Y, Q, \Theta, \rho)$ consisting of the following components:
  \begin{itemize}
    \item $V = \{u_1, u_2, \ldots, u_n\}$ is a finite set of typed {\bf state variables}\symin{State variables}{V}{V}, over finite domains $\{\V(u_1), \V(u_2), \ldots, \V(u_n)\}$ respectively. Let $\V(V) = \prod_{u \in V} \V(u)$ denote the set of all possible valuations of all variables in $V$, called the {\bf alphabet}\index{Alphabet} of $\M$.
    \item $\X \subseteq V$ is the set of {\bf environment variables}\symin{Environment variables}{X}{\X} that are controlled by the environment. They are also called {\bf inputs}. 
    \item $\Y = V \backslash \X$ is the set of {\bf system variables}\symin{System variables}{Y}{\Y} that are controlled by the system (i.e.\ the actuator set points). They are also called {\bf outputs}.
    \item $Q \subseteq \V(V) \times \mathbb{N}_0$ is the set of {\bf states}\index{Q@$Q$}. A state $q \in Q$ is a valuation with a unique identifier, so there can be two distinct states corresponding to the same valuation. We write $q[u]$ to denote the value of the variable $u \in V$ in the state $q$, i.e.\ $q[u] \in \V(u)$. A state $q$ {\bf satisfies} a propositional formula $\v$, written $q \models \v$ iff $\v$ holds when for each occurrence of $u$, the corresponding value $q[u] \in \V(u)$ is substituted.
    \item $\Theta$ is the {\bf initial condition}\index{Theta@$\Theta$}, a propositional formula characterizing the initial states of $\M$. A state $q \in Q$ is initial iff $q \models \Theta$.
    \item $\rho$ is the {\bf transition relation}\symin{Transition relation}{rho}{\rho} defining the behavior of the FDS. It is a function $\rho: Q \rightarrow 2^Q$ that takes the current state $q \in Q$ and produces a set of new states $Q' \subseteq Q$. Note that this formulation allows for nondeterministic transition relations by producing sets containing two or more states. If $q' \in \rho(q)$ this is also written as $q \trans_\rho q'$\index{ trans@$\trans_\rho$|see{Transition relation}}.
  \end{itemize}
\end{definition}

An FDS describes how the values of the variables in $V$ change over discrete time. The concept of how time evolves and relates to real time is described in detail in \sec{TL}. As an FDS proceeds from state to state, a sequence over $Q$ is generated. In order to reason about the behavior of an FDS, we introduce the concept of a computation: 
\begin{definition}
  A {\bf computation}\index{Computation} of an FDS $\M = (V, \X, \Y, Q, \Theta, \rho)$ is a (possibly infinite) sequence of states $q = q_0q_1q_2\ldots$ over $Q$ satisfying
  \begin{itemize}
    \item[\alabel{C1}] {\bf Initiality}: $q_0 \models \Theta$.
    \item[\alabel{C2}] {\bf Consecution}: $\forall j \in \mathbb{N} \qdot j \leq |q| \Rightarrow  q_{j-1} \trans_\rho q_j$.
  \end{itemize}
\end{definition}
Here $|q|$ denotes the {\bf length} of the sequence $q$: it is $\aleph_0$ if the sequence is infinite, and $n$ for a finite sequence $q = q_0q_1\ldots q_n$. Thus, the length of a computation is the number of transitions made. We also introduce the {\bf concatenation}\index{Concatenation (of two sequences)} of two sequences $\sigma_1 = s_0s_1\ldots s_n$ and $\sigma_2 = s_{n+1}s_{n+2}s_{n+3}\ldots$, where $\sigma_1$ must be finite, to be the sequence $\sigma_1\sigma_2 = s_0s_1s_2\ldots$, i.e.\ concatenations of sequences are written just by concatenating the symbols.\\

We are only interested in reactive systems that maintain an ongoing interaction with their environment, hence only infinite computations will be considered. FDS's for which all maximal length computations are infinite are called {\bf nonblocking}\index{Nonblocking} and are characterized by the additional requirement that in each state the transition relation allows the FDS to enter a next state:
\begin{itemize}
  \item[\alabel{NB}] {\bf Nonblocking Property}: $\forall q \in Q \qdot \rho(q) \neq \emptyset$.
\end{itemize}

Moreover, any physical system has an initial state. To reflect this in our definition of an FDS, the following property is added:
\begin{itemize}
  \item[\alabel{SP}] {\bf Startup Property}: $\exists q_0 \in Q \qdot q_0 \models \Theta$.
\end{itemize}

\begin{definition}
Define the (overloaded) operator $\V_W : Q \rightarrow \V(W)$\index{V@$\V$} to map states to their corresponding valuations. That is, for a state $q = (v, n) \in Q$, $\V_W(q) = v \in \V(W)$. If $W$ in $\V_W$ is understood from context, we simply write $\V$.
\end{definition}

The $\V$-operator can also be used for sets, sequences and sets of sequences of states, mapping each element to its corresponding valuation. Moreover, $\V$ is used to map variables to their domains, so it is ``overloaded'' to yield the domain $\V(u)$ for a variable $u$. The sets of valuations of sets of variables $V$, $\X$ and $\Y$ are also defined using $\V$ as $\V(V)$, $\V(\X)$ and $\V(\Y)$.\\


\newcommand{\defseqsets}{Definition~4}
\begin{definition}
The set of all finite nonempty sequences over a set $S$ is denoted by $S^+$, and the set of all infinite sequences over $S$ is denoted by $S^\infty$.
\end{definition}

Note that by \aref{SP} all computations of an FDS $\M$ are nonempty, as $\M$ must have an initial state satisfying~\aref{C1}. The set of infinite computations of an FDS $\M$ is then denoted by $\C(\M) \subseteq Q^\infty$\index{C@$\C(\M)$}, and the corresponding set of infinite sequences of valuations, called the {\bf language}\index{Fair Discrete System!Language} of $\M$ is $\V(\C(\M)) \subseteq \V(V)^\infty$. To simplify notation, we denote the language of $\M$ by $\seq{\M} = \V(\C(\M))$\index{L@$\seq{\M}$}.\\

A transition of an FDS can either be triggered by a change of the environment variables, called {\bf environment-triggered}\index{Fair Discrete System!Environment-triggered}, or by some external events such as the ticks of a system clock on a computer, called {\bf clock-triggered}\index{Fair Discrete System!Clock-triggered}. In a purely environment-triggered FDS, no transitions can be made that merely change the system variables if no changes in the inputs are observed, limiting the set of behaviors (the language) expressible by such FDS's. However, in a purely clock-triggered FDS, changes in the environment variables might be overlooked if several such changes occur during one tick of the clock. Since the FDS's that we consider should be implementable on a microprocessor, we adopt a clock-triggered approach and account for the complications in the specifications, see in particular \sec{stutter} and \sec{communication}.\\

If $\X$ is nonempty, a change in the any of the environment variables in $\X$ requires a transition to be made. Since $\X$ is not controlled by the FDS, the nonblocking condition~\aref{NB} is no longer sufficient to ensure that there is always a valid transition available. The stronger requirement that in every state a transition is possible independently of the inputs is sufficient to ensure that an FDS always maintains an ongoing interaction with its environment. However, often this requirement is too strong for the specification to be realizable. This issue is addressed by introducing assumption/guarantee specifications in \sec{AG}.\\

By moving from state to state, an FDS can represent the behavior of a microprocessor. While an FDS can be considered as the ``program'' running on the microprocessor, it abstracts away from any specific hardware model. As long as a microprocessor {\bf implements}\index{Fair Discrete System!Implementation} the behavior of an FDS, it is sufficient to reason about the FDS --- the results will hold for the actual implementation as well. Indeed, we will later show how to simulate an FDS and use this to test the synthesized implementations.\\

While the behavior of reactive systems can be accurately modeled using FDS's, only very basic reactive systems can be hand-coded in a manageable amount of time. In order to handle large systems, a high-level system specification language is used to describe the operation of an FDS. Synthesis from a correct high-level description yields a correct-by-construction controller in the form of an FDS. Correctness of a system can then be methodically verified in the specification language without having to explicitly reason about its semantics in the foundation logic.\\

\subsection{Temporal Logic} \label{sec:TL}

Temporal Logic (TL) is a well-established formalism that can be traced back even to the ancient Greeks~\cite{TempTutor}. It has been formalized as Tense Logic by Prior in 1962~\cite{TenseLogic} and introduced as a formal specification language for reactive systems by Pnueli in 1977~\cite{LTL}. A statement in propositional logic is interpreted to make a statement over the state at one given time (the present). However, with TL also statements about the past and the future can be made.\\

\subsubsection{The Concept of Time}

The natural concept of time is that it proceeds continuously and at a constant rate. In temporal logic, this concept of real-time is abstracted away in two ways.\\

Firstly, time in TL proceeds in discrete steps, that might be aligned to clock ticks or other external events. The present is considered to be at time zero. A propositional formula $p$ just talks about the present:
\begin{displaymath}
    \xymatrix@R=0.5pc@C=1pc{{} \ar@{|->}[rrrrr]^>>{t} & {} & {} & {} & {} & {} \\ {p} \ar[u] & {\hphantom{p}} & {\hphantom{p}} & {\hphantom{p}} & {\hphantom{p}} & {\hphantom{p}}}
\end{displaymath}

However, with a temporal formula , statements about several points in time can be made. For example, with the ``always'' operator, $\always p$ we can express that $p$ is an invariant over time (in the future):
\begin{displaymath}
    \xymatrix@R=0.5pc@C=1pc{{} \ar@{|->}[rrrrr]^>>{t} & {} & {} & {} & {} & {} \\ {p} \ar[u] & {p} \ar[u] & {p} \ar[u] & {p} \ar[u] & {p} \ar[u] & {\ldots}}
\end{displaymath}
And with the ``eventually'' operator, $\eventually p$ it can be expressed that the proposition $p$ will hold at some unspecified future point in time:
\begin{displaymath}
    \xymatrix@R=0.5pc@C=1pc{{} \ar@{|->}[rrrrr]^>>{t} & {} & {} & {} & {} & {} \\ {\hphantom{p}} & {\hphantom{p}} & {\hphantom{p}} & {p} \ar[u] & {\hphantom{p}}}
\end{displaymath}

Other temporal operators express similar properties. Several forms of TL are in use, but here we only consider Linear Temporal Logic (LTL). It is called ``linear'' because LTL formulae are interpreted over sequences of states, where each point in time is associated with an element in the sequence. Thus for a sequence of states $s_0s_1s_2\ldots$, the element $s_0$ is said to occur at time zero, $s_1$ is said to occur at time one and so on. These times can be the ticks of a clock. However, there is no correspondence to a real-time axis, which is the second abstraction step. Thus only the order and perhaps the number of repetitions of states in a sequence has semantic significance.\\

After discretizing and abstracting away from real time, the ``past'', ``present'' and ``future'' get their meaning from the part of the sequence under consideration. A {\bf propositional formula}\index{Propositional formula} talks about a particular instance in time, a {\bf past formula}\index{Past formula} talks about (non-strictly) smaller time indices, and a {\bf future formula}\index{Future formula} talks about (non-strictly) greater time indices. Time moves strictly forward, so the elements of a sequence are considered one after the other, without repetitions.\\

By interpreting the abstract time in our real-world notion of continuous time, it can be defined what it means to execute an FDS on a microprocessor. The processor clock generates a sequence of ticks $t_0t_1t_2\ldots$ that trigger the transitions of an FDS. A clock-triggered FDS $\M$ is {\bf executed}\index{Fair Discrete System!Execution} by starting in an initial state $q_0 \models \Theta$ at clock tick $t_0$. Then, before every clock tick $t_i, i>0$, the current values $x_{i-1} \in \V(\X)$ of the input variables $\X$ are read from the environment's state and a transition is made from $q_{i-1}$ to $q_i$ s.t. $\V(q_i)[\X] = x_{i-1}$, where $[\X]$ denotes the {\bf restriction}\index{Restriction (to variables)} to the variables in $\X$. Note that the current values of the environment are only visible in the next state.\\

\subsubsection{Syntax}

The syntax of an LTL formula $\v$ over the variables $V$ is given by the BNF definition
\begin{equation}
\v \Coloneqq p \mid \neg \v \mid \v \vee \v \mid \next \v \mid \v\until\v, \label{eq:ltlsyntax}
\end{equation}
where $p$ is an {\bf atomic proposition}\index{Atomic proposition} relating a variable $u \in V$ to a subset of its domain, $v \subseteq \V(u)$. That is, $p = (u, v) \in V \times \V(u)$. An atomic proposition of the form $p = (u, \{v \in \V(u) \mid v \Join w\})$ is written simply as $u \Join w$, where $w \in \V(u)$ is a constant and $\Join$ is one of $=$, $\neq$, $\leq$, $<$, $\geq$ or $>$. For a boolean variable $u$, the atomic propositions ``$u = \mathrm{True}$'' and ``$u = \mathrm{False}$'' are abbreviated simply by $u$ and $\neg u$ respectively.\\

The operator $\next$ is called {\bf weak next} or just {\bf next} and $\until$ is called {\bf strong until}\index{U@$\until$} or just {\bf until}. Next and until are called {\bf temporal operators}\index{Temporal operators}, while $\neg$ and $\vee$ are called {\bf propositional operators}\index{Propositional Operators}. Other propositional operators can be defined in the usual way:
\begin{align*}
\mathrm{True} &\defines \v \vee \neg \v \\
\v \wedge \psi &\defines \neg (\neg \v \vee \neg \psi) \\
\v \rightarrow \psi &\defines \neg \v \vee \psi \\
\v \leftrightarrow \psi &\defines \v \rightarrow \psi \wedge \psi \rightarrow \v
\end{align*}
A formula only containing atoms and propositional operators is called a {\bf propositional formula}\index{Propositional formula}. The following additional temporal operators can be defined:
\begin{align*}
\eventually\v &\defines \mathrm{True}\until\v \\
\always\v &\defines \neg\eventually\neg\v \\
\v \weakuntil \psi &\defines \v\until\psi \vee \always\v \\
\strongnext\v &\defines \neg\next\neg\v
\end{align*}
The operator $\eventually$ is called {\bf eventually}, $\always$ is called {\bf always}, $\weakuntil$ is called {\bf weak until}\index{U@$\weakuntil$} and $\strongnext$ is called {\bf strong next}. Additionally, the temporal {\bf past operators}\index{Past operators} $\pastalways$\index{ pastalways@$\pastalways$} and $\pasteventually$\index{ pasteventually@$\pasteventually$} can be introduced, which will be used in \sec{compspec}. However, it can be shown that past operators do not add expressivity~\cite{PastGlory}. A formula that contains no past operators is called a {\bf future formula}, while a formula that contains only propositional and past operators is called a {\bf past formula}\index{Past formula}. Operator precedence and associativity are summarized in \tab{ops}.

\begin{table}
\centering
\begin{tabular}{c|c|l}
\hline
Operator(s) & Precedence & Associativity\\
\hline
$\neg$, $\next$, $\strongnext$, $\always$, $\eventually$, $\pastalways$, $\pasteventually$ & $1$ & right \\
$\until$, $\weakuntil$ & $2$ & right \\
$\Join$ & $3$ & right \\
$\wedge$ & $4$ & left\\
$\vee$ & $5$ & left\\
$\rightarrow$, $\leftrightarrow$ & $6$ & right \\
\hline
\end{tabular}
\caption{Operator precedence and associativity.}
\label{tab:ops}
\end{table}

\subsubsection{Semantics}

Let $\L(V)$ denote the set of syntactically correct LTL formulae over the variables $V$, i.e.\ the {\bf language}\index{L@$\L(V)$} generated by \eqref{eq:ltlsyntax}. We now define a compositional semantics of LTL via the satisfaction relation $\models$ between a tuple $(\sigma, j) \in \V(V)^\infty \times \mathbb{N}_0$ and an LTL formula $\v \in \L(V)$. Since the semantics is defined over sequences of valuations $\V(V)$, LTL formulae are not tied to a particular FDS with its corresponding state space $Q$. If two states $q$ and $q'$ correspond to the same valuation, i.e.\ $\V(q) = \V(q')$, then there is no way to tell them apart in an LTL formula.\\

Given a finite or infinite sequence $\sigma = s_0s_1s_2\ldots \in \V(V)^\infty \cup \V(V)^+$ and an index $j \in \mathbb{N}_0$, the relation $\models$ is defined as follows:
\begin{align*}
&(\sigma, j) \models p & \Leftrightarrow &\hspace{0.5cm} s_j \models p \\
&(\sigma, j) \models \neg \v & \Leftrightarrow &\hspace{0.5cm} (\sigma, j) \nmodels \v \\
&(\sigma, j) \models \v \vee \psi & \Leftrightarrow &\hspace{0.5cm} (\sigma, j) \models \v \mathrm{~~or~~} (\sigma, j) \models \psi \\
&(\sigma, j) \models \next \v & \Leftrightarrow &\hspace{0.5cm} (\sigma, j + 1) \models \v \\
&(\sigma, j) \models \v\until\psi & \Leftrightarrow &\hspace{0.5cm} \exists k \geq j . (\sigma, k) \models \psi \wedge (\forall i . j \leq i < k \Rightarrow (\sigma, i) \models \v).
\end{align*}
The semantics of the past operators $\pastalways$ and $\pasteventually$ is given by
\begin{align*}
&(\sigma, j) \models \pastalways \v & \Leftrightarrow &\hspace{0.5cm} \forall k \qdot 0 \leq k \leq j \Rightarrow (\sigma, k) \models \v \\
&(\sigma, j) \models \pasteventually \v & \Leftrightarrow &\hspace{0.5cm} \exists k \qdot 0 \leq k \leq j \wedge (\sigma, k) \models \v.\\[-3ex]
&\hphantom{(\sigma, j) \models \v\until\psi} & \hphantom{\Leftrightarrow} &\hphantom{\hspace{0.5cm} \exists k \geq j . (\sigma, k) \models \psi \wedge (\forall i . j \leq i < k \Rightarrow (\sigma, i) \models \v).}
\end{align*}
The formula $\v$ is said to {\bf hold} at position $j$ of the sequence $\sigma \in \V(V)^\infty$ iff $(\sigma, j) \models \v$.\\


In this semantics $\next \v$ means that if there is a next state, it satisfies $\v$. In contrast, $\strongnext \v$ means that there exists a next state and it satisfies $\v$. If only infinite sequences are considered, these two operators are equivalent. In this thesis we will always use $\next$, since only nonblocking FDS's are considered that can always react to a well-behaved environment. Note also that for infinite sequences $\next$ commutes with $\neg$ and distributes over binary propositional operators.\\

Further, $\v \until \psi$ means that $\v$ holds until $\psi$ holds, and $\psi$ will eventually hold, justifying the name strong until. In contrast, $\v \weakuntil \psi$ does not require $\psi$ to eventually hold (i.e.\ $\v$ holds forever) and is therefore called weak until.\\

A formula $\v \in \L(V)$ is said to be {\bf satisfiable} iff there exists a sequence $\sigma \in \V(V)^\infty$ and an index $j \leq |\sigma|$ s.t.\ $(\sigma, j) \models \v$ holds.\\

A sequence $\sigma \in \V(V)^\infty$ is said to {\bf satisfy} a formula $\v \in \L(V)$ iff for all indices $j \leq |\sigma|$, $(\sigma, j) \models \v$ holds. It is said to {\bf initially satisfy} $\v$ iff $(\sigma, 0) \models \v$ holds. The latter is written $\sigma \models \v$.\\

An FDS $\M$ is said to (initially) satisfy a formula $\v \in \L(V)$ iff all sequences $\sigma \in \seq{\M}$ (initially) satisfy $\v$. A formula $\v \in \L(V)$ is called ({\bf initially}) {\bf valid} for an FDS $\M$ iff every sequence $\sigma \in \seq{\M}$ (initially) satisfies $\v$. If $\M$ initially satisfies $\v$, or $\v$ is initially valid for $\M$, then this is written as $\M \models \v$.\\

A formula $\v \in \L(V)$ is said to be ({\bf initially}) {\bf valid} iff all sequences $\sigma \in \V(V)^\infty$ (initially) satisfy $\v$. If $\v$ is initially valid, then this is written as $\models \v$.\\


An LTL formula thus can be seen as specifying a set of sequences that initially satisfy it. LTL is called ``linear'' because it can only make statements about paths along the transitions in an FDS.\\

\subsubsection{Other Temporal Logics in Use}

In contrast to the linear-time LTL, branching time logics such as CTL can make statements about entire transition trees in an FDS~\cite{CTL}. in particular, this includes nondeterministic choices. There are properties that can be expressed in CTL but not in LTL and vice versa. Both specification languages are subsumed in a more general temporal logic called CTL*~\cite{CTLStar}.\\

Several other logics for specifying reactive systems exist. Some are tailored for hardware, others for software. Interval Temporal Logic makes reasoning about periods of time explicit~\cite{Moszkowski}. Atomic propositions in LTL make statements about variables at one point in time, but in the Temporal Logic of Actions introduced by Lamport in 1994 atomic propositions can be made over changes in variables that are termed ``actions''~\cite{TLA}.\\


\subsection{Model Checking and Deductive Proof Systems} \label{sec:forward}


\begin{figure}[t]
\begin{displaymath}
    \xymatrix@R=0.6pc{{\begin{array}{c}\text{Specification:} \\ \v \end{array}} \ar[rd]!L & & \\ & *++[F-]{\text{\sc Model Checker}} \ar[r] & {\begin{array}{c}\text{Result:} \\ \text{True / False} \end{array}} \\ {\begin{array}{c}\text{FDS:} \\ \M \end{array}} \ar[ru]!L }
\end{displaymath}
\caption{Outline of Model Checker}
\label{fig:checker}
\end{figure}
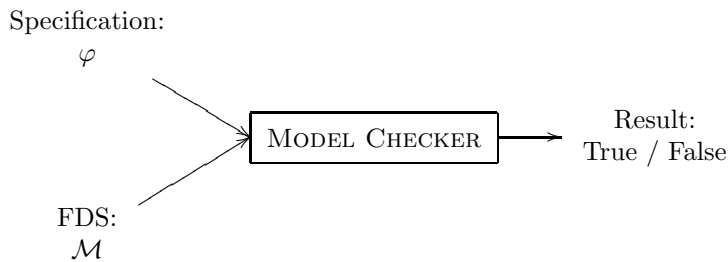

\begin{figure}[t]
\begin{displaymath}
    \xymatrix{{\begin{array}{c}\text{Specification:} \\ \v \end{array}} \ar[r] & *++[F-]{\text{\sc Theorem Prover}} \ar[r] & {\begin{array}{c}\text{Result:} \\ \text{True / False} \end{array}}}
\end{displaymath}
\caption{Outline of Theorem Prover.}
\label{fig:prover}
\end{figure}
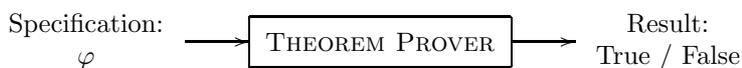

In order to make use of specifications, it must be possible to determine whether a reactive system, in form of an FDS $\M$, satisfies a given specification $\v$. Model checking determines whether $\v$ is (initially) satisfied by $\M$. The outline of this process is shown in \fig{checker}.\\

Moreover, it is often useful to prove properties \emph{about} a specification, without even considering the reactive system it will be applied to. Deductive proof systems require an LTL specification $\v$ to be supplied and will attempt to derive whether $\models \v$ holds independently of a particular FDS, i.e.\ $\forall \M \qdot \M \models \v$. This process is outlined in \fig{prover}.\\


\subsubsection{Model Checking}

In order to check the (initial) validity of an LTL formula $\v$ over a fixed FDS $\M$, a model checker can be employed. The model checker in \fig{checker} is provided with $\v$ and $\M$ and returns true if\footnote{Assuming soundness of the model checker.} all sequences $\sigma \in \seq{\M}$ in the language of $\M$, (initially) satisfy the specification $\v$. The techniques and algorithms for model checking are manifold and can be found for example in~\cite{ModelChecking}.\\

\fig{FDS} shows an FDS $\M = (V, \X, \Y, Q, \Theta, \rho)$ with three variables and five states. Each state is represented as a circle, with the valuations of the state variables $V = \{x, y, z\}$ shown (the unique indices of the states are omitted). In this example, the partition of $V$ into $\X$ and $\Y$ is not relevant, and we may (rather arbitrarily) choose to have $\X = \emptyset$ and $\Y = V$. The arrows represent the transition relation: There is an arrow from a state $q$ to a state $q'$ iff $q \trans_\rho q'$. There is an arrow without a source to a state $q$ iff $q \models \Theta$. Then, model checking could for example establish that the following specifications hold for this FDS:
\begin{align*}
&\always y, \\
&\always \eventually \neg x, \\
&\neg x \wedge \neg z \rightarrow \always(\neg x \wedge y).
\end{align*}

\newcommand{\cellG}[1]{*++++[o][F-]{#1}}
\begin{figure}[h]
\begin{displaymath}
    \xymatrix@dr{ & {} \ar[d]\\
    & \cellG{\phantom{\neg}x, y, \neg z} \ar@/_/[d] & \cellG{\neg x, y, \neg z} \ar[d] \\
    & \cellG{\neg x, y, z\phantom{\neg}} \ar[ur] \ar@/_/[u] & \cellG{\neg x, y, z\phantom{\neg}} \save !R(1) \ar@(r,dr)[] \restore \\
    & \cellG{\phantom{\neg}x, y, z\phantom{\neg}} \ar[u] \\
    {} \ar[ur]}
\end{displaymath}
\caption{Example of an FDS with three variables and five states.}
\label{fig:FDS}
\end{figure}
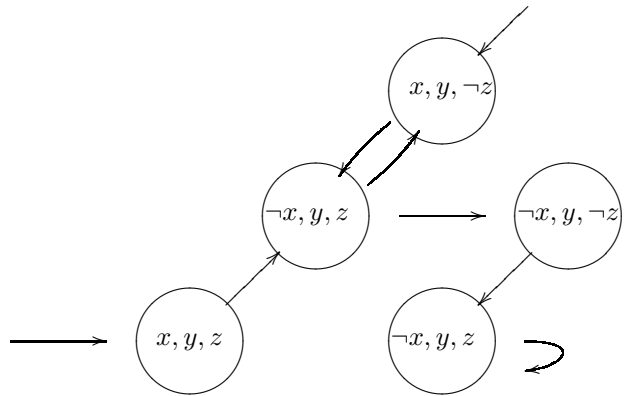

Model checking can also be useful when synthesizing an FDS from its specification. First, the correctness of the synthesis can be verified. Also, if the synthesis algorithm does not support a particular specification (e.g.\ the until operator, $\until$ is not supported by the synthesis procedure we are using), then it can still be verified post-hoc whether the unsynthsesizable properties are satisfied by the resulting FDS, even though not taken into account explicitly during synthesis.\\

\subsubsection{Sequent Calculus for LTL} \index{Sequent calculus}

Sequent calculus is a deductive proof system that allows to prove tautologies in the specification language independent of a particular FDS. Proofs can even be automated up to a certain extent, e.g.\ using tools such as Isabelle~\cite{Isabelle}.\\

This can be employed for example to prove the equivalence of LTL formulae, allowing for simplifications and translations between syntactic forms. Also, when modeling assumptions on the environment, it may be proved that they are satisfiable, avoiding the ``false implies everything'' problem. However, the satisfiability problem for the fragment of LTL that we consider is PSPACE complete, and is NP complete for the fragment of LTL supported by the synthesis procedure that we use~(\cite{Satisfiability}, Figure 3, see the entries for $L(F, X)$ and $\tilde{L}(F, X)$ respectively).\\

Sequent calculus for LTL allows to derive temporal formulae by applying axioms and transformation rules~\cite{TempTutor}. A temporal formula $\v$ that can be derived in such a way is called a {\bf theorem}, written $\derives \v$. The proof system consisting of axiomatic sequents and rules should be {\bf sound}\index{Soundness}, meaning that it is only possible to derive valid LTL formulae. Thus the soundness of a proof system can be expressed by the implication $\derives \v \Rightarrow \models \v$. It is also desirable that the proof system is {\bf complete}\index{Completeness}, meaning that it is possible to derive all valid formulae. This can be expressed by the implication $\models \v \Rightarrow \derives \v$.\\

Theorems can be derived by purely syntactic means using sequent calculus~\cite{LTLDeduction,LTLSequent,TempTutor}. This establishes a set of rules to derive theorems. Usually these rules are applied backwards, starting at the theorem to be proved and working towards axioms such as $\derives \mathrm{True}$.\\

The modus ponens is an example of a sequent calculus rule:
\begin{equation*}
    \frac{\derives p \hspace{1cm} \derives p \Rightarrow q}{\derives q}.
\end{equation*}
This rule says that $q$ can be derived by first deriving $p \Rightarrow q$ and $p$ separately. Both $p$ and $p \Rightarrow q$ must be similarly derived by rules in the sequent calculus, until one arrives at axioms such as the already presented one.\\

\section{Synthesis} \label{sec:synth}

Synthesis is the automatic construction of programs and (digital) designs from logical specifications. It has been investigated already since the 1960's for example by Church, B\"uchi, Landweber and Rabin. The choice of the synthesis algorithms depend on which specification language is used. For example, synthesis of open reactive systems from LTL has been considered by Pnueli and Rosner in 1989~\cite{ReactiveSynth}, and again by Piterman et al.\ in 2006~\cite{Pnueli} and Klein and Pnueli in 2011~\cite{SynthesisRevisited}.\\

Other specification languages have been considered. So far, synthesis from branching time logic is impractical, as the general problems of synthesis from CTL and CTL* specifications are in the complexity classes $\mathrm{2EXPTIME}$ and $\mathrm{3EXPTIME}$\footnote{The complexity classes $\mathrm{2EXPTIME}$ and $\mathrm{3EXPTIME}$ contain algorithms with worst-case complexity $2^{2^{p(n)}}$ and $2^{2^{2^{p(n)}}}$ respectively, where $p(n)$ is a polynomial in the input size $n$.} respectively~\cite{CTLSynth}, and to our knowledge, no methods are available that perform better on a reduced set of specifications, although an interesting approach is suggested by Antoniotti and Mishra~\cite{CTLSynthesis}. Synthesis of open controllers from Probabilistic CTL~\cite{PCTL, PCTLSynthesis} is in $\mathrm{NP} \cap \mathrm{co\!\!-\!\!NP}$~\cite{PCTLComplexity}.\\

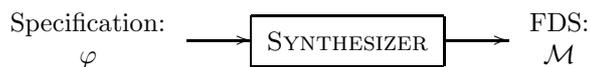
\begin{figure}[ht]
\begin{displaymath}
    \xymatrix{{\begin{array}{c}\text{Specification:} \\ \v \end{array}} \ar[r] & *++[F-]{\text{\sc Synthesizer}} \ar[r] & {\begin{array}{c}\text{FDS:} \\ \M \end{array}}}
\end{displaymath}
\caption{Outline of Synthesizer.}
\label{fig:synthesis}
\end{figure}

We consider the {\bf synthesis problem}\index{Synthesis} to be to find an FDS $\M$ that initially satisfies a given LTL specification $\v$, i.e.\ $\forall \sigma \in \seq{\M} \qdot \sigma \models \v$. This is shown in \fig{synthesis}, where the synthesizer is only provided with the specification.\\

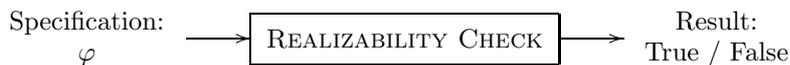
\begin{figure}[hb]
\begin{displaymath}
    \xymatrix{{\begin{array}{c}\text{Specification:} \\ \v \end{array}} \ar[r] & *++[F-]{\text{\sc Realizability Check}} \ar[r] & {\begin{array}{c}\text{Result:} \\ \text{True / False} \end{array}}}
\end{displaymath}
\caption{Outline of Realizability Check.}
\label{fig:realizability}
\end{figure}

Typically, synthesis is preceded by deciding {\bf realizability}\index{Realizability} of the LTL specification, which amounts to checking whether there exists an FDS $\M$ initially satisfying $\v$, see \fig{realizability}.

\subsection{Game Theoretic Approach} \label{sec:gametheorappr}

We use the synthesis tool by Piterman et al.~\cite{Pnueli} which solves the synthesis problem by viewing it as the solution of a two-player game. The synthesized FDS is the controller of the system and acts against its adversary, the environment. The output variables of the FDS directly control the system and thus it is viewed as part of the system. The game is then between the controller and the environment with the utility function being formulated in terms of the behavior of the system: The goal of the game for the system is to satisfy its specification, which the environment tries to prevent. For an introduction into game theory see for example the book by Nisan et al.\ \cite{GameTheor}.\\

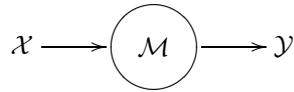
\begin{figure}
\begin{displaymath}
    \xymatrix{ {\X} \ar[r] & *++++[o][F]{\M} \ar[r] & {\Y}}
\end{displaymath}
\caption{The synthesized controller as an FDS}
\label{fig:system}
\end{figure}

Consider the FDS $\M$ in \fig{system}. It represents the synthesized controller with inputs $\X$ (the environment variables\index{Environment variables}) and outputs $\Y$ (the system variables\index{System variables}). The environment and the controller are allowed to influence exclusively the environment and system variables respectively. From the point of view of the controller, these can be seen as the measured state of the system and the inputs to the system respectively. Thus the controller can ``react'' to its observed changes in the environment variables by changing the system variables.\\

The controller and the environment are taking turns in the game. Due to the adversarial nature of the environment it is assumed to always take the worst possible move evaluated against the utility function of the controller. The environment is also thought of as having perfect knowledge about the controller's state and strategy, i.e.\ its transition relation. However, the controller only knows the current observable state of the environment, and possibly the history of environment states that the controller has stored.\\

The utility function of the controller can only take one of two values, corresponding to satisfying and violating its specification. The synthesis procedure therefore tries to find a winning strategy\index{Strategy} for the controller, implemented as an FDS, that maximises this utility by ensuring that the specification is never violated. Such a winning strategy only exists if the specification is realizable. The possible solutions are equilibria in which both the controller and the environment cannot improve their outcome by changing their respective strategy (i.e.\ Nash equilibria). However, these strategies are by no means unique, and thus several FDS's implement the same specification.\\

\subsubsection{Closed and Open Systems}

A system that does not interact with its environment is called {\bf closed}\index{System!Closed}. The computations of such a system do not depend on the environment, and therefore this is modelled by having an empty set of environment variables $\X$. A system that generates its inputs nondeterministically by itself can also be regarded as a closed system.\\

In contrast, a system interacting with its environment is called {\bf open}\index{System!Open}. This is modelled by having a nonempty set of environment variables $\X$. For the system to be open it is necessary that it does not control its environment, in the sense that it cannot directly influence its environment variables.\\

\begin{figure}
\begin{displaymath}
    \xymatrix{ *[o]{} \ar[r] & *+++++[o][F]{\M} \ar@{-}[r] & *[o]{} \ar@{-}[d]_{\Y} \\ *[o]{} \ar@{-}[u]^{\X} & *+++++[o][F]{\P} \ar@{-}[l] & *[o]{} \ar[l]}
\end{displaymath}
\caption{The controller $\M$ with its environment, $\P$. Note that $\P$ does not need to be an FDS.}
\label{fig:closedsystem}
\end{figure}
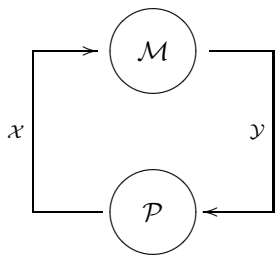

For the synthesis of an open system, it is irrelevant how the variables in $\X$ are controlled -- they might be controlled partially or entirely by another FDS as shown by $\P$ in \fig{closedsystem}. However, all the variables that are not controlled by the system are considered to be part of one single environment. One way to close an open system, which is necessary for simulation (see \sec{statarch}), is by synthesizing an FDS (or a set of FDS's) obeying the environment assumptions and connecting it to the system. This is explained in detail in~\sec{distrsynth}\\

\subsection{Computational Considerations} \label{sec:compsingle}

Synthesis of LTL formulae for general specifications is 2EXPTIME-complete~\cite{Pnueli}. However, for a subset of assume-guarantee specifications called Generalized Reactivity[1] (GR[1]), synthesis can be performed in cubic time in the number of conjuncts in the specification and in the number of possible valuations of the system and environment variables (Theorem 1, \cite{Pnueli}). GR[1]\index{Specification!GR[1]} formulae are explained in detail in \sec{GR1}.\\

While the implementation of an FDS would typically be run on an embedded processor with lower computational power, a high-performance computer can be used for synthesis. Thus, efficiency of executing an FDS is often more important than how much time and memory it takes to synthesize. Once an FDS is loaded into memory, transitions can be performed in $\mathcal{O}(1)$ time.\\

One of the main considerations is therefore the space requirements of the synthesized FDS. The larger the state space, the more memory is required for storage. In the worst case, the space requirements grow quadratically with the size of the state space. This is the case when the graph representing the FDS's transition relation is dense, i.e.\ when there are transitions between almost any two states. Sometimes this can be alleviated by strengthening the assumptions on the environment, since then the transitions can be removed that are precluded by the assumptions, and would never be taken under correct execution.\\

Note that if an environment variable changes so that no transition is available, the FDS blocks. This can only be the case if the environment assumptions are violated, and so the specification still holds even in that case. While this is important to be able to reduce the amount of available transitions during synthesis, such behavior of course is not useful in practice.\\

\subsection{Synthesis Software}

The synthesis tool used in this thesis is based on the work by Piterman et al. \cite{Pnueli}. Synthesis of an LTL specification proceeds in several steps.\\

First, a Game Structure is constructed from the specification, separating the initial conditions, the formulae specifying the transition relation, and the formulae for the game's goals. Then, the realizability of the specification is checked by solving a $\mu$-calculus formula~\cite{MuCalc} that corresponds to the Game Structure~\cite{GR1Solve}. This can be done by the Temporal Verification System (TLV)\index{TLV}, a software tool for computer-aided verification~\cite{TLV}. Once the $\mu$-calculus formula has been successfully solved (which can only be done if the specification is realizable), the intermediate values from the solution process can be used to construct an FDS satisfying the specification, which is again done by invoking TLV. Piterman et al.\ describe the details of these back-end activities~\cite{Pnueli}.\\

The front-end that we use is TuLiP\index{TuLiP}, a Python-based software toolbox for the synthesis of embedded controllers developed by Wongpiromsarn et al.~\cite{TuLiP}. Given an LTL specification, it provides the appropriate inputs for the synthesis to TLV. Its backend is based on JTLV, a Java-based implementation of TLV~\cite{JTLV}. TuLiP supports the synthesis of FDS's from GR[1] specifications. This restricts the set of realizable formulae to those that are in this particular syntactic form. Also, the JTLV backend works with Binary Decision Diagrams that sometimes cannot resolve formulae that are in the form $\always\eventually p$.\\

The specifications that we develop in this thesis were encoded into the appropriate format for TuLiP to process them. TuLiP is then invoked simply by Python function calls.\\

\section{Compositional Synthesis} \label{sec:distrsynth}


In the previous section, the synthesis of a GR[1] formula into a single FDS has been considered. In this case, the synthesis is based on a two-player game of the system against the environment. However, it may sometimes be necessary to synthesize several reactive systems that \emph{together} satisfy one {\bf global specification}\index{Global specification} $\v$. This is called {\bf compositional synthesis}\index{Synthesis!Compositional Synthesis}. It is required when several players need to coordinate with each other in order to ``beat'' the environment.\\

\subsubsection{Motivations for Compositional Synthesis}

\begin{figure}	
\begin{displaymath}
    \xymatrix@R=0.6pc{{\begin{array}{c}\text{Specification:} \\ \v \end{array}} \ar[rd]!L & & \\ & *++[F-]{\text{\sc Distributed Synthesis}} \ar[r] & {\begin{array}{c}\text{FDS's:} \\ \mathbb{M} = \left\{\M_1, \M_2, \ldots\right\} \end{array}} \\ {\begin{array}{c}\text{Architecture:} \\ \A \end{array}} \ar[ru]!L }
\end{displaymath}
\caption{Outline of Compositional Synthesis}
\label{fig:distrsynth}
\end{figure}
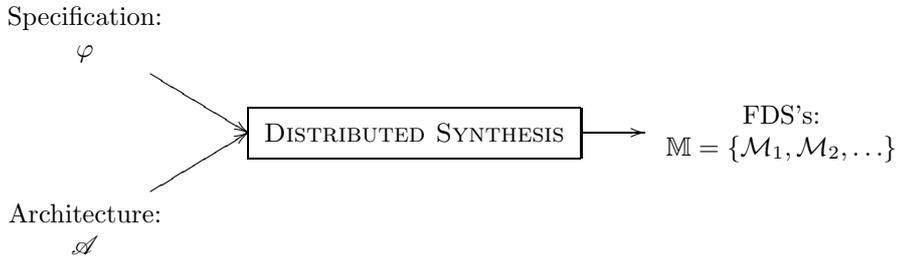

An FDS represents the program of a physical or logical entity. For example, an FDS might be executed as a thread on a processor. If there are several FDS's that are connected with each other, they are also called {\bf components}\index{Component}. The connections between the components are described by an {\bf architecture}\index{Architecture}, which is formally defined in \sec{arch}. When considering one component in the architecture, all other components will be called its {\bf peer components}\index{Peer components} or simply its {\bf peers}.\\

Each component might be executed on a different processor, or as a different thread on the same processor. This imposes restrictions on which variables can be accessed and modified by each component. Often the physical topology naturally imposes restrictions on the architecture supplied to the synthesis procedure.\\

The specification $\v$ provided to the synthesis procedure is an LTL formula over variables in some set $V$ which is partitioned into $\X$ and $\Y$, the {\bf global environment variables}\index{Global environment variables} and the {\bf global system variables}\index{Global system variables} respectively. Synthesizing a single FDS means that all system variables in $\Y$ can be freely manipulated without any constraints. If there are several components, then the components may only manipulate disjoint subsets of $\Y$. This avoids inconsistencies by allowing each variable in $\Y$ to be manipulated by at most one component. However, each variable in $V$ may be read by several components.\\

A specification $\v$ need not refer to all variables in $\Y$. It might be necessary to introduce additional variables that are only relevant locally to the individual components (cf.\ \sec{decomp}) in order to implement communication protocols between the components. Such protocols are of no concern in the global specification and hence are abstracted away. Still, it is necessary that these additional variables are treated as system variables, and thus are in the set $\Y$.\\

\subsection{Game Theoretical Approach} \label{sec:distrgametheorappr}

Compositional synthesis can be seen as solving a multi-player game by finding a set of strategies that obey both the game's rules (the global specification) and the topology of communication (the architecture). The players should cooperate with each other in order to beat an adverse environment.\\

Synthesizing such cooperative strategies is hard and in general undecidable~\cite{HardSynth}. Another way to approach this problem is to recast it into a set of two-player games in which each component acts against all other components and the environment. Since cooperation\index{Cooperation} can be achieved by negotiating\index{Negotiation} according to predefined protocols, the set of realizable global specifications is limited.\\

Negotiation\index{Negotiation} is part of a strategy and thus can be treated using the same kind of specifications as in the rest of the strategy. It is required to establish cooperation patterns such as deciding which robots should rescue a target and which ones should keep searching. When synthesizing strategies for multiple two-player games, the negotiation channels, i.e.\ who negotiates with whom and for what, must be explicitly encoded in each component's specification. This limits the scope of negotiation, but will lead to a substantial reduction of the state space. This is because the components can make local decisions that together satisfy the global specification, and thus lead to the system maximizing its utility. This is ensured by decomposing the specification appropriately \emph{before} synthesis. Thus the decomposition into multiple two-player games is not done by the synthesis procedure, as, in fact, this would still be undecidable.\\

\subsection{Computational Considerations}

The two main considerations in compositional synthesis are realizability and the size of the state space. A specification might be realizable for a single FDS, but not for a distributed architecture. This results from the restrictions on which variables can be controlled by each individual component, which may severely limit the extent of coordination between the components that can be mustered to beat the environment.\\

\subsubsection{State Space Reduction}

Compositional synthesis is a reasonable approach to the state space explosion problem. A single FDS with $n$ boolean variables may contain up to $\mathcal{O}(2^n)$ states. If the system can be decomposed into $N$ subsystems of $n/N + m$ variables, where $m$ is the number of variables needed for communication in each component, then each component would contain only up to $\mathcal{O}(\sqrt[N]{2^n}2^m)$ states.\\

If $m$ is small enough, this can result in fewer states per component. However, if each component receives full information about the state of each of its peers, $m$ has to be large and this might cause even \emph{more} states in each component. This is because in this way each component has full global information and additional variables are required for reliable transmission, and hence the decomposition does not result in a reduction of the state space. By fixing negotiation patterns, it is actually possible to satisfy the global specification by using only local information.\\

Therefore, it might sometimes be desirable to decompose an FDS even if no physical communication restrictions are present. In fact, this might be the only way to practically synthesize a specification within reasonable time and for a given amount of memory.\\

\subsubsection{Realizability}\index{Realizability}

When reasoning about the realizability of a specification in compositional synthesis, the required architecture has to be taken into account. The architecture imposes constraints on the variables that each component can manipulate and thus not all global system variables can be arbitrarily influenced, as is the case in a centralized design.\\

Realizability of a specification depends primarily on the architecture and how the components communicate. If the architecture is in form of a ``pipe'', i.e. a purely serial interconnection in which only the first component receives inputs from the environment, then realizability can be decided. However, even for architectures not much more complicated, for example those involving feedback, realizability is undecidable~\cite{DistLoc, HardSynth}.\\

That does not mean that no feedback communication is realizable, but that no algorithm can be expected to decide realizability for any given specification and architecture. By performing the decomposition into specifications satisfying the architecture before synthesis, many global specifications can be realized, even if distributed synthesis of the global specification is undecidable.\\

\subsection{Architectures} \label{sec:arch}

Compositional synthesis requires not only a global specification $\v$, but also information on how to split up the tasks between the players and their means of communication. This is captured by an architecture $\A$. The synthesizer therefore gets both the specification $\v$ and the architecture $\A$, and outputs a whole set of FDS's $\mathbb{M} = \{\M_1, \M_2, \ldots \}$\index{M@$\mathbb{M}$} that together satisfy the specification and the architecture. This is illustrated in \fig{distrsynth}.\\

\begin{definition}
An {\bf architecture}\symin{Architecture}{A}{\A} $\A$ is a set of pairs $\{(\X_i, \Y_i) \mid i \in I \}$, where $I \subseteq \mathbb{N}$ is a finite set, indexing the components, for which the following restrictions hold:
\begin{itemize}
	\item [\alabel{A1}]: $\forall i \in I \qdot (\X_i \subseteq V) \wedge (\Y_i \subseteq V)$, all variables in the architecture are global.
    \item [\alabel{A2}]: $\forall i, j \in I \qdot (i \neq j) \Rightarrow (\Y_i \cap \Y_j = \emptyset$), no variable is controlled by more than one FDS.
    \item [\alabel{A3}]: $\forall i \in I \qdot (\X_i \cap \Y_i = \emptyset$),  no component controls its environment.
\end{itemize}
\end{definition}

The sets $\X_i$ and $\Y_i$ represent the environment and system variables respectively of each FDS in $\mathbb{M}$. Still, each component may have some of its environment variables controlled by the global environment, and some by peer components. The variables controlled by the global environment will be denoted $\E_i \subseteq \X_i$\index{E@$\E$}. The environment variables of a component $\M_i$ controlled by its peer $\M_j$ will be denoted by $\T_{i,j}$ and are called {\bf transmission variables}\symin{Transmission variables}{T}{\T}.\footnote{Clearly $\T_{i,i} = \emptyset$ for all $i$, because no component controls its own environment.} Given an architecture, the state variables are $V = \bigcup_{i \in I}(\X_i \cup \Y_i)$, the global environment variables can be obtained from $\X = \bigcup_{i \in I} \X_i \backslash \bigcup_{i \in I} \Y_i$, and the global system variables are $\Y = V \backslash \X$. Also, $\E_i = \X_i \backslash \bigcup_{j \in I} \Y_j$ and $\T_{i,j} = \X_i \cap \Y_j$. The variables in $\S_i$\index{S@$\S$} are controlled by $\M_i$ and represent the system variables that are relevant in the global specification.\\

The set of components $\mathbb{M} = \{\M_i \mid i \in I \}$ is said to {\bf satisfy} a given architecture $\A = \{(\X_i, \Y_i) \mid i \in I \}$ if the input variables of $\M_i$ are $\X_i$, its output variables are $\Y_i$ and its state variables are $\X_i \cup \Y_i$ for all $i \in I$. If $\mathbb{M}$ satisfies $\A$, then this is written as $\mathbb{M} \models \A$.\\

\begin{figure}
\begin{displaymath}
    \xymatrix{ {} \ar[r]^<{\E_1} & *++++[o][F]{\M_1} \ar@<1ex>[r]|{\T_{2,1}} \ar[d]^>{\S_1} & *++++[o][F]{\M_2} \ar@<1ex>[l]|{\T_{1,2}} \ar[d]^>{\S_2} & {} \ar[l]_<{\E_2} \\ & {} & {}}
\end{displaymath}
\caption{Composition of two components into a composite FDS.}
\label{fig:asynccomp}
\end{figure}
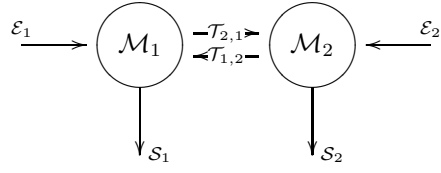

Consider for example the distributed architecture in \fig{asynccomp}. It shows two components $\M_1$ and $\M_2$ and how they are connected by the choice of variables: $\M_1$ can control the variables in $\Y_1 = \S_1 \cup \T_{1,2}$, where $\S_1$ and $\T_{1,2}$ may have a nonempty intersection. The variables in $\X_1 = \E_1 \cup \T_{1,2}$ cannot be controlled by $\M_1$. Instead, $\M_1$ considers them to be controlled by its environment. $\E_1$ and $\T_{1,2}$ are disjoint, where $\T_{1,2}$ is controlled by $\M_2$. However, from the point of view of $\M_1$, it is irrelevant how the variables in $\X_1$ are controlled. Thus,
\begin{equation}
\M_1 = (\underbrace{\E_1 \cup \S_1 \cup \T_{1,2} \cup \T_{2,1}}_{V_1}, \underbrace{\E_1 \cup \T_{1,2}}_{\X_1}, \underbrace{\S_1 \cup \T_{2,1}}_{\Y_1}, Q_1, \Theta_1, \rho_1)
\end{equation}
and symmetrically
\begin{equation}
\M_2 = (\underbrace{\E_2 \cup \S_2 \cup \T_{1,2} \cup \T_{2,1}}_{V_2}, \underbrace{\E_2 \cup \T_{1,2}}_{\X_2}, \underbrace{\S_2 \cup \T_{2,1}}_{\Y_2}, Q_2, \Theta_2, \rho_2).
\end{equation}

\subsection{Composition}

An architecture merely describes the connections between the individual components. As stated above, compositional synthesis has to produce a set of components that satisfy both the architecture and the global specification $\v$. So far, we have only defined what it means for a \emph{single} FDS to satisfy its specification. Therefore a formal description of the behavior of a set of components is required. The idea is to compose the FDS's in $\mathbb{M}$ into one larger FDS $\M$ that generates computations over the variables $V$. This FDS is called the {\bf composite FDS}\index{Composite FDS}\index{Composite FDS}.\\

In a {\bf synchronous composition}\index{Synchronous Composition}, the FDS's proceed in lock-step, i.e.\ they all make a transition at the same time. This would be reflected by the composite FDS $\M$ making a transition if and only if \emph{all} FDS's in $\mathbb{M}$ make a transition. Considering that typically each FDS is executed on an individual processor, this requires their clocks to be precisely synchronized and communication to be implemented with negligible latency. These assumptions can usually not be satisfied in practice, but still this form of composition is often a topic of theoretical interest~\cite{SynchCtrl, HardSynth}.\\

Instead, in an {\bf asynchronous composition}\index{Asynchronous composition} the execution does not have to be in lock-step and therefore no synchronization is required. Communication is still assumed to be with negligible latency, but this can often be relaxed when adapting the local specifications accordingly. The composite FDS $\M$ then proceeds even if only a single FDS in $\mathbb{M}$ makes a transition.\footnote{The standard definition of asynchronous composition prohibits two FDS's to make a synchronous transition.} In order to simplify notation, the following definition of an asynchronous composition is only stated for two FDS's, but it can easily be extended for an arbitrary number of FDS's.\\

\begin{definition}
The  {\bf asynchronous composition}\index{Asynchronous composition} of the FDS's
\begin{align*}
\M_1 &= (\underbrace{\E_1 \cup \S_1 \cup \T_{1,2} \cup \T_{2,1}}_{V_1}, \underbrace{\E_1 \cup \T_{1,2}}_{\X_1}, \underbrace{\S_1 \cup \T_{2,1}}_{\Y_1}, Q_1, \Theta_1, \rho_1),\\
\M_2 &= (\underbrace{\E_2 \cup \S_2 \cup \T_{2,1} \cup \T_{1,2}}_{V_2}, \underbrace{\E_2 \cup \T_{2,1}}_{\X_2}, \underbrace{\S_2 \cup \T_{1,2}}_{\Y_2}, Q_2, \Theta_2, \rho_2)
\end{align*}
is the composite FDS
\begin{equation*}
\M_1 \comp \M_2 =
(\underbrace{V_1 \cup V_2}_{V}, \underbrace{\E_1 \cup \E_2}_{\X}, \underbrace{\Y_1 \cup \Y_2}_{\Y}, \underbrace{Q_1 \times Q_2}_{Q}, \underbrace{\Theta_1 \wedge \Theta_2}_{\Theta}, \rho),
\end{equation*}
where the transition relation is a function $\rho: Q_1 \times Q_2 \rightarrow 2^{Q_1 \times Q_2}$ such that for all $(q_1, q_2), (q_1', q_2') \in Q_1 \times Q_2$,
\begin{eqnarray*}
(q_1', q_2') \in \rho(q_1, q_2) \Leftrightarrow &q_1' \in \rho_1(q_1) \wedge (q_2' = q_2)\,\vee\\ &q_2' \in \rho_2(q_2) \wedge (q_1' = q_1)\,\vee\\ &q_1' \in \rho_1(q_1) \wedge q_2' \in \rho_2(q_2),
\end{eqnarray*}
or equivalently
\begin{eqnarray*}
(q_1, q_2) \trans_\rho (q_1', q_2') \Leftrightarrow &(q_1 \trans_{\rho_1} q_1') \wedge \V_{\T_{1,2}}(q_1') = \V_{\T_{1,2}}(q_2) \wedge (q_2 = q_2')\,\vee\\ &(q_1 = q_1') \wedge (q_2 \trans_{\rho_2} q_2)\,\vee\\ &(q_1 \trans_{\rho_1} q_1') \wedge \V_{\T_{2,1}}(q_2') = \V_{\T_{2,1}}(q_1) \wedge (q_2 \trans_{\rho_2} q_2').
\end{eqnarray*}
\end{definition}

Note that in order to model asynchrony, the values of the transmission variables $\T_{1,2}$ and $\T_{2,1}$ are only equated between $q_1$ and $q_2$ in a transition of $\M_1$ and $\M_2$ respectively.\\

The asynchronous composition is in this thesis also referred to simply as {\bf composition}. Composition is both associative and commutative. That is, $\M_1 \comp \M_2 = \M_2 \comp \M_1$ and $(\M_1 \comp \M_2) \comp \M_3 = \M_1 \comp (\M_2 \comp \M_3)$ for any components $\M_1$, $\M_2$ and $\M_3$ with appropriate environment and system variables. Thus, the parentheses are usually dropped and an arbitrary (finite) number of components can be composed to obtain a composite FDS.\\

Defining the composition of FDS's to be an FDS in its own right has one major advantage: Since LTL specifications are statements over sequences, the global specification $\v$ can directly be used to make statements about the composition. For example, for the composite FDS $\M = \M_1 \comp \M_2 \comp \cdots \comp \M_n$, the statement $\M \models \v$ is already defined to mean that $\forall \sigma \in \seq{\M_1 \comp \M_2 \comp \cdots \comp \M_n} \qdot \sigma \models \v$.\\

The definition of an architecture allows some variables to be system variables of one component and environment variables of its peers. These transmission variables are used in communication protocols and the behavior of the composite FDS $\M$ depends on when a change in these variables is seen by the peer components.\\

\subsection{Decomposition} \label{sec:decomp} \index{Decomposition}

Distributed synthesis is based on a single global specification $\v$ and an architecture $\A$. The specification describes how the asynchronous composition of all components in the architecture has to behave, but makes no explicit statements about the behavior of the individual FDS's.\\

The compositional synthesis problem is to find a set of FDS's so that their asynchronous composition satisfies the global specification. If the architecture defines decoupled or weakly coupled components (i.e.\ there are no or few transmission variables), this can be done without introducing too many additional states in the synthesized components. Since general compositional synthesis is very hard and might even be undecidable, compositional synthesis algorithms are typically built around the synthesis of a single FDS from its local specification. A separate decomposition algorithm is used to obtain the FDS's satisfying $\A$ from the global specification.\\

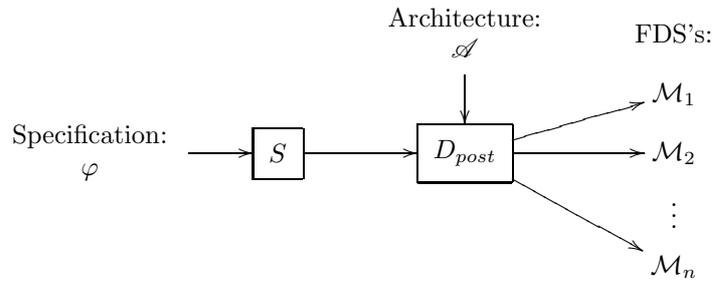
\begin{figure}
\begin{displaymath}
    \xymatrix@R=0pc{ & & {\begin{array}{c}\text{Architecture:} \\ \A \end{array}} \ar[dd] & {\text{FDS's:}} \\
    & & & {\M_1} \\
    {\begin{array}{c}\text{Specification:} \\ \v \end{array}} \ar[r]!L      & *++[F-]{S} \ar[r] & *++[F-]{D_{post}} \ar[ru] \ar[r] \ar[rdd]& {\M_2}  \\
  & & & {\vdots} \\
  & & & {\M_n}
      }
\end{displaymath}
\caption{Outline of Postdecomposition. $S$ is the synthesizer and $D_{post}$ performs the postdecomposition.}
\label{fig:postdecomp}
\end{figure}

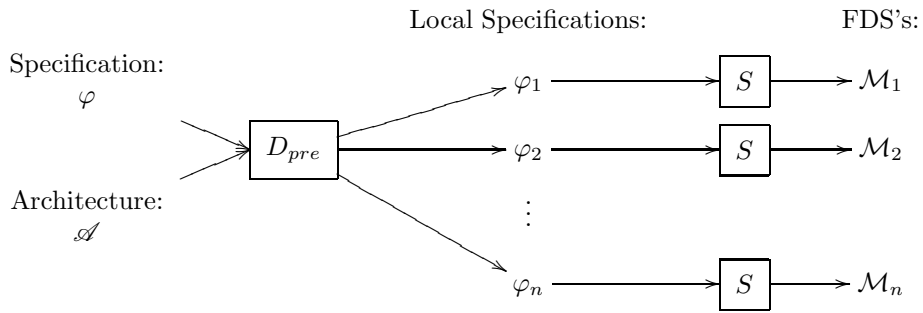
\begin{figure}
\begin{displaymath}
    \xymatrix@R=0pc{ & & {\text{Local Specifications:}} & & {\text{FDS's:}}\\
    {\begin{array}{c}\text{Specification:} \\ \v \end{array}} \ar[rd]!L & & {\v_1} \ar[r] & *++[F-]{S} \ar[r] & {\M_1} \\
     & *++[F-]{D_{pre}} \ar[ru] \ar[r] \ar[rdd] & {\v_2} \ar[r] & *++[F-]{S} \ar[r] & {\M_2}  \\
      {\begin{array}{c}\text{Architecture:} \\ \A \end{array}} \ar[ru]!L & & {\vdots} \\
      & & {\v_n}  \ar[r] & *++[F-]{S} \ar[r] & {\M_n}
      }
\end{displaymath}
\caption{Outline of Predecomposition. $D_{pre}$ performs the predecomposition, and then the same synthesizer $S$ is invoked several times.}
\label{fig:predecomp}
\end{figure}

Two principal approaches for decomposition exist: In {\bf postdecomposition}\index{Decomposition!Postdecomposition}, the global specification is synthesized into a single FDS which is then decomposed according to the architecture (see \fig{postdecomp}). In {\bf predecomposition}\index{Decomposition!Predecomposition}, the global specification is decomposed first into several local specifications that are then synthesized separately (see \fig{predecomp}). Both post- and predecomposition require the resulting synthesized components to satisfy the architecture.\\

The approach that is adopted in this thesis is predecomposition. Let $\v$ be the global specification, and $\A$ be an architecture. A set of local specifications $\{\v_i \mid i \in I\}$ is a predecomposition of a global specification $\v$ iff any set of FDS's $\mathbb{M} = \{\M_i \mid i \in I\}$ that satisfies the architecture $\A$ and for which $\M_i \models \v_i$ for all $i \in I$, the asynchronous composition $\M_1 \comp \M_2 \comp \cdots \comp \M_n$ satisfies the global specification $\v$. That is, the local specifications $\v_1, \v_2, \ldots, \v_n$ must satisfy
\begin{itemize}
	\item[\alabel{PD}]: $\forall \mathbb{M} = \{\M_i \mid i \in I \} \qdot \mathbb{M} \models \A \wedge (\forall i \in I \qdot \M_i \models \v_i) \Rightarrow \M_1 \comp \M_2 \comp \cdots \comp \M_n \models \v$.
\end{itemize}


The decomposition of a global specification into such local specifications is discussed in \sec{specs}. To our knowledge no general systematic approach has yet been developed. The undecidability of compositional synthesis for almost all architectures makes a fully automated approach unlikely.\\

This is why in the next chapter we develop approaches to manually decomposing the global specification. While we consider the restrictions imposed by the architecture on the global system variables that occur in the global specification, the architecture may be enhanced with additional transmission variables if this is needed.\\

\chapter{Specification} \label{sec:specs}

The reactive controllers of the SAR robots decide on the actions of the robots in order to find and rescue the targets. As described in \sec{gametheorappr} and \sec{distrgametheorappr}, the controllers are thought of as players in a game that is subject to a set of rules, has a goal and an initial configuration, similar to a board game. However, board games are usually described informally in prose, which is prone to ambiguities, contradictions and omissions.\\

Instead, formal specifications can be used to precisely describe the behavior of software and hardware systems~\cite{HWPSL,Hoare,WPEFF}. The two most common scenarios are to use the specification to \emph{verify}\index{Verification} a given system (see Model Checking in \sec{forward}) or to \emph{synthesize}\index{Synthesis} a system that satisfies a specification by construction (see \sec{synth}).\\

A specification language should facilitate the development of specifications satisfying the following requirements~\cite{SysSpec}:
\begin{description}
\item[Incremental Modification] Once a specification is finalized, it should be easy to add another requirement, or relax an existing one.
\item[Consistency] There should exist at least one system that can actually satisfy the implementation. This property does not, however, state that such a system can be found in a straightforward manner.
\item[Completeness]\index{Completeness} A specification should include all important properties the system has to satisfy.
\item[Validation]\index{Validation} From a specification, it should be possible to check whether an implementation satisfies the designer's intentions. This is not the same as verification, since an implementation that satisfies a formal specification might still not do what the designer wants. Validation is an informal process, and so the specification language should be semantically closely related to the specification-relevant subset of natural language.
\item[Scalability]\index{Scalability} The specification language should be suitable for large systems.
\end{description}

Incremental modification, consistency and validation are all viable in LTL. However, the completeness of an LTL specification is very difficult to verify, which also restricts the scalability of specifications. Notwithstanding these limitations, LTL has become a common way to express properties of reactive systems. We therefore investigate how to use LTL as a specification language for controllers that fit into the game-theoretic paradigm, as well as taking into account the input-to-output characteristics of standard controllers.\\



A {\bf specification}\index{Specification} for an FDS $\M = (V, \X, \Y, Q, \Theta, \rho)$ is an LTL formula $\v \in \L(V)$ over the variables $V$. Similarly, a program then satisfies the specification if $\v$ is valid for any FDS $\M$ describing the program. Note that there might be several FDS's describing the same program and vice versa.\\

A {\bf property}\index{Property} over the variables $V$ is a subset of the sequences in $\V(V)^\infty$. An LTL formula $\v \in \L(V)$ then expresses the property $\{\sigma \in \V(V)^\infty \mid \sigma \models \v \}$. Not every property can be expressed by an LTL formula, since the properties expressible by LTL are countable, but the set of all properties, $2^{\V(V)}$ is not.\footnote{Here $2^S$ denotes the {\bf powerset} of a set $S$.} Also, the expressivity of LTL is limited by the (bounded) nesting depth of the next operator~\cite{NestExpr}.\\

Since we are only concerned with properties expressible in LTL, we use the terms property and formula fairly interchangeably if the context is clear. Also, given a formula $\v$, the corresponding property can be found by forming the set $\{\sigma \in \V(V)^\infty \mid \sigma \models \v \}$.\\

In this and the following chapters, intervals of integers are denoted in the following way: $[a, b] = \{ x \in \mathbb{Z} \mid a \leq x \leq b\}$ and $[a, b) = \{ x \in \mathbb{Z} \mid a \leq x < b\}$ for integers $a$ and $b$ such that $a \leq b$. Also, the boolean domain $\{0, 1\}$ is denoted by $\mathbb{B}$.\\

Often, integer variables are used to represent data storage. To represent that a storage is empty, the integer $-1$ may be used. For better readability we use $\e$\index{e@$\e$} as a synonym for $-1$, and so the intervals $[\e, n)$ and $[\e, n]$ are the same as $[-1, n)$ and $[-1, n]$.\\

\section{Classification of Formulae}

Specifications for reactive systems typically consist of both safety and liveness formulae. This classification was first introduced by Lamport in 1977~\cite{Lamport}. Later, Manna and Pnueli gave an entire taxonomy in the form of a ``hierarchy'' to classify different formulae~\cite{TempHier}.\\

This classification also contains elaborate formulae such as assumption/guarantee (A/G) and GR[1] specifications, which are introduced in \sec{AG} and \sec{GR1} respectively. One important attribute ``orthogonal'' to the classification between safety and liveness is whether the sequences satisfying a formula are invariant under finite repetitions of elements, which is discussed in~\sec{stutter}.\\

\subsection{Safety and Liveness Properties}

Informally, a {\bf safety property}\index{Safety} states that ``nothing bad will ever happen.'' In the specification of control systems, safety properties ensure that constraints are never violated. A formula $\v_S$ expressing a safety property then states that all finite prefixes of a computation $\sigma$ satisfy some propositional formula $p$.\\


According to Sistla~\cite{Sistla}, $\v_S$ is a {\bf safety formula} iff all sequences $\sigma$ for which all finite prefixes can be extended to an infinite sequence satisfying $\v_S$, satisfy $\v_S$. Denoting the set of finite prefixes of a sequence $\sigma$ by $\pref{\sigma}$\index{pref@$\namefont{pref}$}, this statement can be written as
\begin{equation*}
	(\forall \sigma \in \V(V)^\infty \qdot \forall \bar{\bar{\sigma}} \in \pref{\sigma} \qdot \exists \bar{\sigma} \in \V(V)^\infty \qdot \bar{\bar{\sigma}} \bar{\sigma} \models \v_S) \Rightarrow \sigma \models \v_S.
\end{equation*}
Any safety formula can be written in the form $\always p$, where $p$ is a past formula.\\

Deciding whether a given set of sequences is a safety property is PSPACE-complete (\cite{Sistla}, Theorem 5.5). However, it has been shown that safety formulae are closed under $\wedge$, $\vee$, $\next$, $\until$ and $\always$ (\cite{Sistla}, Theorem 3.1). Of course, then safety is also preserved under weak until $\weakuntil$, but not necessarily under strong next $\strongnext$ for finite sequences. These syntactic properties provide an efficient way to decide whether a given formula expresses a safety property.\\

A {\bf liveness property}\index{Liveness} states that ``something good will eventually happen.'' Therefore, a {{\bf liveness formula} $\v_L$ expressing a liveness property requires that every finite squence $\sigma$ can be extended to satisfy $\v_L$. That is, if $\sigma \nmodels \v_L$ then there exists a possibly infinite sequence $\bar{\sigma}$ so that the concatenation of the two sequences $\sigma\bar{\sigma}$ satisfies $\v_L$. Formally, $\v_L$ is a liveness formula iff
\begin{equation*}
	\forall \sigma \in \V(V)^+ \qdot (\sigma \nmodels \v_L \Rightarrow \exists \bar{\sigma} \in \V(V)^\infty \qdot \sigma\bar{\sigma} \models \v_L).
\end{equation*}

Liveness formulae are necessary in the specification of reactive control systems\index{Reactive System} in order to ensure that the controller maintains an ongoing interaction with its environment. If no liveness formula is included in a specification then it can be satisfied by sequences consisting only of the initial state (or a finite number of states.) This becomes an issue when synthesizing an FDS from such a specification not containing a liveness formula, since the synthesis algorithm will try to generate the simplest possible FDS, which does not need to be nonblocking in this case.\\


\subsection{Derived Classes} \label{sec:derprop}

Several important and widely-used classes can be derived from the safety and liveness classification~\cite{TempHier,RecedingControl}. We present these classes using past formulae $p$ and $q$. However, since we only consider future formulae for synthesis, $p$ and $q$ are propositional formulae (which are both past and future formulae).\\

A {\bf guarantee} formula is of the form $\eventually p$. It is therefore a liveness formula and expresses that the system must eventually satisfy the formula $p$, e.g. termination in a total correctness specification of a program.\\

An {\bf obligation} formula is a combination of safety and liveness formulae of the form $\always p \vee \eventually q$. Such formulae can specify the behavior in exceptional circumstances. For example, $\always \neg p \vee \eventually(p \leftrightarrow q)$ ensures that either the exception $p$ never happens and if it does, it is handled correctly by $q$.\\

The class of obligation formulae strictly contains the safety and liveness formulae and so a general obligation formula cannot be specified by either a safety or a liveness formula. Every boolean combination of safety and guarantee formulae is an obligation formula.\\

A {\bf progress}\index{Progress formula} or {\bf recurrence} formula is of the form $\always\eventually p$ and expresses that $p$ has to be satisfied at infinitely many positions. If $p$ is a propositional formula, then a progress formula is a liveness formula. However, if $p$ is allowed to be any past formula, then the class of progress properties is strictly greater than that of obligation, safety or liveness properties.\\

A {\bf response}\index{Response formula} formula is of the form $\always (p \rightarrow \eventually q)$ and ensures that the formula $q$ will hold at some point in the future whenever $p$ is satisfied. Response formulae are progress formulae, and if both $p$ and $q$ are propositional formulae then they are even liveness formulae, since they ensure that $q$ is always asserted as a reaction to $p$.\\

A {\bf stability} or {\bf persistence} formula ensures that once a proposition $p$ holds, it will never be violated in the future. Such a formula is of the form $\eventually\always p$. It can be understood as requiring the system to enter into a stable state characterized by $p$, and remaining there. The class of stability properties is strictly larger than the class of obligation, recurrence and persistence properties.\\

A {\bf reactivity} formula is of the form $\always \eventually p \vee \eventually \always q$ and is a combination of a progress and a stability formula. The class of reactivity properties is strictly larger than any of the abovementioned classes. An even larger class of properties is that of the {\bf general reactivity} properties, which contains sequences satisfying finite conjunctions of reactivity formulae. Thus a general reactivity formula is of the form $\bigwedge_{i=1}^n\left[\always\eventually p_i \vee \eventually\always q_i\right]$ where $p_i$ and $q_i$ are past formulae. Every formula specifiable in LTL can be expressed as a general reactivity formula~\cite{TempHier}.\\

The format of general reactivity formulae gives the idea for the classification of GR[1] specifications introduced below in \sec{GR1}. Since every LTL formula is expressible as a general reactivity formula, also GR[1] formulae are surprisingly expressive, even if the restrictions imposed on $p_i$ and $q_i$ are quite severe.\\

\subsection{Assumption/Guarantee Specifications} \label{sec:AG}

When synthesizing open controllers from LTL specifications, a two-player game is considered, in which the controller plays against an adversarial environment. The controller tries to win by satisfying its specification while it assumes that the environment always tries to falsify the specification with the moves it is allowed to make.\\

Since the environment is considered to be adverse\index{Adversary}, it may always do the worst possible move from the controller's point of view, open controllers\index{System!Open} cannot be expected to satisfy their specification in all environments. An open controller interacts with an environment it cannot influence, so the assumptions it has about the environment do not actually constrain the environment. The controller cannot prevent the environment from violating the assumptions but in this case it is allowed to also violate its goals.\\

The synthesis procedure only supports the synthesis of an FDS satisfying a single specification $\v$, without any explicit assumptions on the environment. However, since the environment and system variables play different roles in the synthesis algorithm, the specification $\v$ can be written in a format that makes explicit what is assumed of the environment and what is guaranteed by the controller.\\

Let $\v^e$ be the {\bf environment assumptions}\index{Environment assumption}, i.e.\ the formula describing exactly what the system may assume about the environment. Let $\v^s$ be the {\bf system goals}\index{System goals}, i.e.\ the specification that the controller is supposed to satisfy \emph{given} that the environment satisfies its assumptions. It is then sufficient for the synthesized FDS to satisfy the specification $\v^e \rightarrow \v^s$.\\

Such a specification is called an {\bf assumption/guarantee specification}\index{Specification!A/G} (A/G specification for brevity,) and was first introduced by Misra and Chandy in 1981~\cite{Networks}. The controller is implemented to satisfy $\v^s$ given $\v^e$, but has no knowledge about the actual implementation of the environment. It cannot assume that an actual FDS is controlling $\X$ according to $\v^e$. If there was such an FDS, $\v^e$ would never be violated, so it would not be necessary to even include it in the original specification. In contrast, the system only needs to satisfy $\v^s$ as long as the environment satisfies $\v^e$ and if $\v^e$ is violated, the system is allowed to exhibit any behavior.\\

A/G specifications are mainly used to specify open controllers that communicate with each other. The environment assumptions of components that communicate with each other can be thought of as specifying a communication protocol. A component assumes that its peers comply with the protocol, and only if this is the case is it required to also satisfy the protocol itself. If the environment violates the protocol, the component also is not required to be able to satisfy it.\\

\subsection{GR[1] Specifications} \label{sec:GR1}


A/G specifications are very useful in defining open reactive systems. However, since we are concerned with synthesizing FDS's from their specifications, the class of formulae that we can use is restricted by the capabilities of the synthesizer. Thus, even a realizable specification (i.e.\ one for which an FDS \emph{exists}) might not be synthesizable.\\

Synthesis of general reactivity formulae is 2EXPTIME-complete~\cite{HardSynth}, but there exist smaller classes of formulae that can still express a large class of useful properties but exhibit better computational complexity for synthesis. The class of A/G formulae supported by the JTLV\index{JTLV} synthesizer is called {\bf Generalized Reactivity[1]}\index{Specification!GR[1]} (GR[1]). Synthesis and checking realizability of such formulae can be done in polynomial time, see \sec{compsingle}. A GR[1] formula is of the form
\begin{equation*}
\v_{init}^e \wedge \v_t^e \wedge \v_g^e \rightarrow \v_{init}^s \wedge \v_t^s \wedge \v_g^s .
\end{equation*}
where the superscripts $e$ and $s$ stand for the environment assumptions and the system goals respectively. A GR[1] formula thus states that only if some assumptions on the environment $\v^e = \v_{init}^e \wedge \v_t^e \wedge \v_g^e$ are satisfied, then it is required that the controller satisfies the specification $\v^s = \v_{init}^s \wedge \v_t^s \wedge \v_g^s$. In particular, if the environment violates $\v^e$, then no restrictions on the controller apply. Moreover, once a safety property\index{Safety} is violated, it cannot be satisfied in any future state.\\

In the above definition, $\v_{init}^e$ and $\v_{init}^s$ are propositional formulae over $\X \cup \Y$ and characterize the initial states of the environment and the controller respectively. Synthesizing an FDS from a GR[1] specification will determine the initial condition $\Theta$, assuming that the environment initially satisfies $\v_{init}^e$. Since by the startup property \aref{SP} each FDS has an initial state satisfying $\Theta$, it is necessary to have $\models \Theta \rightarrow (\v^e_{init} \wedge \v^s_{init})$.\\

Further, $\v_t^e$ and $\v_t^s$ are safety formulae of the form $\bigwedge_{i\in I}\always B_i$, where $B_i$ is a boolean combination of atomic propositions $p$ over variables in $\X \cup \Y$ and of expressions of the form $\next p$, where $p$ is an atomic proposition over variables in $\X$ for $\v_i^e$ and over variables in $\X \cup \Y$ for $\v_i^s$. With this part of the GR[1] formula it is possible to constrain or force transitions of an FDS by using the next operator. Note however, that in the environment assumptions nothing can be said about the next values of the system variables $\Y$.\\

Lastly, $\v_g^e$ and $\v_g^s$ are liveness\index{Liveness} formulae of the form $\bigwedge_{i\in I}\always\eventually B_i$ where the $B_i$ are propositional formulae over variables in $\X \cup \Y$. These properties can be used to ensure that the controller is not idle as long as the environment is making transitions.\\

While GR[1] formulae might seem restrictive at first sight, they allow the expression of many useful properties of reactive systems. Let
\begin{align*}
\phi^s &= \bigwedge_{j\in J_1}\always p^s_{1,j} \wedge \bigwedge_{j\in J_2}\eventually p^s_{2,j} \wedge \bigwedge_{j\in J_3}(\always p^s_{3,j} \vee \eventually q^s_{3,j})\\
 &\wedge \bigwedge_{j\in J_4}\always\eventually p^s_{4,j} \wedge \bigwedge_{j\in J_5}\always (p^s_{5,j} \rightarrow \eventually q^s_{5,j}) \wedge \bigwedge_{j\in J_6}\eventually\always p^s_{6,j}
\end{align*}
be a formula composed of safety, guarantee, obligation, progress, response and stability formulae, where $p_{k,j}$, $q_{3,j}$ and $q_{5,j}$ are propositional formulae over the variables in $V$. Further, let
\begin{align*}
\phi^e &= \bigwedge_{i\in I_1}\always p^e_{1,i} \wedge \bigwedge_{i\in I_2}\always\eventually p^e_{2,i}
\end{align*}
be a formula composed of safety and progress properties, where $p_{k,i}$ are propositional formulae over the variables in $\X$, and let $\phi_{init}$ be a propositional formula over $V$. It has been shown that any formula of the form $(\phi_{init} \wedge \phi^e) \rightarrow \phi^s$ can be translated into a GR[1] formula (\cite{RecedingControl}, Proposition 1). For example, a response\index{Response formula} property $\always(p \rightarrow \eventually q)$ that has to be satisfied by the system can be translated into a GR[1] formula by introducing an auxiliary boolean variable $t$ and letting
\begin{align*}
\v^s_{init} &= \neg t \rightarrow (p \wedge \neg q)\\
\v^s_t &= \always(((\neg t \vee \next p) \wedge \next \neg q) \rightarrow \next \neg t)\\
            &\wedge \always(\neg t \wedge  \next t \rightarrow \next q)\\
            &\wedge \always(t \wedge \next \neg t \rightarrow \next p)\\
\v^s_g &= \always\eventually t.
\end{align*}
The variable $t$ is called a {\bf trigger}\index{Trigger}, because it triggers $q$ to be asserted eventually. The trigger is asserted as long as there is no obligation to assert $q$. Thus, when $t$ is lowered, the system must assert $q$. Initially, ``$t = \mathrm{True}$'' holds unless $p \wedge \neg q$, the only case in which there is an obligation to eventually raise $q$. The three conjuncts of $\v_t^s$ ensure that $t$ is only changed appropriately. $\always(((\neg t \vee \next p) \wedge \next \neg q) \rightarrow \next \neg t)$ ensures that $t$ is lowered when $p$ is asserted but $q$ is not, and that $t$ \emph{stays} low as long as ``$q = \mathrm{False}$'' holds. The formulae $\always(\neg t \wedge \next t \rightarrow \next q)$ and $\always(t \wedge \next \neg t \rightarrow \next p)$ ensure that the trigger is only raised when $q$ is high and lowered only when $p$ is high. Finally, the progress formula $\v_g^s$ ensures that the system will always eventually reach a state where it is no longer under an obligation to raise $q$.\\

\subsection{Stutter-Invariance} \label{sec:stutter}

Stuttering is a phenomenon connected with the repetition of states in the computations of an FDS. Two sequences $a = a_0a_1a_2\ldots$ and $b = b_0b_1b_2\ldots$ are said to be {\bf stutter-equivalent} if there exist two infinite sequences of strictly increasing indices $0 = i_0 < i_1 < i_2 < \cdots$ and $0 = j_0 < j_1 < j_2 < \cdots$ such that for every $k \geq 0$, the sequences
\begin{align*}
	&a_{i_k}a_{i_k+1}\ldots a_{i_{k+1}-1}\\
	\mathrm{and}\hspace{0.5cm}&b_{j_k}b_{j_k+1}\ldots b_{j_{k+1}-1}
\end{align*}
are identical~\cite{StutterNext}. An LTL formula $\v$ is {\bf stutter-invariant}\index{Stutter-Invariance} if for any two stutter-equivalent sequences $a$ and $b$, either both $a \models \v$ and $b \models \v$ or both $a \nmodels \v$ or $b \nmodels \v$.\footnote{That is, $\v$ is satisfied by the union of stutter-equivalence classes.} Informally, if a sequence satisfies a stutter-invariant formula, the same sequence with any element duplicated any number of times also satisfies the same formula.\\


Stutter-invariance can be an important feature of specifications of an FDS. In a clock-triggered FDS\index{Fair Discrete System!Clock-triggered}, at each clock tick the values of the environment variables are read, and the next state is selected so that the environment variables have exactly these values. However, it might be the case that the values of the environment variables have not changed from one clock tick to the next. This is called {\bf environment stuttering}.\\

In an asynchronous composition of several components, it is possible that one component proceeds more than once while all its peers as well as its environment stay in the same state. The inputs are observed to be repeated. When the components are executed on different processors or as different threads on the same processor, this situation occurs naturally from thread scheduling, different clock speeds and clock rate drift~\cite{Kopetz}.\\

A clock-triggered FDS must therefore be invariant to environment stuttering, requiring the environment assumptions to be stutter-invariant. However, even if the system goals are stutter-invariant, the synthesis algorithm always tries to minimize {\bf system stuttering}, since this would increase the state space.\\

Peled and Wilke have shown that stutter-invariant properties in (future) LTL are exactly those that can be expressed without the next operator (i.e.\ the only allowed temporal operator is the until operator) and that stutter-invariance is decidable (\cite{StutterNext}, Lemma 2 and Corollary 6). Moreover, PSPACE methods exist for determining stutter invariance~\cite{StutterCheck}.\\

To obtain stutter-invariant environment assumptions, it is therefore sufficient to avoid the next operator. However, the  synthesis method used in this thesis only supports the temporal operators always, eventually and next. Since the until operator is not supported, it is sometimes necessary to express formulae using the next operator. However, careful construction of these formulae might still render them \emph{expressible} without the next operator.\\

For example, consider the formula
\begin{equation*}
    \always(x \rightarrow \next(\neg x \rightarrow y))
\end{equation*}
expressing that $x$ may only be cleared if $y$ is asserted in the next state. This is a synthesizable property as it can be included in either $\v_t^e$ or $\v_t^s$ in a GR[1] specification. Moreover, it can be expressed without using the next operator\footnote{Note that $\always\v$ is equivalent to $\neg (\mathrm{True}\until\neg\v)$ and therefore expressible without using $\next$.} by\index{U@$\weakuntil$}\index{U@$\until$}
\begin{align*}
    \always(x \rightarrow x\weakuntil y) \Leftrightarrow \always(x \rightarrow (x\until y \vee \always x)),
\end{align*}
and is therefore stutter-invariant.\\

\subsection{Specification Example}

In order to illustrate the specification concepts above, a toy example is presented. A GR[1] specification is given that is synthesized by TuLiP\index{TuLiP} into an FDS.\\

The controller has one boolean environment variable $x$ and two boolean system variables $y_1$ and $y_2$. First the guarantee part of the specification is given. The system should start with both $y_1$ and $y_2$ cleared and then respond to an input $x$ by raising $y_1$:
\begin{equation*}
	\always(x \rightarrow \eventually y_1).
\end{equation*}
This can be translated into a GR[1] formula by the method shown in \sec{GR1}. Note that this will introduce another boolean system variable, called a {\bf trigger} that we denote by $t$ in this example. We demand that $y_1$ may only be raised if $y_2$ is high. This is expressed by the guarantee 
\begin{equation*}
	\always(\neg y_1 \wedge \next y_1 \rightarrow y_2). \\
\end{equation*}
To prevent $y_2$ from being raised arbitrarily, $x$ is required to be cleared as a prerequisite:
\begin{equation*}
	\always(\neg y_2 \wedge \next y_2 \rightarrow \neg x). \\
\end{equation*}

This specification consisting of just these three guarantees is unrealizable. This is because $x$ might be asserted initially, requiring $y_1$ to be raised eventually. But if the environment never clears $x$, then $y_2$ cannot be raised, which is required for $y_1$ to fulfill the response\index{Response formula} formula.\\

In order to make the specification realizable, an assumption on the environment has to be included. Assuming that $x$ is always cleared eventually turns out to be sufficiently strong. It is expressed by the progress\index{Progress formula} formula
\begin{equation*}
	\always\eventually \neg x. \\
\end{equation*}
The symmetric assumption that $x$ will always be asserted eventually is included as well, because the environment should never just keep $x$ low at all times:
\begin{equation*}
	\always\eventually x. \\
\end{equation*}

Thus, the complete GR[1] specification is
\begin{align*}	
\v = (&\always\eventually x\,\wedge\\
	&\always\eventually \neg x) \\
	&\rightarrow \\
	(&\neg y_1 \wedge \neg y_2\,\wedge\\
	&\always(\neg y_1 \wedge \next y_1 \rightarrow y_2)\,\wedge\\
	&\always(\neg y_2 \wedge \next y_2 \rightarrow \neg x)\,\wedge\\
	&\always(x \rightarrow \eventually y_1))
\end{align*}

This specification is synthesizable and gives rise to an FDS $\M_\v$ with seven states, shown in \fig{exafds}. Recall that the extra system variable $t$ has been introduced as a trigger to express the response property as a GR[1] formula.\\

\newcommand{\cellE}[1]{*++++[o][F-]{#1}}
\newcommand{\cellEa}{\cellE{\begin{array}{c}\boxed{q_0}\\\neg y_1, x, \\ \neg y_2, \neg t \end{array}}}
\newcommand{\cellEb}{\cellE{\begin{array}{c}\boxed{q_1}\\\neg y_1, \neg x, \\ \neg y_2, \neg t\end{array}}}
\newcommand{\cellEc}{\cellE{\begin{array}{c}\boxed{q_2}\\\neg y_1, \neg x, \\ y_2, \neg t\end{array}}}
\newcommand{\cellEd}{\cellE{\begin{array}{c}\boxed{q_3}\\\neg y_1, x, \\ y_2, \neg t\end{array}}}
\newcommand{\cellEe}{\cellE{\begin{array}{c}\boxed{q_4}\\y_1, \neg x, \\ \neg y_2, t\end{array}}}
\newcommand{\cellEf}{\cellE{\begin{array}{c}\boxed{q_5}\\\neg y_1, \neg x, \\ \neg y_2, t\end{array}}}
\newcommand{\cellEg}{\cellE{\begin{array}{c}\boxed{q_6}\\\phantom{\neg}y_1, x,\\ \neg y_2, t\phantom{\neg}\end{array}}}
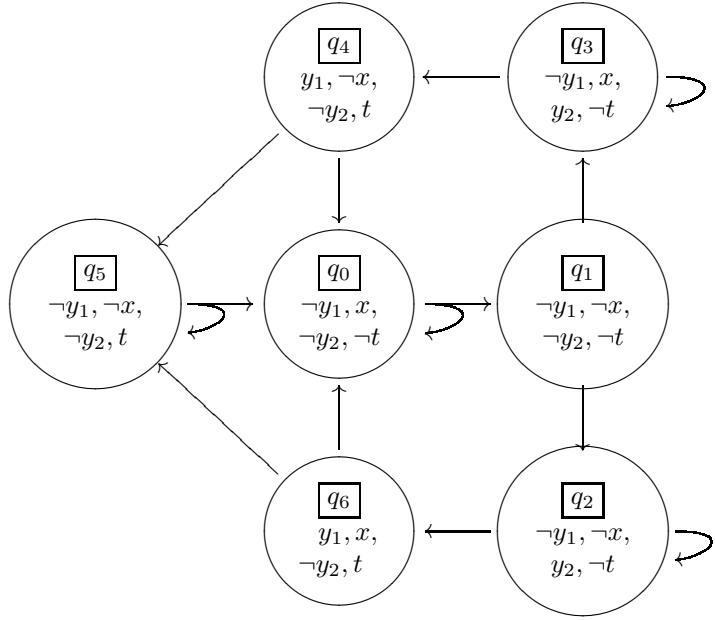
\begin{figure}
\begin{displaymath}
\SelectTips{cm}{}
    \xymatrix { {} & \cellEe \ar[d] \ar[dl] & \cellEd \ar[l] \save !R(1) \ar@(r,dr)[] \restore \\
    			\cellEf \ar[r] \save !R(1) \ar@(r,dr)[] \restore & \cellEa \save !R(1) \ar@(r,dr)[] \restore \ar[r]  & \cellEb \ar[d] \ar[u]\\
    			{} & \cellEg \ar[u] \ar[ul] & \cellEc \ar[l] \save !R(1) \ar@(r,dr)[] \restore }
\end{displaymath}
\caption{The FDS $\M_\v$ synthesized from $\v$. States referred to in the text are boxed, and the valuations of the system and environment variables are given below.}
\label{fig:exafds}
\end{figure}

The initial states of $\M_\v$ are all those for which $\neg y_1$, $\neg y_2$ and $\neg t$ hold (compare with the translation of the response formula into a GR[1] formula). The restrictions on the rising edges of $y_1$ and $y_2$ can be verified by tracing the corresponding arrows in the FDS. Thus $\always(\neg y_1 \wedge \next y_1 \rightarrow y_2)$ is verified by noting that $q_1 \trans  q_2$ and $q_1 \trans q_3$ are the only transitions for which $\neg y_1 \wedge \next y_1$ holds and that ``$y_2 = \mathrm{True}$ '' holds in $q_0$. $\always(\neg y_1 \wedge \next y_1 \rightarrow y_2)$ can be similarly verified.\\

The response property $\always(x \rightarrow \eventually y_1)$ is also satisfied, but only provided that both the environment assumptions $\always\eventually x$ and $\always\eventually \neg x$ hold.\footnote{The guarantees are realizable given only $\always\eventually\neg x$, but the synthesis procedure also makes use of $\always\eventually x$ to simplify $\M_\v$.} To see this, consider all states in which ``$x = \mathrm{True}$'' holds: In $q_6$, ``$y_1 = \mathrm{True}$'' holds, so the property is immediately satisfied. In $q_3$, there is a transition to $q_4$ which will satisfy the property. But there is also a self-transition that keeps ``$y_1 = \mathrm{False}$''. However, since by $\always\eventually \neg x$ the environment has to clear $x$ eventually, this self-transition cannot be taken infinitely many times, and so $q_4$ must be entered eventually. In $q_0$, the same assumption $\always\eventually \neg x$ forces $\M_\v$ to eventually enter $q_1$, but there ``$y_1 = \mathrm{False}$'' still holds. However, using both $\always\eventually \neg x$ and $\always\eventually x$, $\M_\v$ must eventually enter either $q_4$ or $q_6$, both satisfying ``$y_1 = \mathrm{True}$'' and thus the response property holds.\\

\section{Specifying Distributed Systems} \label{sec:distrsped}

As explained in \sec{distrsynth}, it is often necessary to synthesize a set of reactive systems that \emph{together} satisfy a given global specification. Compositional synthesis requires both a global specification and an architecture. It has already been mentioned that compositional synthesis for almost all architectures is undecidable. In this section we present principles that facilitate designing synthesizable\index{Synthesizability} local specifications and reasoning about the composition of the corresponding components in the architecture in order to establish that the global specification is met.\\

We consider predecomposition\index{Decomposition!Predecomposition}, i.e.\ decomposing the global specification into several local specifications that satisfy the architecture \emph{before} passing each of them to the synthesizer separately. The main challenge is to perform this decomposition in such a way that the asynchronous composition of the FDS's synthesized from the local specifications satisfies the global specification. It is therefore necessary to have a mathematically sound way of reasoning about the asynchronous composition of the components in an architecture. Neglecting accuracy at this stage would render the efforts in synthesizing correct-by-construction controllers useless.\\

Several attempts have been made at providing tools for the decomposition and reasoning about the composition of specifications. Misra and Chandy were among the first to reason about the composition of A/G specifications~\cite{Networks}. Abadi and Lamport provide a formalism for reasoning about decomposing and composing A/G specifications in TLA~\cite{Composing,Conjoining}. In response to this, a purely syntactic formalism for LTL specifications was developed by Jonsson and Tsay~\cite{AGSpec}, providing a stronger composition rule. The work of Ozay et al.\ provides examples of decomposing GR[1] specifications and how to reason about their compositions~\cite{AsyncComp,PTZ}.\\

\subsection{Composing Specifications} \label{sec:compspec}

Before principles for decomposing specifications can be stated, it is necessary to introduce the necessary framework to reason about their composition. We present the preconditions that have to be satisfied for a set of local specifications to satisfy a global specification. For notational convenience, these conditions are stated only for two components, but the extension to more components is straightforward.\\

We present the Feedback Interconnection Refinement Rule\index{Feedback Interconnection Refinement Rule} stated by Ozay et al.\ \cite{AsyncComp}. It is a specific case of the Composition Theorem proved by Abadi and Lamport~\cite{Conjoining}. We employ this version because it uses LTL and GR[1] specifications, although it is not the most general composition rule available. The proof of Abadi and Lamport's theorem covers several pages but the simplified version by Ozay et al. is straightforward to prove by induction on the length of the computations of the relevant FDS's (\cite{AsyncComp}, see the proof of Proposition 3). However, they do not take into account that the environment assumptions have to be stutter-invariant\index{Stutter-Invariance} for the rule to hold for an asynchronous composition\index{Asynchronous composition}. The Composition Theorem by Abadi and Lamport implicitly includes this assumption, since every property expressible in TLA is stutter-invariant.\\

\subsubsection{Architecture}

Let $\X$ and $\Y$ be the global environment and system variables respectively, i.e.\ the variables that occur in the global specification $\v$. These sets are necessarily disjoint, because the environment variables $\X$ cannot be controlled by any component which controls $\Y$.\\

The local environment and system variables of the synthesized components $\M_1$ and $\M_2$ are then subsets of $\X \cup \Y$ so that even feedback between the components can be introduced (see \fig{asynccomp}). Since all global system variables\index{Global system variables} $\Y$ must be controlled by exactly one component, $\Y$ is partitioned over the two components\index{Component}. Thus, the system variables of $\M_1$ are $\Y_1 \subseteq \Y$, and the system variables of $\M_2$ are $\Y_2 = \Y \backslash \Y_1$. \\

The environment variables of $\M_1$ can consist of variables from the global environment, but also from the system variables of $\M_2$. Thus $\X_1 \subseteq \X \cup \Y_2$. Symmetrically, $\X_2 \subseteq \X \cup \Y_1$. In \fig{asynccomp}, we would have $\E_1 = \X_1 \backslash \Y_2$ and $\T_{2,1} = \X_1 \cap \Y_2$ and the symmetrical relations $\E_2 = \X_2 \backslash \Y_1$ and $\T_{1,2} = \X_2 \cap \Y_1$.\\

This choice of variables defines the architecture for which we now develop the specification format and the hypotheses for the composition rule.\\

\subsubsection{Specification Format}

Suppose that a global specification\index{Global specification} $\v^e \rightarrow \v^s$ is a GR[1]\index{Specification!GR[1]} formula, where $\v^e \in \L(\X)$ and $\v^s \in \L(\X \cup \Y)$. This is the specification that must be satisfied by the asynchronous composition of $\M_1$ and $\M_2$.\\

Let $\v_1^e \in \L(\E_1)$ and $\v_2^e \in \L(\E_2)$ be stutter-invariant\index{Stutter-Invariance} LTL formulae that will be part of the environment assumptions\index{Environment assumptions} of $\M_1$ and $\M_2$ respectively. One of the conditions that these two formulae must satisfy is that any sequence $\sigma$ that satisfies the global environment assumptions also satisfies both $\v_1^e$ and $\v_2^e$. Thus
\begin{itemize}
	\item [\alabel{P1}] $\forall \sigma \in \V(\X)^\infty \qdot \sigma \models \v^e \Rightarrow \sigma \models \v_1^e \wedge \v_2^e$.
\end{itemize}
Note that $\wedge$ distributes over $\models$, so when verifying this property it can alternatively be shown that $\models \v^e \rightarrow \v_1^e$ and $\models \v^e \rightarrow \v_2^e$ hold individually.\\

In an asynchronous composition, the transitions of the components $\M_1$ and $\M_2$ may be triggered by two independent clocks and thus may react to changes in environment variables at different times. In particular, the transitions are not synchronized to the environment's changes of the variables it controls. It is therefore necessary for $\v_1^e$ and $\v_2^e$ to be stutter invariant so that \aref{P1} is sufficient in an asynchronous composition of $\M_1$ and $\M_2$.\\

The specifications $\v_1^e$ and $\v_2^e$ only allow to assume properties over the global environment variables $\X$ and not over the other component in the composition. We therefore introduce the safety formulae $\phi_1^e \in \L(\X_1)$ and $\phi_2^e \in \L(\X_2)$, called {\bf refinement formulae}\index{Refinement formula}. They are added to the local specifications and so the environment assumptions of $\M_1$ and $\M_2$ become $\phi_1^e \wedge \v_1^e \in \L(\X_1)$ and $\phi_2^e \wedge \v_2^e  \in \L(\X_2)$ respectively.\\

The intuition behind separating these assumptions is that the refinements $\phi_1^e$ and $\phi_2^e$ can also contain variables from $\Y_2$ and $\Y_1$ respectively, while $\v_1^e$ and $\v_2^e$ can only contain variables from $\X_1 \backslash \Y_2$ and $ \X_2 \backslash \Y_1$ respectively. The refinements will thus make environment assumptions about a component's peers and will therefore somehow have to be satisfied by the system goals\index{System goals} of the peers\index{Peer components}.\\

Let $\v_1^s \in \L(\X_1 \cup \Y_1)$ and $\v_2^s \in \L(\X_2 \cup \Y_2)$ be LTL formulae for the local guarantees. When these two specifications are satisfied by $\M_1$ and $\M_2$ respectively, they should together imply the global guarantee, that is
\begin{itemize}
	\item [\alabel{P2}] $\forall \sigma \in \V(V)^\infty \qdot \sigma \models (\v_1^s \wedge \v_2^s) \Rightarrow \sigma \models \v^s$.
\end{itemize}

We now also refine the guarantee part of the local specifications by the safety formulae $\phi_1^s \in \L(\X_1 \cup \Y_1)$ and $\phi_2^s \in \L(\X_2 \cup \Y_2)$. Thus the components $\M_1$ and $\M_2$ must satisfy their respective specifications:
\begin{itemize}
	\item [\alabel{P3}] $\M_1 \models \v_1^e \wedge \phi_1^e \rightarrow \v_1^s \wedge \phi_1^s$ and $\M_2 \models \v_2^e \wedge \phi_2^e \rightarrow \v_2^s \wedge \phi_2^s$
\end{itemize}

Note again that $\phi_1^e$, $\phi_2^e$, $\phi_1^s$ and $\phi_2^s$ are safety formulae and $\v_1^e$, $\phi_1^e$, $\v_2^e$ and $\phi_2^e$ are stutter-invariant.\\

\subsubsection{Circular Reasoning}\index{Circular reasoning}

Including the system variables of the peer component in the refinement formulae gives rise to circular dependencies: $\M_1$ only has to guarantee $\phi_1^s$ if its environment assumptions $\v_1^e \wedge \phi_1^e$ are satisfied. This requires $\M_2$ to satisfy $\phi_2^s$ which in turn only has to be guaranteed if $\v_2^e \wedge \phi_2^e$ is satisfied. The circularity comes from $\phi_2^e$ requiring $\phi_1^s$ to be satisfied by $\M_1$ (or symmetrically, $\phi_1^e$ requiring $\phi_2^s$ to be satisfied by $\M_2$).\\

In order to establish sufficient conditions on the local specifications\index{Local specification} so that their composition satisfies the global specification, it is necessary to resolve this circularity. This is done by imposing the appropriate conditions on the refinement formulae $\phi_1^e$, $\phi_2^e$, $\phi_1^s$ and $\phi_2^s$. We require that for any sequence $\sigma$ satisfying the global environment assumptions $\v^e$, if a prefix can be extended to satisfy the local guarantee part of one component, then the same prefix and the next element in $\sigma$ still must be able to be extended to satisfy the local environment assumptions of the other component.\\

It is important to require that the prefixes of sequences satisfying the global guarantees can be \emph{extended} to satisfy the local guarantees, because the components must be prevented from entering a ``one-way street'' in their computations.\footnote{This is comparable to entering a tree-subgraph of the graph-representation of the transition relation, in which the leaf-nodes have no out-edges. Once a leaf of the tree is reached, no further transitions are possible.} Consider for example a sequence $s_0s_1s_2\ldots = \sigma \models \v^e$ and a prefix $s_0s_1\dots s_n = \bar{\sigma} \in \pref{\sigma}$ of $\sigma$ that can be extended to satisfy $\phi_1^s$, the local guarantee refinement of the component $\M_1$. That means that $\M_1$ can make a finite number of transitions in order to satisfy $\phi_1^s$. If $\M_1$ is currently in state $q_n$ with $\V(q_n) = s_n$ and it makes a transition $q_n \trans_{\rho_1} q_{n+1}$ s.t.\  $\V(q_{n+1}) = s_{s+1}$ then the requirement that $\bar{\sigma}$ extended by the element $s_{n+1}$ in $\sigma$ can be extended to satisfy $\phi_1^s$ simply means that by choosing to make this transition, $\M_1$ actually makes such a transition in the ``right direction''.\\

Also, the assumption must hold for one further element in the sequence, because it allows to prove properties about the transition from one state to the next in a component without needing to assume anything about the next state of its peers. This resolves the circular reasoning. The resolution in the proof is based on induction, so for the base case the first element in $\sigma$ has to be able to be extended to satisfy the local guarantee parts of both components.\\

\subsubsection{Bad Prefixes and Closure}

The extensibility property introduced above can be formalized by defining the concept of bad prefixes, i.e.\ finite sequences that cannot be extended to satisfy a given formula.\\

\begin{definition}
A finite sequence of states $\bar{\sigma} \in \V(V)^+$ is a {\bf bad prefix}\index{Bad prefix} of an LTL formula $\v \in \L(V)$ iff for all infinite sequences $\sigma \in \V(V)^\infty$, the concatenation $\bar{\sigma}\sigma$ does not satisfy $\v$, i.e. $\bar{\sigma}\sigma \nmodels \v$. The set of bad prefixes of $\v$ is denoted by $\badpref{\v}$\index{Bad prefix!badpref@$\namefont{badpref}$}. Then for a sequence $\bar{\sigma} \in \V(V)^+$,
\begin{equation*}
	\bar{\sigma} \not\in \badpref{\v} \Leftrightarrow \exists \sigma \in \V(V)^\infty \qdot \sigma \models \v \wedge \bar{\sigma} \in \pref{\sigma}.
\end{equation*}
\end{definition}

We call a formula $\v_1$ {\bf stronger} than another formula $\v_2$ iff $\models \v_1 \rightarrow \v_2$.\\

\begin{definition}
For an LTL formula $\v$, its {\bf closure}\index{Closure}\index{C@$\CC$|see{Closure}} $\CC(\v)$ is the strongest safety formula implied by it. That is, $\CC(\v)$ is the strongest safety property s.t.\ $\models \v \rightarrow \CC(\v)$.
\end{definition}

The sequences that satisfy $\CC(\v)$ are exactly those sequences $\sigma$ s.t.\ each prefix of $\sigma$ is a prefix of some sequence satisfying $\v$. In other words, $\sigma$ satisfies $\CC(\v)$ iff any of its prefixes can be extended to an infinite sequence satisfying $\v$. Note that
\begin{equation*}
	\sigma \models \CC(\v) \Leftrightarrow \forall \bar{\sigma} \in \pref{\sigma} \qdot \bar{\sigma} \not\in \badpref{\v}.
\end{equation*}
Jonsson and Tsay give a syntactic definition of closure in LTL using past operators~\cite{AGSpec}:\index{ pastalways@$\pastalways$}\index{ pasteventually@$\pasteventually$}
\begin{equation*}
	\CC(\v) = \always\left[ \exists \bar{x} \in \V(\X) \qdot \pastalways(\bar{x} = x) \wedge \pasteventually(\v[\bar{x}/x]) \right],
\end{equation*}
where $\v[\bar{x}/x]$ is the formula $\v$ with the values $\bar{x}$ substituted for the variables $x$, and the existential quantifier $\exists$ over finite sets $\V(\X)$ is defined by
\begin{align*}
&(\sigma, j) \models \exists \bar{x} \in \V(\X) \qdot \v & \Leftrightarrow &\hspace{0.5cm} (\sigma, j) \models \bigvee_{\bar{x} \in \V(\X)} \v[\bar{x} / x]. \\
\end{align*}
The existential quantification in the definition of the closure is over \emph{all} valuations of the variables in the environment variables $\X$, or in other words, the {\bf free}\index{Free variables} variables. This definition is used to prove the four-phase handshake protocol which is introduced below in \sec{4PHP}.\\

Using the definition of bad prefixes, circular reasoning\index{Circular reasoning} can be resolved if we impose the following restrictions on the refinement formulae:
\begin{itemize}
	\item [\alabel{P4}] For all sequences $s_0s_1s_2\ldots \in \V(\X \cup \Y)^\infty$ that satisfy the global environment assumptions $\v^e$, for any index $n \in \mathbb{N}_0$, if the prefix $s_0s_1\ldots s_n \not\in \badpref{\phi_1^s}$ then also the prefix $s_0s_1\ldots s_ns_{n+1} \not\in \badpref{\phi_2^e}$; similarly if the prefix $s_0s_1\ldots s_n \not\in \badpref{\phi_2^s}$ then also the prefix $s_0s_1\ldots s_ns_{n+1} \not\in \badpref{\phi_1^e}$; moreover, $s_0 \not\in \badpref{\phi_1^s}$ and $s_0 \not\in \badpref{\phi_2^s}$.
\end{itemize}

\subsubsection{Feedback Interconnection Refinement Rule}

We can now state the Feedback Interconnection Refinement Rule\index{Feedback Interconnection Refinement Rule}:\\

{\bf Proposition 1}. (\cite{AsyncComp}, Proposition 3) Given the architecture and the restrictions on the specifications outlined above, it holds that
\begin{eqnarray*}
	\frac{(P3)\left\{\begin{array}{l}\M_1 \models \v_1^e \wedge \phi_2^e \rightarrow \v_1^s \wedge \phi_1^s \\ \M_2 \models \v_2^e \wedge \phi_1^e \rightarrow \v_2^s \wedge \phi_2^s\end{array}\right.}{\M_1 \comp \M_2 \models \v^e \rightarrow \v^s} & \mathrm{given~(P1), (P2), (P4)}.
\end{eqnarray*}

{\bf Proof sketch}. Take a computation $\sigma \in \seq{\M_1 \comp \M_2}$ of $\M_1 \comp \M_2$. If $\sigma \nmodels \v^e$, then the specification $\v^e \rightarrow \v^s$ is automatically met. So we assume that the environment assumptions $\v^e$ are satisfied. Then, by \aref{P1}, $\sigma$ also satisfies both local environment assumptions $\v_1^e$ and $\v_2^e$. Now \aref{P4} ensures that the environment refinement formulae $\phi_1^e$ and $\phi_2^e$ are satisfied for the initial state and both components can proceed infinitely from this initial state (because of the no bad prefix property). Therefore, for the initial state, $\v_1^e \wedge \phi_1^e$ and $\v_2^e \wedge \phi_2^e$ are satisfied. By \aref{P3}, the local guarantees $\v_1^s \wedge \phi_1^s$ and $\v_1^s \wedge \phi_1^s$ are satisfied for the initial state as well. Then \aref{P2} allows to conclude that $\v^s$ is satisfied by the initial state.\\

Because of \aref{P4}, the environment refinement formulae $\phi_1^e$ and $\phi_2^e$ are always satisfied for one further state than the system refinement formulae $\phi_1^s$ and $\phi_2^s$. Since $\v_1^e$ and $\v_2^e$ are satisfied by $\sigma$ anyway, $\sigma$ also satisfies $\v_1^e \wedge \phi_1^e$ and $\v_2^e \wedge \phi_2^e$, not just for the initial state. By a similar argument as for the initial state using \aref{P3} and \aref{P2}, $\sigma$ satisfies $\v^s$.\qed\\

\subsubsection{Verifying the Hypotheses}

In order to be able to use the Feedback Interconnection Refinement Rule, techniques are required to verify the hypotheses (P1)--(P4) that are required to deduce that the composition of the individually synthesized components satisfies the global specification.\\

The hypotheses give \emph{semantic} conditions on the GR[1] specifications, which are harder to verify than syntactic conditions that are used e.g.\ in the work of Jonsson and Tsay~\cite{AGSpec}. The latter approach is not adopted because it requires the introduction of past operators\index{Past operators} and existential quantification into the temporal logic, which are not supported by the synthesis procedure based on JTLV\index{JTLV}.\\

Given $\v_1^e$, $\v_2^e$, $\v_1^s$ and $\v_2^s$, the properties \aref{P1} and \aref{P2} can be derived syntactically by using sequent calculus\index{Sequent calculus} rules for LTL, see \sec{forward}. The property \aref{P3} is satisfied by construction from synthesis.\\

Verifying \aref{P4} in its full generality is more challenging. The simplest case is to just have $\models \v^e \wedge \phi_1^s \rightarrow \phi_2^e$ and $\models \v^e \wedge \phi_2^s \rightarrow \phi_1^e$, which can also be verified by sequent calculus.\\


\subsection{Communication} \label{sec:communication} \index{Communication}

For a coordinated action of the components to satisfy the global specification, it is often necessary to enable some form of information exchange~\cite{CommNeed}. One way to handle communication is proposed by Kloetzer and Belta~\cite{LTLPlanning,DistrImpl}, but their approach does not involve the encoding of the communication protocols directly in LTL, allowing  it to fit in the framework used in this thesis.\\

An approach more appropriate for our purposes is to define communication protocols directly in LTL~\cite{HWPSL,AgentComm}. The refinement of the local specifications by safety formulae that can contain system variables of the peer component allows some limited form of information exchange.\\

However, in order to obtain reliable communication, it is required to introduce handshake protocols between the components. Their specification requires system variables to occur in the environment assumptions as well, which is not handled by the composition rule presented above. The stronger composition rules by Abadi and Lamport~\cite{Conjoining} and Jonsson and Tsay~\cite{AGSpec} support this but have the disadvantage that they introduce operators into LTL that the synthesis does not handle.\\

We will therefore identify where the Feedback Interconnection Refinement Rule fails to hold when system variables are allowed in the local environment assumptions and resolve them for the particular case of the four phase handshake protocol that is also used by Bloem et al.\ \cite{HWPSL}.\\

\subsubsection{Glitches}

\newcommand{\ard}{\ar@{.}[d]}
\newcommand{\ardd}{\ar@{.}[dd]}
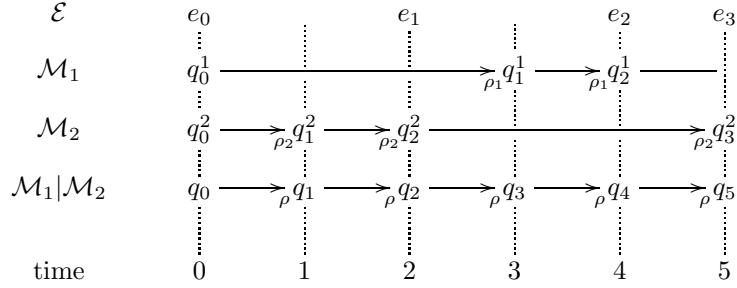
\begin{figure}
\begin{displaymath}
    \xymatrix@R=0.2cm{\E & e_0\ard & {}\ard & e_1\ard & {}\ard & e_2\ard & e_3\ardd \\
\M_1 & q_0^1\ard \ar[rrr]_>{\rho_1} & {}\ard & {}\ard & q_1^1\ard \ar[r]_>{\rho_1}& q_2^1\ard \ar@{-}[r] & {}\ard\\
\M_2 & q_0^2\ard \ar[r]_>{\rho_2} & q_1^2\ard \ar[r]_>{\rho_2} & q_2^2\ard \ar[rrr]_>{\rho_2} & {}\ard & {}\ard & q_3^2\ard \\
\M_1|\M_2 & q_0\ardd \ar[r]_>{\rho} & q_1\ardd \ar[r]_>{\rho} & q_2\ardd \ar[r]_>{\rho} & q_3\ardd \ar[r]_>{\rho} & q_4\ardd \ar[r]_>{\rho} & q_5\ardd \\ \\
\mathrm{time} & 0 & 1 & 2 & 3 & 4 & 5}
\end{displaymath}
\caption{Stuttering and Glitches}
\label{fig:stutglitch}
\end{figure}

The symmetric situation to stuttering is when one FDS misses one or several transitions of its peers or the environment. When viewing shared variables as communication lines, then this may cause information to be lost and environment assumptions to be violated, even though the environment behaves appropriately. This is illustrated in \fig{stutglitch}. The FDS $\M_1$ is started in $q_0^1$ while the environment is in state $e_0$. The transition of $\M_1$ to $q_1^1$ follows after only one transition of the environment to $e_1$, but $\M_1$ observes $\M_2$ to directly transition from $q_0^2$ to $q_2^2$. This could cause problems if $q_1^2$ needs to be observed in order to not violate the environment assumptions of $\M_1$. Similarly, $\M_2$ observes the environment to directly transition from $e_1$ to $e_3$.\\

These problems occur when the environment assumptions contain the next and the eventually operator. In particular, environment assumptions like
\begin{equation*}
    \always(x \rightarrow \next y),
\end{equation*}
where $x$ and $y$ are system and environment variables respectively, are not permissible. No environment can guarantee that it will have asserted $y$ the next time an FDS is observing it. This is because there can be a delay between $x$ being asserted and this information reaching the environment. The above formula is neither stutter nor glitch invariant. Also,
\begin{equation*}
    \always(x \rightarrow \eventually y)
\end{equation*}
is not glitch invariant, even though it is stutter invariant. The problem is that the environment may satisfy this assumption by asserting $y$ for only one step and then clearing it again. This might be missed by an FDS that makes this environment assumption.\\

It is therefore important to formulate specifications to be robust to such glitches. If a finite number of transitions are missed, correct operation must still be guaranteed.

\subsection{The Four-Phase Handshake Protocol} \label{sec:4PHP} \index{Communication!Protocol}

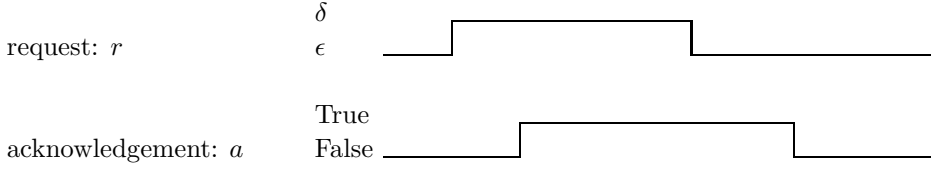
\begin{figure}
\setlength{\unitlength}{0.45cm}
\begin{centering}
\begin{picture}(23,6)
\put(0,0){acknowledgement: $a$}
\put(9,0){$\mathrm{False}$}
\put(9,1){$\mathrm{True}$}
\put(11,0){\line(1,0){4}}
\put(15,0){\line(0,1){1}}
\put(15,1){\line(1,0){8}}
\put(23,1){\line(0,-1){1}}
\put(23,0){\line(1,0){4}}

\put(0,3){request: $r$}
\put(9,3){$\e$}
\put(9,4){$\delta$}
\put(11,3){\line(1,0){2}}
\put(13,3){\line(0,1){1}}
\put(13,4){\line(1,0){7}}
\put(20,4){\line(0,-1){1}}
\put(20,3){\line(1,0){7}}
\end{picture}
\end{centering}
\caption{Four-Phase Handshake Protocol.}
\label{fig:4phs}
\end{figure}

Consider the case of transmitting an integer $\delta \in [0, n)$ from a sender $\M_S$ to a receiver $\M_R$. The sender controls a request signal $r \in [\e, n)$. If $r = \e$\index{e@$\e$}, this indicates that no data should be transmitted. If $r \neq \e$, then the data that is to be transmitted is $\delta = r$. The receiver controls a boolean acknowledgement signal $a$. The request signal $r$ appears as an environment variable $r \in [\e, n)$ at the receiver. Similarly, the acknowledgement signal $a$ appears as a boolean environment variable $a$ at the sender.\\

Thus, the architecture that contains the sender $\M_S$ and the receiver $\M_R$ can be visualized as in \fig{protocol}. Comparing this with the generic composite architecture in \fig{asynccomp} yields $\E_S = \{t\}$, $\S_S = \emptyset$, $\T_{R,S} = \{r\}$, $\T_{S,R} = \{a\}$, $\E_R = \emptyset$ and $\S_R = \{s\}$. Thus the composite FDS $\M_S \comp \M_R$ has the environment variables $\X = \{t\}$ and the system variables $\Y = \{a, r, s\}$. When using the protocol in a larger specification, the trigger $t$ is likely to also be a system variable of $\M_S$.\\

\begin{figure}
\begin{displaymath}
    \xymatrix@C=1.5cm{ {} \ar[r]^<{\{t\}} & *++++[o][F]{\M_S} \ar@<1ex>[r]|{\{r\}} & *++++[o][F]{\M_R} \ar@<1ex>[l]|{\{a\}} \ar[r]^>{\{s\}} & {}}
\end{displaymath}
\caption{Composition of the sender $\M_S$ and receiver $\M_R$.}
\label{fig:protocol}
\end{figure}
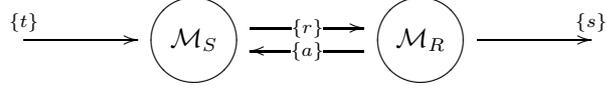

The four-phase handshake protocol can be informally described as follows. The sender $\M_S$ initiates a transfer by setting $r = \delta$. Once $\M_R$ detects this, it may use the transferred data somehow, and then raises $a$. When $\M_S$ sees that $a$ has been asserted, it resets $r = \e$. After $\M_R$ detects this, it clears $a$, completing the transmission of $\delta$. This is outlined in \fig{4phs}.\\

\subsubsection{Sender Specification}

The sender assumes one liveness and two safety properties about the receiver. It is assumed that if the request is reset, the acknowledgement will be cleared eventually, allowing for a new request to be made:
\begin{equation*}
	\mathrm{I}\hspace{0.5cm}\always(r = \e \rightarrow \eventually \neg a).
\end{equation*}

The assumed safety properties assert that the acknowledgement signal does not change in a way that is not allowed by the protocol. There is no acknowledgement without a request:
\begin{equation*}
	\mathrm{II}\hspace{0.5cm}\always(\neg a \wedge \next a \rightarrow r \neq \e).
\end{equation*}
Also, the acknowledgement may only be cleared if the request has been reset before:
\begin{equation*}
	\mathrm{III}\hspace{0.5cm}\always(a \wedge \next \neg a \rightarrow r = \e).
\end{equation*}

The sender guarantees three safety properties. No liveness property is guaranteed, because the sender may just remain idle. However, a system using this protocol will have to contain a liveness property that implicitly guarantees liveness in the protocol. Also, as long as the request is not acknowledged, it must stay at the same value:
\begin{equation*}
	\mathrm{VII}\hspace{0.5cm}\bigwedge_{i \in [0,n)} \always(\neg a \wedge r = i \rightarrow \next(r = i)).
\end{equation*}
The request may only be reset to $\e$:
\begin{equation*}
	\mathrm{VIII}\hspace{0.5cm}\bigwedge_{i \in [0,n)} \always(r = i \wedge \next(r \neq i) \rightarrow \next(r = \e)).
\end{equation*}
Lastly, the sender responds to an acknowledgement by resetting the request:
\begin{equation*}
	\mathrm{IX}\hspace{0.5cm}\always(a \rightarrow \next (r = \e)).
\end{equation*}
Note that this last guarantee also ensures that the sender may only initiate a transmission if the acknowledgement signal is low. This could be expressed by the additional guarantee $\always(r = \e \wedge \next(r \neq \e) \rightarrow \neg a)$, but clearly this is already implied by IX.\\

\subsubsection{Receiver Specification}

On the receiver side, one liveness and two safety assumptions are made. The receiver assumes that each acknowledgement eventually results in the request to be cleared:
\begin{equation*}
	\mathrm{IV}\hspace{0.5cm}\always(a \rightarrow \eventually (r = \e)).
\end{equation*}

It also assumes that the request reacts appropriately to the acknowledgement. Similar to guarantee VII, no transmission is initiated if the acknowledgement is still high:
\begin{equation*}
	\mathrm{V}\hspace{0.5cm}\always(r = \e \wedge \next(r \neq \e) \rightarrow \neg a).
\end{equation*}
Also, the request may only be cleared if it has been acknowledged:
\begin{equation*}
	\mathrm{VI}\hspace{0.5cm}\always(r \neq \e \wedge \next(r = \e) \rightarrow a).
\end{equation*}

The receiver guarantees that each request will be acknowledged:
\begin{equation*}
	\mathrm{XI}\hspace{0.5cm}\always(r \neq \e \rightarrow \next a).
\end{equation*}
And it guarantees that the resetting of the request results in the acknowledgement to be cleared:
\begin{equation*}
	\mathrm{XII}\hspace{0.5cm}\always(r = \e \rightarrow \next \neg a).
\end{equation*}
Moreover, the acknowledgement may only be raised on a request:
\begin{equation*}
	\mathrm{XIII}\hspace{0.5cm}\always(\neg a \wedge \next a \rightarrow r \neq \e).
\end{equation*}
Lastly, the acknowledgement may not be cleared unless the request is reset:
\begin{equation*}
	\mathrm{XIV}\hspace{0.5cm}\always(a \wedge \next \neg a \rightarrow r = \e).
\end{equation*}

\subsubsection{Reliability of the Four-Way Handshake Protocol}

It now remains to show that the four-phase protocol\index{Communication!Protocol} reliably transmits data from the sender to the receiver. To show this, we introduce a boolean variable $t$ into the environment variables of the sender $\M_S$, which will act as a ``trigger'' to a message being sent. The data will just be a single constant $\delta \in [0,n)$. Similarly, introduce an integer variable $s \in [0,n)$ into the system variables of the receiver, which will act as a ``sink'' where the sent data will be stored for later use. We add the response property
\begin{equation*}
	\mathrm{X}\hspace{0.5cm}\always(t \rightarrow \eventually(r=\delta))
\end{equation*}
to the guarantees of the sender, which ensures that the data will eventually be sent, once the transfer has been triggered by ``$t = \mathrm{True}$''. Similarly, the response property
\begin{equation*}
	\mathrm{XV}\hspace{0.5cm}\always(r=\delta \rightarrow \eventually(s=\delta))
\end{equation*}
will be added to the guarantees of the receiver. Note that while in XV the eventually operator might be substituted by a next operator, this cannot be done in X, since changing $r$ is restricted by $a$, which is not controlled by $\M_S$.\\

The full local specifications are shown in \fig{commspec}. The global specification is $\always(t \rightarrow \eventually(s = \delta))$, requiring that once a transmission is triggered by $t$ being asserted, the data $\delta$ eventually arrives correctly at the sink $s$. We thus want to prove the following:\\

{\bf Proposition 2}. If $\M_S \models \v_S^e \rightarrow \v_S^s$ and $\M_R \models \v_R^e \rightarrow \v_R^s$ then $\M_S \comp \M_R \models \always(t \rightarrow \eventually(s = \delta))$.\\

\begin{sidewaysfigure}
\begin{center}
\begin{equation*}
	\boxed{\begin{array}{l}
	\M_S \models \underbrace{\left(
	\begin{array}{ll}
	\mathrm{I} &\always(r = \e \rightarrow \eventually \neg a)\,\wedge\\
	\mathrm{II} &\always(\neg a \wedge \next a \rightarrow r \neq \e)\,\wedge\\
	\mathrm{III} &\always(a \wedge \next \neg a \rightarrow r = \e)
	\end{array} \right) }_{\v_S^e}
	\rightarrow \underbrace{\left(
	\begin{array}{ll}
	\mathrm{VII} &\always(r = \e \wedge \next(r \neq \e) \rightarrow \neg a)\,\wedge\\
	\mathrm{VII} &\bigwedge_{i \in [0,n)} \always(\neg a \wedge r = i \rightarrow \next(r = i))\,\wedge\\
	\mathrm{VIII} &\bigwedge_{i \in [0,n)} \always(r = i \wedge \next(r \neq i) \rightarrow \next(r = \e))\,\wedge\\
	\mathrm{IX} &\always(a \rightarrow \next (r = \e))\,\wedge\\
	\mathrm{X} &\always(t \rightarrow \eventually(r=\delta))
	\end{array} \right)}_{\v_S^s}\\[2cm]
	
	\M_R \models \underbrace{\left(
	\begin{array}{ll}
	\mathrm{IV} &\always(a \rightarrow \eventually (r = \e))\,\wedge\\
	\mathrm{V} &\always(r = \e \wedge \next(r \neq \e) \rightarrow \neg a)\,\wedge\\
	\mathrm{VI} &\always(r \neq \e \wedge \next(r = \e) \rightarrow a)
	\end{array} \right) }_{\v_R^e}
	\rightarrow \underbrace{\left(
	\begin{array}{ll}
	\mathrm{XI} &\always(r \neq \e \rightarrow \next a)\,\wedge\\
	\mathrm{XII} &\always(r = \e \rightarrow \next \neg a)\,\wedge\\
	\mathrm{XIII} &\always(\neg a \wedge \next a \rightarrow r \neq \e)\,\wedge\\
	\mathrm{XIV} &\always(a \wedge \next \neg a \rightarrow r = \e)\,\wedge\\
	\mathrm{XV} &\always(r=\delta \rightarrow \eventually(s=\delta))
	\end{array} \right) }_{\v_R^s}
	\end{array}}
\end{equation*}
\end{center}
\caption{Specification of the four-phase handshake protocol.}
\label{fig:commspec}
\end{sidewaysfigure}


{\bf Proof}. We make use of the Composition Theorem of Abadi and Lamport (\cite{Conjoining}, Theorem 3), which in our particular case is:
\begin{equation}
	\frac{\begin{array}{ll}(A1) &\models \CC(\v_S^s) \wedge \CC(\v_R^s) \rightarrow \v_S^e \wedge \v_R^e \\
	(A2) &\models \CC(\v_S^s) \wedge \CC(\v_R^s) \rightarrow \CC(\always(t \rightarrow \eventually(s = \delta))) \\
	(A3) &\models \v_S^s \wedge \v_R^s \rightarrow \always(t \rightarrow \eventually(s = \delta)) \end{array}}{(\M_S \models \v_S^e \rightarrow \v_S^s) \wedge (\M_R \models \v_R^e \rightarrow \v_R^s) \Rightarrow \M_1 \comp \M_2 \models \always(t \rightarrow \eventually(s = \delta))}
	\label{eq:compthm}
\end{equation}
Note that we only use a nesting depth of the next operator of one in the GR[1] specifications, so the explicit translation to TLA is omitted. For the closure\index{Closure} required for the Composition Theorem, the above LTL definition by Jonsson and Tsay will be used.\\

Precondition (A3) can be established by taking X $\always(t \rightarrow \eventually(r = \delta))$ and XV $\always(r = \delta \rightarrow \eventually (s = \delta))$. From these specifications it follows that $\always(t \rightarrow \eventually(s = \delta))$. This is intuitively clear, but can also established by a sequent calculus proof.\\

For the preconditions (A1) and (A2), consider the closure of $\v_S^e$, which is $\always(\exists \bar{t}, \bar{a} \qdot \pastalways(\bar{t} = t \wedge \bar{a} = a) \wedge \pasteventually (\v_S^e [\bar{t}/t][\bar{a}/a]))$ and the closure of $\v_R^e$, which is $\always(\exists \bar{r} \qdot \pastalways(\bar{r} = r) \wedge \pasteventually (\v_R^e [\bar{r}/r]))$. We consider the individual cases for values of $\bar{t}$, $\bar{a}$ and $\bar{r}$ to deal with the existential quantification, which, by definition, is just a shorthand for disjunction.\\

Note that $\models \always\pasteventually\always \v \leftrightarrow \always \v$  and $\models \always\pastalways \v \leftrightarrow \always \v$ for any LTL formula $\v$, which helps to simplify the closures of the formulae we are considering.\\

{\bf Case 1}. $\bar{t} = \mathrm{True}$, $\bar{a} = \mathrm{True}$:\\

Evaluating the conjuncts for these valuations yields $\always\next(r = \e)$ from IX and $\always\eventually(r=\delta)$ from X. These two formulae can never hold together and thus in this case the closure of $\v_S^s$ is simply false. As false implies everything, both (A1) and (A2) hold in this case.\\

{\bf Case 2}. $\bar{t} = \mathrm{True}$, $\bar{a} = \mathrm{False}$, $\bar{r} = \e$:\\

Here the conjunct resulting from X is $\always\eventually(r = \delta)$, but we also have $\always(r = \e)$ due to the existential quantification. Thus, again the antecedent is false and so both (A1) and (A2) are implied.\\

{\bf Case 3}. $\bar{t} = \mathrm{True}$, $\bar{a} = \mathrm{False}$, $\bar{r} = \delta$:\\

Evaluating the conjuncts in the closures for these variables yields the conjunct $\always\eventually(s = \delta)$ for XV. Moreover, the conjuncts $\always(t = \mathrm{True})$, $\always(a = \mathrm{False})$ and $\always(r = \delta)$ are in the closures from the existential quantification.\\

For (A1) we need to deduce I-VI from these conjuncts. I and VI follow from $\always(r = \delta)$ and II--V follow from $\always(a = \mathrm{False})$.\\

For (A2) we need to verify the individual terms of the closure of $\always(t \rightarrow \eventually(s = \delta))$, which is $\always(\exists \bar{\bar{t}} \qdot \pastalways(\bar{\bar{t}} = t) \wedge \pasteventually\always(\bar{\bar{t}} \rightarrow \eventually(s = \delta)))$.\\

In order to resolve the existential quantification, we again consider both truth values of $\bar{\bar{t}}$. However, since we are now proving the closure as a consequent, we only need to exploit one value of $\bar{\bar{t}}$ for which the closure of $\always(t \rightarrow \eventually(s = \delta))$ is implied by the antecedents.\\

Let $\bar{\bar{t}} = \mathrm{True}$. Then the closure becomes $\always\pastalways(t = \mathrm{True}) \wedge \always\pasteventually\always\eventually(s = \delta)$. The first conjunct follows simply from $\always (t = \mathrm{True})$ from the closure of $\v_S^s$. The second conjunct follows from XV, i.e.\ $\always\eventually(s = \delta)$.\\

{\bf Case 4}.  $\bar{t} = \mathrm{True}$, $\bar{r} \neq \e$ and $\bar{r} \neq \delta$:\\

Evaluating the conjuncts for these variables yields the conjuncts$\always(a \vee \next\neg a)$ from XIII and $\always(\neg a \vee \next a)$ from XIV. These two conjuncts already yield a contradiction, and so (A1) and (A2) are both satisfied.\\

{\bf Case 5}. $\bar{t} = \mathrm{False}$, $\bar{a} = \mathrm{True}$, $\bar{r} = \e$:\\

The conjunct in the closure resulting from XII is $\always\next\neg a$. This stands in contradiction with the conjunct $\always(a = \mathrm{True})$ from the existential quantification. Thus both (A1) and (A2) are satisfied.\\

{\bf Case 6}. $\bar{t} = \mathrm{False}$, $\bar{a} = \mathrm{True}$, $\bar{r} \neq \e$:\\

The conjunct in the closure resulting from X is $\always\next(r=\e)$, but the existential quantification requires $\always(r\neq\e)$. This leads to an immediate contradiction, so both (A1) and (A2) are satisfied.\\

{\bf Case 7}. $\bar{t} = \mathrm{False}$, $\bar{a} = \mathrm{False}$, $\bar{r} = \e$:\\

To prove (A1) we again need to be able to deduce I-VI from the conjuncts of the closure. I--IV follow from $\always(a = \mathrm{False})$ while V and VI follow from $\always(r = \e)$.\\

For (A2) consider again the closure $\CC(\always(t \rightarrow \eventually(s = \delta)) = \always(\exists \bar{\bar{t}} \qdot \pastalways(\bar{\bar{t}} = t) \wedge \pasteventually\always(\bar{\bar{t}} \rightarrow \eventually(s = \delta)))$. By taking $\bar{\bar{t}} = \mathrm{False}$, this closure is implied simply by $\always(t = \mathrm{False})$.\\

{\bf Case 8}. $\bar{t} = \mathrm{False}$, $\bar{a} = \mathrm{False}$, $\bar{r} = \delta$:\\

For the preconditions (A1) and (A2) the conjuncts from the existential quantification are sufficient: I--V follow from $\always(a = \mathrm{False})$ and VI follows from $\always(r = \delta)$. Moreover, in the same way as in Case 6, (A2) follows from $\always(t = \mathrm{False})$.\\

This covers all cases and so the preconditions (A1) and (A2) have been established. The result follows from the Composition Theorem \eqref{eq:compthm} stated at the beginning of this proof. \qed\\

Note that we had to use \eqref{eq:compthm}, a stronger result than the Feedback Interconnection Refinement Rule in order to be able to incorporate the system variables\index{System variables} in the environment assumptions\index{Environment assumption}, and to include liveness\index{Liveness} formulae in the specifications.\\

When using the four-phase handshake protocol\index{Communication!Protocol} in the form just proved several things should be noted. It is always possible to use weaker assumptions, as long as the specification remains synthesizable\index{Synthesizability}. In particular, omitting conjuncts in $\v_S^e$ or $\v_R^e$ does not change the correctness of the protocol. However, the guarantees $\v_S^s$ and $\v_R^s$ must be provided in a specification. They need not occur in the same syntactic form, but can be strengthened to an arbitrary degree without affecting correctness of the protocol.\\

\chapter{Searching} \label{sec:searching}


One of the main challenges in SAR is to locate the targets that need to be rescued. In order to avoid ambiguities, the physical environment of the robot, such as a building in USAR or a landscape in WiSAR, is referred to as {\bf topology}. The controllers are specified and synthesized with a two-player game in mind, which then can be thought of as taking place on a graph representing the topology. Assuming that a SAR robot can detect the presence of a target if it is in the same vertex of the graph-representation of the topology, the problem is to detect the vertex the target is in.\\ 

The solution depends on the capabilities of the robots and the targets, as well as the amount of information they have. Here we differentiate searching for a stationary and a moving target. When the target is stationary, it is sufficient to ensure that each vertex in the graph is eventually visited by at least one robot.\footnote{This can be extended to require any number of robots to visit the vertex at the same time.} When the target can move, we formulate a {\bf pursuit-evasion game}, in which an evader (the target) tries to avoid being captured by one or more pursuers (the robots). The controller specifications are developed from the winning strategies for the cops of this game.\\

\section{Preliminaries}

Before presenting the search strategies for the different scenarios, we make precise the concept of the graph representation of the topology and the way sensors are incorporated in our specification framework from \sec{specs}.\\

\subsection{Graph Representation of the Topology} \label{sec:maprepresentation} \index{Topology}

In order to reason about the movement of the robots in an SAR task, the topology must be represented internally in a way that admits LTL specifications about the positions of the robots. The topology is partitioned into discrete ``cells'' that are typically polygonal shapes. The robots move from one cell into another cell using continuous control, which is discussed e.g.\ in the work of Kloetzer and Belta~\cite{LTLPlanning}, Kress-Gazit et al.\ \cite{KressFainekos} and Topcu et al.\ \cite{RecedingControl}. After subdividing the topology into, say $n$ such cells $\{C_0, C_1, \ldots, C_{n-1}\}$, we are only concerned with controlling the discrete moves between cells, represented by changing the integer system variable $\cellID \in [0, n)$ of the robot, which thusly holds its position.\\

The changes of $\cellID$ are constrained by the topology, which can be represented by a directed graph $G = (V_G, E_G)$. Each vertex in $V_G = \{v_0, v_1, \ldots, v_{n-1}\}$ corresponds to a cell. The possible transitions between the cells are the edges $E_G$. An edge $e$ going from the vertex $u \in V_G$ to the vertex $v \in V_G$ is an ordered pair $(u, v) \in V_G \times V_G$. The edge $e = (u, v)$ is an {\bf in-edge} of $v$ and an {\bf out-edge} of $u$.\\

\begin{definition}
A {\bf path} is a sequence of vertices $p = v^0v^1\ldots v^{l-1}$ over $\V_G$ s.t.\  $(v^i, v^{i+1}) \in E_G$ is an edge for all indices $i \in [0, l-1)$. A {\bf simple path} is a path with unique vertices, i.e.\ it contains no cycles. The $i^\mathrm{th}$ element of the path $p$ is also denoted by $p(i) = v^i$, and we say that $p$ is a path {\bf from} $p(0)$ {\bf to} $p(l-1) = p(|p|-1)$.
\end{definition}

\begin{definition}
A graph $G = (V_G, E_G)$ is {\bf strongly connected}\index{Strongly connected graph} iff there exists a path from every vertex $v \in V_G$ to every other vertex $u \in V_G$. Denote the set of all strongly connected directed graphs with $n$ vertices by $\mathfrak{S}_n$\index{S@$\mathfrak{S}$|see{Strongly connected graph}}.
\label{def:strongconn}
\end{definition}

We require the graph $G$ representing the topology to be strongly connected, that is, it is possible to get from every vertex to any other vertex of the graph. If a graph is not strongly connected, there may be cells that the robots are not able to access, depending on their initial positions. Targets located in such inaccessible cells could never be found with any strategy.\\

\newcommand{\cellF}[1]{*++[o][F-]{v_{#1}}}
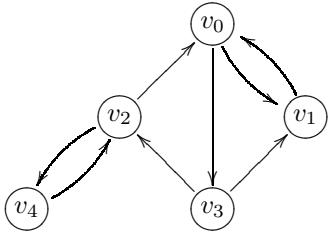
\begin{figure}
\begin{displaymath}
    \xymatrix@dr{ \cellF{0} \ar@/_/[r] \ar[dr] & \cellF{1} \ar@/_/[l] \\
    \cellF{2} \ar@/_/[d] \ar[u] & \cellF{3} \ar[u] \ar[l] \\
    \cellF{4} \ar@/_/[u]}
\end{displaymath}
\caption{Graph representing a topology with five cells. Each vertex stands for a cell and the arrows are the allowable moves between cells, which can be thought of e.g.\ as doors or corridors between rooms in a building. The graph implicitly contains a self-loop for each vertex.}
\label{fig:map}
\end{figure}

The constraints on $\cellID$ are incorporated in the guarantee part of a robot's GR[1] specification. Consider for example the graph shown in \fig{map}, which is strongly connected and therefore qualifies as a valid basis for synthesis. A robot is always allowed to stay at a vertex, so the self loops are included implicitly. For notational convenience, we use the boolean variables $X_i$ instead of explicitly writing $\cellID = i$, i.e.\ $X_i = \mathrm{True} \Leftrightarrow \cellID = i$. Further let $n = |V_G|$ be the number of vertices in $G$. Then, for the graph in \fig{map} the corresponding specification is
\begin{align*}
  \v_G = &\always(X_0 \rightarrow \next(X_0 \vee X_1 \vee X_3)) \wedge \\
       &\always(X_1 \rightarrow \next(X_0 \vee X_1)) \wedge \\
       &\always(X_2 \rightarrow \next(X_0 \vee X_2 \vee X_4)) \wedge \\
       &\always(X_3 \rightarrow \next(X_1 \vee X_2 \vee X_3)) \wedge \\
       &\always(X_4 \rightarrow \next(X_2 \vee X_4)) \wedge \\
       &\always(\bigvee_{h \in [0, n)} (X_h \wedge \bigwedge_{i \in [0, n) \backslash\{h\}} \neg X_i)),
\end{align*}
where the last line is a mutual exclusion formula, preventing the robot to think that it is in two cells at the same time. This ensures that $\always(\cellID = i \leftrightarrow X_i)$.\\

\subsection{Generating Random Strongly Connected Graphs} \label{sec:graphgen}

The methods developed in this chapter are tested on randomly generated graphs. We use the Erd\H{o}s-R\'enyi Model in order to generate random directed graphs that are strongly connected~\cite{Erdos}.\\

For this purpose, we represent a graph $G = (V_G, E_G)$ with $n$ vertices $V_G = \{v_0, v_1, \ldots, v_{n-1}\}$ by its adjacency matrix $\Adj{G} \in \mathbb{Z}^n \times \mathbb{Z}^n$, where the $i,j^\mathrm{th}$ entry of $\Adj{G}$ is nonzero exactly if there is an edge $(v_i, v_j) \in E_G$ from $v_i$ to $v_j$. We require that a graph contains all self-loops, so the diagonal of $\Adj{G}$ for any graph $G$ that we are considering contains only nonzero elements. Since any nonzero entry in $\Adj{G}$ represents an edge, the adjacency matric can also be used to store state information of edges by using different nonzero values. This is used for example for the moving target search, see \sec{graphstate}.\\

To generate a random graph $G$ with $n$ vertices, we select a constant $P$ such that
\begin{equation*}
	P > \frac{1}{n-1}.
\end{equation*}
Further, let $\mathbf{X}$ be a continuous random variable that is uniformly distributed over the range $\{ x \mid 0 \leq x < 1\} \subseteq \mathbb{R}$. Next, we draw one sample for each pair of vertices from $\mathbf{X}$, yielding the $n^2$ samples $e_{1,1}, e_{1,2}, \ldots, e_{1,n}, e_{2,1}, e_{2,2}, \ldots, e_{n,n}$. For all $i$ and $j$, if $e_{i,j} < P$ then let the $i,j^\mathrm{th}$ element of $\Adj{G}$ be one and zero otherwise. Thus, a matrix of ones and zeros is generated that defines the graph $G$.\\

With this choice of $P$, the resulting graph $G$ almost surely contains one ``giant'' component, i.e.\ a subset of vertices covering almost the entire graph in which each vertex is reachable by a path from each other vertex in the same set. However, we still need to check, whether $G$ is indeed a strongly connected graph. This can be done by self-multiplying $\Adj{G}$ $n$ times and testing whether or not all elements in $\Adj{G}$ are nonzero. The process of generating a random graph is repeated until this test yields a positive result, confirming that the graph is strongly connected.\\

\subsection{Sensors} \label{sec:sensors}

So far, the input variables of an FDS were considered to be controlled by an abstract environment. When an FDS is executed on a microprocessor, the input variables are abstractions e.g.\ from inputs received from sensors or data transfer lines connected to the processor. For example, a boolean environment variable may be the abstraction of the input received from a sensor detecting the state of a switch.\\

When considering an FDS to be the controller of a robot that moves on a graph $G$, sensors can be differentiated between gathering information from the entire graph or just a subgraph of $G$. In the former case we speak of {\bf global} information, while information from a subgraph in the vicinity of the robot is called {\bf local}. In the extreme case, a robot may only obtain sensor inputs from the vertex it is on.\\

Thus, sensor inputs, e.g.\ represented by an environment variable $s$, depends on the system variable $\cellID$. However, the robot does not know about this dependency. Therefore, when simulating the FDS, we must ensure that the state of the graph is maintained globally, but only the appropriate local information is available to the robot by changing $s$ accordingly. This is further discussed in Section \ref{sec:globstore} and is implemented through connectors\index{Connector} as explained in \sec{statarch}.\\

\section{Stationary Target Search} \label{sec:search}

In an USAR\index{USAR} application, taking place for example in a collapsed building, the targets are usually assumed to be stationary, and thus the search only involves visiting each vertex by a robot until the target is detected. A global specification therefore requires that once a vertex contains a target, it is eventually visited by at least one robot. Here we present the concept of stationary target search by considering only a single target at a time in the graph. In the implementation presented in \sec{statSAR}, several targets can be rescued at the same time.\\

\subsection{Global Specifications} \label{sec:globsearchspec}

Let $G = (V_G, E_G)$ be a directed graph with $n = |V_G|$ vertices. In order to formulate the global specification, we introduce the global integer environment variable $t \in [\e,  n)$, which indicates whether and where a target needs to be rescued. If $t = \e$ then no target is in the graph, but if $t \neq \e$ then a target is located on vertex $v_t$, i.e.\ the vertex with index $t$. This represents the actual state of the target, but the robots have no direct information about the value of $t$. Rather, the robots are equipped with sensors that provide local information about the vertex the robot is currently on. Thus the robot only sees the value of $t$ when it is on the vertex $v_t$.\\

Assume there are $m$ robots $\{R_0, R_1, \ldots, R_{m-1}\}$ searching for the targets, each with a unique index $j \in [0, m)$. The position of robot $R_j$ is controlled via the system variable $\cellID_j \in [0, n)$. For notational convenience, we use the boolean variable $X_i^j$ instead of explicitly writing $\cellID_j = i$.\\

The main global system goal therefore can be stated as follows:
\begin{equation}
	\v^s = \bigwedge_{i \in [0, n)}\always(t = i \rightarrow \eventually (\bigvee_{j \in [0, m)} X_i^j)).
\label{eq:globstatgoal}	
\end{equation}
This expresses that once a target is at vertex $v_i$, at least one robot will find it, i.e.\ at least one robot will eventually move to the vertex $v_i$. In order for this to be realizable, we have to at least assume that the target stays at the vertex until it is discovered. This is the {\bf stationarity assumption}, and it can be expressed by the global environment assumption
\begin{equation*}
	\v^e = \bigwedge_{i \in [0, n)}\always(t = i \wedge \next(t \neq i) \rightarrow \bigvee_{j \in [0, m)} X_i^j),
\end{equation*}
which says that in order for $v_t$ to change from a value that is not $\e$, at least one robot has to be at the target's vertex. A different way of expressing the global system goal is
\begin{equation*}
	\hat{\v}^s = \always(t \neq \e \rightarrow \eventually(t = \e)),
\end{equation*}
which makes the actions required for the target at $t$ to be rescued implicit. In order for $\hat{\v}^s$ to be realizable, the environment assumptions must indicate how $t$ reacts to the actions for the robots, i.e.\ how the system variables can be changed in order to clear $t$. For example, $t$ can be cleared if a robot moves to the vertex $v_t$, which is expressed by the environment assumption
\begin{equation*}
	\hat{\v}^e = \bigwedge_{i \in [0, n)}\always(t = i \wedge \bigvee_{j \in [0, m)} X_i^j \rightarrow \eventually(t = \e)).
\end{equation*}

Note that the two specifications here might not be synthesizable and not even realizable. They are merely shown to illustrate the principle behind specifying stationary target search globally.\\

\subsection{Compositional Specifications} \label{sec:statsearch}

The specification $\v^e \rightarrow \v^s$ (or $\hat{\v}^e \rightarrow \hat{\v}^s$) might not be complete or realizable, but it expresses the main idea behind globally specifying a distributed search in a graph. If a synthesis algorithm existed for general compositional synthesis, this specification could directly be used to synthesise a reactive controller for each robot in the architecture, s.t.\ together the robots would implement the global specification. However, since no such algorithm exists, the specification must be decomposed manually.\\

\subsubsection{Minimal Specification}

Each robot can have a simple local specification
\begin{equation*}
	\bigwedge_{i \in [0, n)} \always\eventually X_i^j,
\end{equation*}
stated here for the robot $R_j$. Note that the robot knows nothing about the location $t$ of the target and no sensor information is used. This way of searching assumes that the target reacts automatically to the robot being in the same cell by becoming rescued. Note however that in order to actively engage in the rescue of a target, the robot needs to react to a sensor input that indicates that it is in the target's cell.\\

In this specification, there is no coordination between the robots, and so the search might be inefficient. However, we are not so much concerned about efficiency of our controllers as with their correctness.\footnote{Arbitrary efficiency bounds cannot be provided in GR[1] specifications that have a limited nesting depth of the next operator $\next$ and lack the bounded until operator $\until^{\leq k}$. In fact, the allowed nesting depth of $\next$ is just one in GR[1] specifications.} Synthesizing a controller from this specification would indeed satisfy the global specification given above. However, when not one, but several robots are required to find a target, a different approach is needed. With stationary targets this is not hard to implement, since a robot can just stay with the target until a second robot arrives, requiring the robot to have information from a sensor that it is in the target's cell.\\

\subsubsection{Circular Search} \index{Circular search}

Another approach is to explicitly encode the paths that the robots have to take through the graph in order to visit each vertex at least once. This again does not allow for coordination between the robots. It has however one decisive advantage over the previous approach: It is explicitly known to the robot when the search is finished. In the previous approach, only the finiteness of the state space of the synthesized FDS induces a bound on the time to complete searching the graph.\footnote{We assume that the topology of the environment does not change, so it is a valid solution to explicitly encode a path for the robot. However, if the topology is unknown or changing, this approach is no longer appropriate.}\\

The paths are extracted from the graph by finding a shortest cycle that contains all vertices, which can be done by the algorithm {\bf FindCyle} in \tab{findcycles}. The algorithm is provided with a starting vertex $u$ and returns a shortest path $p$ that starts at $u$ and ends at a vertex $v$, such that there exists an edge $(v, u) \in E_G$ and $p$ contains each vertex in $V_G$ at least once.\\

\begin{table}
\begin{tabbing}
 \hspace*{.4cm} \= \hspace*{.4cm} \= \hspace*{.4cm} \= \hspace*{.4cm} \= \kill
\sc{\textbf {FindCycle}} ($u$)\\
\> paths = [[$u$]]\\
\> \textbf{while} True \textbf{do}\\
\>\> extendedPaths = [] \\
\>\> \textbf{for} $p$ \textbf{in} paths \textbf{do} \\
\>\>\> \textbf{if} $p$ does not contain all vertices in $V_G$ or $p$ does not end in $u$ \textbf{then} \\
\>\>\>\> newpaths = \sc{\textbf{ExtendByOne}}($p$) \\
\>\>\>\> append newpaths to extendedpaths \\
\>\>\> \textbf{else} \\
\>\>\>\> take away the last element of $p$ (i.e.\ $p$ no longer ends in $u$) \\
\>\>\>\> \textbf{return} $p$ \\
\>\> paths = extendedpaths\\
\\
\sc{\textbf {ExtendByOne}} ($p$)\\
\> \textbf{if} $p$ is empty \textbf{return} \\
\> paths = [] \\
\> \textbf{for} $v \in V_G$ \textbf{do} \\
\>\> \textbf{if} $(p(|p|-1), v) \in E_G$ \textbf{then} \\
\>\>\> newpath = $p$ extended with $v$\\
\>\>\> append newpath to paths \\
\> \textbf{return} paths
\end{tabbing}
\caption{Algorithm to find cycles containing all vertices in a graph. Calling \textbf{FindCycle}($u$) on some starting vertex $u \in V_G$ yields a list of vertices $p$ that induces a cycle in $G$. Assuming list operations to be $\mathcal{O}(1)$, the runtime complexity of this na\"ive implementation of \textbf{FindCycle} is $\mathcal{O}(n\cdot n!)$ for dense graphs, and $\mathcal{O}(n^4)$ for sparse graphs with $\mathcal{O}(|E_G|) \cong \mathcal{O}(|V_G|)$.}
\label{tab:findcycles}
\end{table}

In the specification, the robot can explicitly be instructed to follow the path $p$ returned by \textbf{FindCycle}. Since the cycle formed by the edges connecting the vertices in $p$ might not be Hamiltonian\index{Hamiltonian Cycle}\footnote{A cycle is Hamiltonian if it contains all vertices of a graph exactly once.}, it is not sufficient to just encode this with guarantees such as $\always(X_{p(0)}^j \rightarrow \next X_{p(1)}^j)$. This is because some vertices might need to be visited several times and at such vertices the choice of the next vertex depends on the current position in the path. By introducing a counter system variable $c \in [0, |p|)$ and letting the robot visit the elements on the path in order it is possible to keep track of the elements in $p$ already visited.\\

The search is started with the counter at zero and the robot being at $p(0) = u$:
\begin{equation*}
	X_{p(0)}^j \wedge c = 0.
\end{equation*}
Note that $X_{p(0)}^j$ is a synonym for $\cellID_j = p(0)$, so all variables $X_i^j$ with $i \neq p(0)$ necessarily valuate to ``$\mathrm{False}$''. The counter is restricted to only increase, unless it has to be reset to zero in order to loop when the path is completed.  Define $\nextval{\gamma} = (\gamma + 1 + |p|) \mathrel\mathrm{mod} |p|$ to be the next value of the counter when the current value is $\gamma$, where $\mathrm{mod}$ is the infix operator for the remainder of integer division. Then this rule can be written as:
\begin{equation}
	\always(\bigvee_{\gamma \in [0, |p|)}(c = \gamma \wedge \next (c = \nextval{\gamma}))).
\label{eq:countincrease}
\end{equation}
The robot moves along the path, increasing the counter by one each step:
\begin{equation}
	\bigwedge_{\gamma \in [0, |p|)}\always(X_{p(\gamma)}^j \wedge c = \gamma \rightarrow \next (X_{p(\nextval{\gamma})}^j \wedge c = \nextval{\gamma})).
\label{eq:pathmove}
\end{equation}
Both \eqref{eq:countincrease} and \eqref{eq:pathmove} are required in a complete specification: \eqref{eq:countincrease} prevents the counter to change to a value not corresponding to a valid position in the path (safety property), which would allow the robot to just remain at the same vertex. \eqref{eq:pathmove} is required for the robot to make any progress at all (liveness property), forcing the counter to increase.\\

This way of searching a graph is used in the implementation of moving target search, since there it is necessary that the robot knows when the search is finished, which is indicated by $c = |p| - 1$. A full specification of this type of search is given in \sec{implmovt} when searching for particular vertices in the topology.\\

\subsubsection{Other Search Methods}

Searching a graph can also be implemented as a depth-first-search or as another exploration algorithm. The specification of such algorithms is not straightforward, but can be done using counters and storing information in the vertices similar to the approach presented in \sec{distrsoln} under the heading ``Global Storage'', which ensures that the state space does not explode.\\

A depth-first search can handle an unknown environment topology of finite extent, if the vertices are able to store information for the robot and provide the robot with information of its neighboring vertices. However, synthesis from an LTL specification might not be the best approach to implement such an algorithm that has already been proved to work in a standard implementation.\\

\section{Moving Target Search} \label{sec:movtsearch}

When the target is moving, such as in a WiSAR\index{WiSAR} application, it is no longer sufficient to visit each vertex eventually, like in the stationary target search. For most topologies, a fixed number of robots might never be able to find a moving target. Consider for example the situation in \fig{evade}, where a single robot ($\bullet$) is trying to find a moving target ($\circ$) by moving on the same vertex in the topology. The target can evade the robot by always going to the opposite vertex at the same time as the robot moves. If the target knows the robot's strategy, it can always evade the robot.\\

\begin{figure}
\centering
\subfigure[Initial configuration]{
    \xymatrix@R=0.18cm@C=0.45cm{ & *++[o][F-]{\bullet v_0} \ar@/_1pc/[ddrr] \\ \\ & & & *++[o][F-]{\circ v_1} \ar@/_1pc/[uull] \\ {}}
}
\subfigure[During the transition]{
    \xymatrix@R=0.18cm@C=0.45cm{ & *++[o][F-]{\phantom{\bullet} v_0} \ar@/_1pc/[ddrr]|{\bullet} \\ \\ & & & *++[o][F-]{\phantom{\circ} v_1} \ar@/_1pc/[uull]|{\circ} \\ {}}
}
\subfigure[Final configuration]{
    \xymatrix@R=0.18cm@C=0.45cm{ & *++[o][F-]{\circ v_0} \ar@/_1pc/[ddrr] \\ \\ & & & *++[o][F-]{\bullet v_1} \ar@/_1pc/[uull] \\ {}}
}
\caption{The target ($\circ$) evading the SAR robot ($\bullet$).}
\label{fig:evade}
\end{figure}
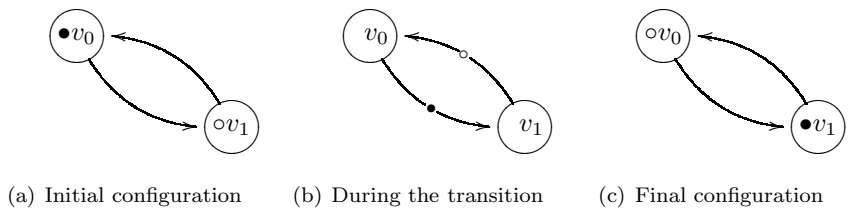

It is therefore necessary to implement a search strategy that ensures that even a moving target cannot escape the robots. This problem has been considered before and successfully implemented, for example in the work of Vidal et al.\ \cite{ProbPurs}. These implementations use probabilistic methods to find the target; we use a different approach in this thesis, and consider the problem of finding a winning strategy for a game of the robots against the targets on the graph representing the topology.\\

\subsection{Game Theoretic Approach}

Trying to find a moving target can be seen as a form of a {\bf pursuit-evasion game}\index{Pursuit-evasion game} in which the target tries to evade its pursuers, the SAR robots. The objective of the game in its original formulation is to find the minimum number of robots necessary to guarantee that the target is eventually found. Depending on what moves are allowed for the robots and the targets, the robots must implement different strategies in order to reliably find the target. A recent survey by Chung et al.\ provides a taxonomy of properties for pursuit-evasion games \cite{SearchBots}.\\

\subsubsection{Cops and Robbers Game} \index{Cops and robbers game}

We consider a particular form of pursuit-evasion game called the {\bf cops and robbers game} first introduced by Nowakowski and Winkler~\cite{CopRobbers}. A number of cops and robbers take turns in moving on a directed graph by traversing edges from vertex to vertex. A cop captures a robber if they find themselves on the same vertex or edge. Cops can move one edge at a time, but the robbers can move with infinite speed, i.e.\ in each turn they can move along any number of edges. However, they cannot pass through vertices that are occupied by cops. To align with standard formulations, a robber may hide either on a vertex or on an edge.\\

The cops {\bf win} the game if the robber is eventually captured, while the cops {\bf lose} the game if any of the robbers are able to evade the cops forever. The robbers are supposed to be adverse and always choose the worst possible move against the cops. They also know exactly where the cops are and their strategies, while the cops have no information on the position or strategy of a robber, unless it is captured. The game is considered as a two-player game between $\Cops$, the central coordinating authority of the cops against the adverse robbers, modelled as a single player $\Robbers$.\\

A {\bf strategy}\index{Strategy} for the cops defines the moves of the cops on the graph depending on the observations about the targets. A {\bf winning strategy}\index{Winning strategy} for the cops is a strategy that guarantees that all robbers are found in a finite number of moves. The game is {\bf solved} by finding the minimum number of cops for which a winning strategy exists for a given graph.\\

Note that the terms robot, cop and pursuer can be used interchangeably, as is the case for target, robber and evader. Unless the context demands otherwise, we use ``robot'' and ``target'' throughout the rest of this section.\\

\subsubsection{Cop Numbers}

Most graphs require more than one robot to capture the targets, see for example \fig{evade}. This concept is formalized by introducing the {\bf cop number}\index{Cop Number} $c(G)$ of a directed graph $G$, which is the minimal number of robots required to win, regardless of the initial positions of robots and targets. For example, the graph in \fig{map} has a cop number of $3$.\\

Immediately the question arises how to compute $c(G)$ for a given graph. Another problem is to give sufficient properties for a graph $G$ so that $c(G)$ has a fixed value. In this thesis we are not concerned with answering these questions but rather how to find a winning strategy for $c(G)$ robots for a class of graphs that is as general as possible. Such a winning strategy clearly exists for $c(G)$ robots by definition.\\

\subsection{Computational Considerations} \label{sec:compconsider}

For general graphs and if the robots have no information about the positions of the targets, finding the cop number is NP-complete (\cite{PEComplex}, Figure 1). Even when both robots and targets have complete information about the positions of all robots and targets, computing $c(G)$ in an undirected graph, which is a simpler problem than the one we are considering here, is EXPTIME-complete (\cite{PursCOMPLEX}, Theorem 3).\\

It was shown before that the synthesis of a GR[1]\index{Specification!GR[1]} specification $\v$ can be done in polynomial time in the size of $\v$. However, we do not try here to solve an NP-complete problem by a polynomial time algorithm. When formulating a GR[1] specification $\v$ for a winning strategy, $\v$ must explicitly contain the system variables that control the robots, i.e.\ $\cellID_j$ for all $j$. Therefore, the number of robots that is available to pursue the targets is already explicitly encoded in $\v$.\\

With such a specification, GR[1] synthesis merely decides whether a given number of robots is sufficient to win the game. However, this does not mean that the cops and robbers game is solved, which consists in \emph{finding} $c(G)$. This must be done at the time of formulating the specification, because the system variables must be able to independently control at least $c(G)$ robots. \\

From this discussion we conclude that the cops and robbers game cannot be \emph{solved} by GR[1] synthesis: An NP-complete problem cannot be solved by the polynomial time algorithm for GR[1] synthesis. However, realizability of a specification $\v$ of a winning strategy that explicitly includes the number of robots, can be decided by GR[1] synthesis. This is shown constructively in \sec{implmovt} by synthesizing controllers implementing a winning strategy.\\
%
%
%

%

Since GR[1] synthesis cannot be used to find $c(G)$ and a GR[1] specification must have the number of robots encoded explicitly, we assume that the cop number $c(G)$ is provided alongside a graph $G$.\\

\section{Moving Target Search Strategies}

Instead of solving the cops and robbers game, we are interested in finding winning strategies for a given set of $m$ robots $\{R_0, R_1, \ldots, R_{m-1}\}$. Clearly, no such strategy exists if $m < c(G)$. We first categorize the winning strategies given that a central authority $\Cops$\index{C@$\Cops$} can directly control the moves of all robots, called {\bf globally coordinated winning strategies}\index{Winning strategy!Globally coordinated}.\\

Thus the player $\Cops$ implements its strategy as an FDS $\M_\Cops$, which is specified by a GR[1] specification $\v_\Cops$. The environment variables of $\M_\Cops$ are controlled by the player $\Robbers$\index{R@$\Robbers$}. However, no FDS implementation $\M_\Robbers$ of the targets' strategy is synthesized. The targets are simply assumed to always perform the worst possible moves for the robots.\\

The central authority $\Cops$ controls the system variables $\cellID_j$ for all $j$. This is the standard formulation found in the literature. However, we go on to identify globally coordinated winning strategies that can also be specified by \emph{several} GR[1] specifications that are synthesized into a set of FDS's that together implement the globally coordinated strategy. This corresponds to reformulating the two-player cops and robbers game into a game where several robots $\Cops_1, \Cops_2, \ldots, \Cops_m$ play against the adverse targets $\Robbers$. The winning strategies of the robots are based on local information in the graph only, and they must negotiate with each other in order to cooperate or share information. Therefore, such strategies are referred to as {\bf locally coordinated winning strategies}\index{Winning strategy!Locally coordinated}.\\

Note that the requirement to provide sufficiently many robots means that the cop number must be known at specification time. It turns out, that the cop number is not even required to be included in the local specifications, but is used only to build the architecture\index{Architecture} for sufficiently many robots.\\

\subsection{The State of a Graph} \label{sec:graphstate}


We define the state of a graph to indicate to the robots whether the target can be said with certainty to not be on a particular edge or vertex. If the target cannot be on an edge, then the edge is contaminated, and the  same notion is introduced for vertices  and graphs in the following definitions:

\begin{definition} An edge $e \in E_G$ in a graph $G = (V_G, E_G)$ can either be {\bf contaminated}\index{Edge state} or {\bf cleared}. Initially, all edges of the graph are contaminated. A contaminated edge $e = (u, v)$ from $u$ to $v$ can be cleared by a robot either by placing a {\bf guard}\index{Guard} robot on $u$ and sliding down the edge from $u$ to $v$ with another robot, or if all in-edges of $u$ are already cleared by simply sliding down the edge from $u$ to $v$ with a robot, without requiring a guard on $u$.
\end{definition}

\begin{definition}
A vertex is {\bf cleared}\index{Vertex state} if all its in-edges and out-edges are cleared, {\bf partially cleared} if all its in-edges are cleared and at least one of its out-edges is contaminated, {\bf contaminated} if at least one of its in-edges and all of its out-edges are contaminated and no robot is guarding it, and {\bf critical} otherwise. A contaminated vertex can be a {\bf start vertex}\index{Start vertex}, a concept which is introduced below.
\end{definition}

The state of an edge $e \in E_G$ is $\estate(e) \in \{\ecl, \eco\}$, where $\ecl$ and $\eco$ correspond to ``cleared'' and ``contaminated'' respectively. Similarly, the state of a vertex $v \in V_G$ is $\vstate(v) \in \{\vcl, \vpc, \vcr, \vco\}$ where $\vcl$, $\vpc$, $\vcr$ and $\vco$ correspond to ``cleared'', ``partially cleared'', ``critical'' and ``contaminated'' respectively. The {\bf state of a graph} is the set of all states of its edges and vertices.\\

\begin{definition}
A graph is {\bf cleared}\index{Graph state} if all its vertices are cleared, partially cleared, or are guarded by  a robot, it is {\bf contaminated} if all vertices and edges are contaminated, and it is {\bf critical} if it is neither cleared nor contaminated.
\end{definition}

\subsection{Graph Clearing Sequences}

A strategy defines a set of moves of the robots depending on the observations about the targets, which are obtained by querying the state of the graph, e.g.\ using sensors, see \sec{sensors}. We define strategies for the cops that assume that a central authority can control the movement of the robots. A list of robots $\freelist$\index{R@$\freelist$} that are available to move is maintained as well as an auxiliary list of cops $\guardlist$\index{G@$\guardlist$} that keeps track of the guarding robots. Given that there are $m$ robots $\{R_0, R_1, \ldots, R_{m-1}\}$, we always have $\freelist \cup \guardlist = \{R_0, R_1, \ldots, R_{m-1}\}$.\\

\begin{definition}
Given a graph $G = (V_G, E_G)$ and sufficiently many cops in $\freelist$, a {\bf clearing move}\symin{Clearing move}{g}{\cmove} $\cmove$ is defined as a sequence of movements of robots that satisfies the following procedure: 
\begin{tabbing}
 \hspace*{1cm} \= \hspace*{.4cm} \= \hspace*{.4cm} \= \hspace*{.4cm} \= \kill
(S1) \> \textbf{if} $\freelist \neq \emptyset$ \textbf{then} select $R \in \freelist$ and remove $R$ from $\freelist$ \\
\> \textbf{else} select a guarding robot $R$ from $\guardlist$ \\
\> Place $R$ as a guard on a contaminated vertex $v \in V_G$ and put $R$ in $\guardlist$ \\
(S2) \>  \textbf{for each} out-edge $(v, u) \in E_G$ of $v$ \textbf{do}\\
(S2.1) \>\> \textbf{if} $\freelist \neq \emptyset$ \textbf{then} select $R' \in \freelist$ \\
\>\> \textbf{else} select a guarding robot $R'$ from $\guardlist$ and remove $R'$ from $\guardlist$ \\
\>\> Place $R'$ on $v$\\
(S2.2)\>\> Slide $R'$ from $u$ to $v$ along $(v, u)$. \emph{This clears the edge $(v, u)$}\\
(S3)\>  \textbf{while} there is a partially cleared vertex $p \in V_G$ \textbf{do}\\
(S3.1) \>\> \textbf{if} $\freelist \neq \emptyset$ \textbf{then} select $R'' \in \freelist$ \\
\>\> \textbf{else} select a guarding robot $R'' \in \guardlist$ and remove $R''$ from $\guardlist$ \\
\>\> Place $R''$ on $p$\\
(S3.2)\> \> \textbf{for each} contaminated out-edge $(p, q) \in E_G$ of $v$ \textbf{do}\\
(S3.2.1)\>  \>\> Slide $R''$ from $p$ to $q$ along $(p, q)$. \emph{This clears the edge $(v, u)$}\\

(S4)\>  \textbf{if} $v$ is cleared \textbf{then} place $R$ back into $\freelist$
\end{tabbing}
This procedure is satisfied by a sequence of robot moves if the statements are completed sequentially, i.e.\ (S1) precedes (S2), (S2) precedes (S3), (S2.1) precedes (S2.2) and so on. Note that the dots indicate a hierarchy, so (S2) is completed only if (S2.1) and (S2.2) are. The contaminated vertex $v \in V_G$ selected in (S1) is called the {\bf start vertex}\index{Start vertex} of the clearing move.
\end{definition}

The procedure presented in the above definition does not deterministically specify the moves of each robot at every time. The sources of nondeterminism are the following:
\begin{itemize}
\item There is no specific order according to which robot should be chosen from $\freelist$ in (S1), (S2.1) or (S3.1).
\item There is no specific order according to which the guarding robot should be chosen from $\guardlist$ in (S1), (S2.1) or (S3.1).
\item Any contaminated vertex may be chosen in (S1).
\item The partially cleared vertices may be chosen in any order in (S3).
\item Robots that are in $\freelist$ or not in $\guardlist$, and not sliding down an edge in (S2.2) or (S3.2.1) may move around on the graph freely.
\item The order in which the edges are cleared on the sliding moves is not specified in (S2) and (S3.2).
\item Only the order of necessary moves is specified, but there could be an arbitrary number of actions interleaved into the given sequence that do not violate the procedure (i.e.\ guards may not move).
\end{itemize}

While a clearing move does not completely specify all robot moves, it completely specifies the change in the state of the graph, depending only on the choice of the start vertex $v$ in (S1) and the choice of the robots $R$, $R'$ and $R''$ in (S1), (S2.1) and (S3.1) respectively. We adopt the convention that $\guardlist$ acts like a first in first out (FIFO)\index{FIFO} queue so that the robot that was first selected as guard will be the first robot to be selected to move away from the vertex it is guarding. The effect of a sequence of clearing moves $\cmove_0\cmove_1\ldots\cmove_n$ on the state of the graph is therefore uniquely specified by the start vertices of the clearing moves.\\

\begin{definition}
	A {\bf graph clearing sequence}\symin{Graph clearing sequence}{Gamma}{\cseq} (GCS) for a graph $G = (V_G, E_G)$ is a sequence of clearing moves $\cseq = \cmove_0\cmove_1\ldots\cmove_l$ such that at the end of the clearing move $\cmove_l$ the graph is cleared.
\end{definition}

A graph is initialized with all its edges and vertices being contaminated. Since we are only considering strongly connected graphs, there can be no partially cleared vertices initially if all edges are contaminated. The winning strategies then consist of sequences of clearing moves in which the robots move together to clear the graph.\\

{\bf Theorem 1} (\cite{SearchStrategies}, Theorem 3.4). Given a contaminated graph $G = (V_G, E_G)$ with cop number $c(G)$ and at least $c(G)$ cops in $\freelist$, there exists a GCS $\cseq$ for $G$. \qed\\

When a guarding robot moves away from a critical vertex (i.e.\ a vertex with some cleared out-edges), there is the possibility of recontamination~\cite{RCNHelp}. An edge $e = (u, v) \in V_G$ is {\bf recontaminated}\index{Recontamination} if it has been previously cleared but the vertex $u$ is no longer guarded, cleared or partially cleared. The recontamination of an edge can cause vertices to change their state, and thus further edges to change their state, possibly causing a cascade of recontaminations in the graph. The state of the graph can be updated using the procedure in \tab{recontaminate}.\\

\begin{table}
\begin{tabbing}
 \hspace*{.4cm} \= \hspace*{.4cm} \= \hspace*{.4cm} \= \hspace*{.4cm} \= \kill
recontaminate = True \\
\textbf{while} recontaminate \textbf{do} \\
\> recontaminate = False \\
\> \textbf{for each} vertex $v \in V_G$ \textbf{do} \\
\>\> \textbf{if} no guarding robot on $v$ and $v$ cleared or contaminated \textbf{then} \\
\>\>\> \textbf{for each} out-edge $e = (v, u) \in E_G$ of $v$ \textbf{do} \\
\>\>\>\> contaminate $e$ \\
\>\>\>\> recontaminate = True
\end{tabbing}
\caption{Procedure ensuring recontamination is propagated appropriately.}
\label{tab:recontaminate}
\end{table}

This procedure is used in the implementation of the global storage, cf.\ \sec{globstore}. It can be shown that no recontaminations are necessary to clear a contaminated graph $G$ with no less than $c(G)$ robots:\\

{\bf Theorem 2} (\cite{RCNHelp}, Theorem 2). Given a contaminated graph $G = (V_G, E_G)$ with cop number $c(G)$, at least $c(G)$ robots in $\freelist$ and a GCS $\cseq$ of $G$, the set $\freelist$ will always be nonempty at steps (S1), (S2.1) and (S3.1) of any clearing move in $\cseq$. That is, no guarding robot ever has to be removed and no recontaminations occur. \qed\\

{\bf Theorem 3}. Given a graph $G = (V_G, E_G)$, $c(G)$ robots in $\freelist$, and a GCS $\cseq$ for $G$, the graph is cleared by $\cseq$ independently of the initial state of $G$.\\

{\bf Proof sketch.} Denote $G$ and $\freelist$ at the beginning of $\cseq$ by $\hat{G}$ and $\hat{\freelist}$ respectively.\footnote{Of course, $c(G) = c(\hat{G})$.} Let $\hat{\guardlist}$ be the set of the guarding robots at the beginning of $\cseq$ (some robots must be guarding, otherwise the graph must be cleared and Theorem 1 yields the result).\\

When executing $\cseq$ on $\hat{G}$, at steps (S1), (S2.1) and (S3.1), the robots in $\hat{\guardlist}$ will be chosen \emph{before} the robots that have been removed previously from $\hat{\freelist}$ in step (S1). On using the guarding robots from $\hat{\guardlist}$, recontamination may happen. By Theorem 2, during the execution of $\cseq$, no guarding robots other than those in $\hat{\guardlist}$ are required in steps (S1), (S2.1) or (S3.1). Therefore, recontamination never affects edges that have been cleared during the execution of $\cseq$ in steps (S2.2) and (S3.2.1) and so $\cseq$ clears $G$ independently of the initial state of $G$. \qed\\

{\bf Corollary 1}. Given a contaminated graph $G = (V_G, E_G)$, $c(G)$ robots in $\freelist$, a GCS $\cseq$ for $G$, and a finite sequence of clearing moves $\cseq' = \cmove_0\cmove_1\ldots\cmove_n$, the concatenation $\cseq'\cseq$ is also a GCS for $G$.\\

{\bf Proof}. Let $\hat{G}$ and $\hat{\freelist}$ be $G$ and $\freelist$ at the end of $\cseq'$ respectively. At the end of $\cseq'$, $\hat{G}$ is either contaminated, critical or cleared, and $|\hat{\freelist}| \leq c(G)$. The result follows from Theorem 3. \qed\\

Corollary 1 states that the set of graph clearing sequences is closed under prefixing with clearing moves. When developing a globally coordinated winning strategy\index{Winning strategy!Globally coordinated} for the robots, it is therefore sufficient to ensure that the sequence of robot moves it entails eventually ends in a GCS\index{Graph clearing sequence}. As mentioned above, it suffices to specify the order of starting vertices of the clearing moves. Thus, a winning strategy can be defined by a function from the state of the graph to the next starting vertex.\\

Since we are ultimately interested in finding a locally coordinated winning strategy\index{Winning strategy!Globally coordinated}, the choice of the next start vertex\index{Start vertex} should not depend on the entire state of the graph, because in a real implementation this information is not available to the robots without a central authority to coordinate them. Therefore, let $V_G^i$ be the set of vertices for which a robot can observe the state before the $i^\mathrm{th}$ clearing move. It is defined to be $V_G$ for $i = 0$ and for $i \leq 0$ by
\begin{align*}
	v \in V_G^i \Leftrightarrow \exists e &= (p, q) \in E_G \qdot \estate(e) = \ecl \,\wedge (p = v \vee q = v \vee \exists f = (r, s) \in E_G \qdot\\& (r = v \wedge (p = s \vee q = s) \vee s = v \wedge (p = r \vee q = r))).
\end{align*}
That is, before the $i^\mathrm{th}$ clearing move\index{Clearing move} $\cmove_i$ in $\cseq$, a robot can observe the state of a vertex $v \in V_G$ iff $v$ is the destination or source of an edge $f$ that goes to a vertex $u \in V_G$ which itself is the destination or source of a cleared edge $e$. Hence, the robots have a limited visibility radius of the graph state (while still only being able to sense a target in the same cell).\\

\begin{definition}
Let $E_G^i$ be the set of cleared edges \emph{before} the $i^\mathrm{th}$ clearing move $\cmove_i$ in $\cseq$ and $\langle E_G^i \rangle$ the graph induced by the edges in $E_G^i$. Let $C_G^i \subseteq \V_G^i$ be the set of critical or contaminated vertices in $V_G^s$ that have at least one contaminated in-edge.
\end{definition}

\begin{definition}
Let $\mu_i$ be the minimum number of contaminated in-edges of any vertex $v$ in $C_G^i$ if $C_G^i$ is nonempty, and let $\mu_i = \infty$ otherwise. A vertex $v \in V_G^i$ is a {\bf $\mu_i$-clear-candidate}\index{mu-clear-candidate@$\mu_i$-clear-candidate} if it has exactly $\mu_i$ contaminated in-edges before the $i^\mathrm{th}$ clearing move $\cmove_i$ in $\cseq$.\\
\end{definition}

\begin{definition}
Let $\nu_i$ be the maximum number of contaminated out-edges to $\mu_i$-clear-candidates of any vertex in $V_G^i$ if $\mu_i < \infty$, and let $\nu_i = -\infty$ otherwise. A vertex is a {\bf $\nu_i$-start-candidate}\index{nu-start-candidate@$\nu_i$-start-candidate} if it has exactly $\nu_i$ out-edges to $\mu_i$-clear-candidate vertices before the $i^\mathrm{th}$ clearing move $\cmove_i$ in $\cseq$.\\
\end{definition}

The start vertex $v^i \in V_G$ of the $i^\mathrm{th}$ clearing move $\cmove_i$ in $\cseq$ is selected according to the following heuristics, denoted by $\mathcal{H}$. The six criteria are listed in order of decreasing priority:
\renewcommand{\labelenumi}{(H\arabic{enumi})}
\begin{enumerate}
	\item The vertex $v^i$ is contaminated.
	\item The graph $\langle E_G^{i+1} \rangle$ must be connected (but not necessarily strongly connected).
	\item If it exists, select a $\nu_i$-start-candidate vertex $v^i \in V_G^i$ which is not a $\mu_i$-clear-candidate.
	\item Otherwise, if it exists, select a $\nu_i$-start-candidate vertex $v^i \in V_G^i$ which is also a $\mu_i$ clear-candidate.
	\item Select the vertex with the minimum number of in-edges.
	\item Select the vertex with the maximum number of out-edges.
\end{enumerate}

Note that even after (H6) there still might be several possibilities of choosing $v^i$ in which case the start vertex is chosen nondeterministically.\\

\subsection{Evaluation of the Strategy}

With these heuristics, the $i^\mathrm{th}$ start vertex is selected from the contaminated vertices in $V_G^i$. This is a local decision in order to solve the global problem of finding the best starting vertex. This requires to make assumptions on the graph that permit the local heuristics to always lead to a global solution, i.e.\ choosing the start vertices according to the local heuristics successfully clears the graph $G$ with exactly $c(G)$ robots.\footnote{The same principle is used when optimizing convex functions using local decisions, since the global assumption of convexity means that any local optimum is a global optimum.}\\

When choosing the start vertices according to the above heuristics $\mathcal{H}$, it is hoped that a GCS results. However, this is not true for all graphs. One possible counterexample is the graph in \fig{cex}. This is a planar graph with just five vertices, but still the heuristics do not necessarily produce a GCS. A GCS would arise from choosing to guard $v_1$ and $v_4$ (in any order). However, the heuristics may choose from any of the following sequences of start vertices: $v_2v_0v_3$ or $v_2v_0v_4$ or $v_4v_1$. However, the former two require four robots, while the cop number of the graph is just three.\\

\begin{figure}
\begin{displaymath}
    \xymatrix{ & \cellF{0} \ar@/_2.5cm/[dd] \ar@/_/[ld] \ar@/_/[dr] \ar[d] \\ \cellF{1} \ar@/_/[ru] \ar@/_/[r] &
    \cellF{2} \ar@/_/[l] \ar@/_/[r] \ar[d] & \cellF{4} \ar@/_/[l] \ar@/_/[lu] \ar@/_/[ld] \\
    & \cellF{3} \ar@/_/[ru]}
\end{displaymath}
\caption{An example graph with five vertices for which the heuristics fail. Its cop number is three, but with the heuristics $\mathcal{H}$, four cops might be required (although not in all cases.)}
\label{fig:cex}
\vspace{-0.2cm}
\end{figure}
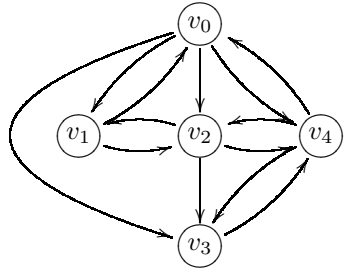

The heuristics would be easy to fix for this graph by adding another criterion
\begin{enumerate}
\setcounter{enumi}{6}
	\item In a tie, sum up the contaminated out-edges of the destinations of each vertex and choose the vertex with the maximum such number.
\end{enumerate}
Even if this would give rise to a GCS for this particular graph, it does not however solve the general problem: We try to find a simple solution to a provably hard problem, i.e.\ solving a global problem using only local information.\\

In general, the larger the graphs get, the more likely it is to encounter graphs that cannot be cleared using the heuristics with just $c(G)$ robots. \tab{stats} presents the results of simulations for a number of small graphs. For various counts of vertices, random graphs have been generated and the number of instances in which $\mathcal{H}$ does not result in a GCS are listed as counterexamples. The larger the graphs, the more likely it is for $\mathcal{H}$ to not yield a GCS. In the worst case of the explored examples in \tab{stats} even $c(G) + 4$ robots are required.\\

\begin{table}
\centering
\begin{tabular}{ccc}
\textbf{Vertices} & \textbf{Samples} & \textbf{Counterexamples} \\
\hline
$5$ & $1000$ & $1$ \\
$6$ & $1000$ & $4$ \\
$7$ & $1000$ & $11$ \\
$8$ & $1000$ & $16$ \\
$9$ & $1000$ & $23$ \\
$10$ & $1000$ & $42$ \\
\end{tabular}
\caption{Counterexamples to the heuristics $\mathcal{H}$. Random graphs generated according to \sec{graphgen}.}
\label{tab:stats}
\vspace{-0.2cm}
\end{table}

\subsubsection{Reformulation as Optimization Problem}

Finding the minimum number of robots required to clear a graph can be formulated as an optimization problem. In order to evaluate the cop number\index{Cop Number} of a graph with $n$ vertices, we consider an objective function $\mathrm{cn}_n(G, \delta) : \mathfrak{S}_n \times V_G^+ \rightarrow \mathbb{N}$ from graphs $G \in \mathfrak{S}_n$ and sequences of starting vertices $\delta \in V_G^+$ to the number of robots required to clear the graph when placing them on the start vertices $\delta$ in (S1) of the clearing moves.\footnote{Recall from definition \ref{def:strongconn} that $\mathfrak{S}_n$ is the set of strongly connected graphs with $n$ vertices and from \defseqsets~that $V_G^+$ is the set of nonempty sequences over $V_G$.} The function $\mathrm{cn}_n$ is potentially a very complicated function, and, as the results presented in \sec{compconsider} show, requires exponential time to compute in the number of vertices $n$.\\

We now minimize $\mathrm{cn}_n$ over the sequences of starting vertices for some given graph $G \in \mathfrak{S}_n$. When all sequences of starting vertices are feasible, then the minimum will be equal to the cop number $c(G)$ by definition:
\begin{equation*}
	\mathscr{P}_1:\hspace{1cm} c(G) \defines \min_{\delta \in V_G^+}~\mathrm{cn}_n(G, \delta).
\end{equation*}
Encoding the graph $G$ as an adjacency matrix $\mathrm{Adj}(G)$\index{Adjacency Matrix} and $\delta$ as a vector may yield a combinatorial optimization problem. Solving this is of course NP-complete, corresponding to the complexity of computing the cop number.\\

Now, the heuristics can be interpreted as constraints on the set of feasible sequences of starting vertices. Here we denote a heuristics simply by a propositional function $h(G, \delta) : \mathfrak{S}_n \times V_G^+ \rightarrow \mathbb{B}$ that is true iff the $\delta$ satisfies the heuristics $h$ in the graph $G$. Finding the number of robots under a heuristics $h$ can thus be reformulated as the optimization problem
\begin{align*}
	\mathscr{P}_2:\hspace{1cm}\min_{\delta \in V_G^+}~&\mathrm{cn}_n(G, \delta) \\
	                         &\mathrm{s.t.}~h(G, \delta) = \mathrm{True}.
\end{align*}
Showing whether a heuristics $h$ is optimal, in the sense that it always chooses the best start vertex so that only $c(G)$ cops are required to clear the graph is equivalent to proving that for all graphs $G$, the solutions to the optimization problems $\mathscr{P}_1$ and $\mathscr{P}_2$ yield the same value. This can be written as:
\begin{equation}
	\begin{array}{ll}h~\mathrm{optimal}~\Leftrightarrow \forall G \in \mathfrak{S}_n \qdot \min_{\delta \in V_G^+}~\mathrm{cn}_n(G, \delta) = &\min_{\delta \in V_G^+}~\mathrm{cn}_n(G, \delta) \\
	                         &\mathrm{s.t.}~h(G, \delta) = \mathrm{True}
\end{array}
\label{eq:checkhoptimal}
\end{equation}
\tab{stats} clearly shows that the heuristics $\mathcal{H}$ we developed is not optimal. However, even if we could develop an optimal strategy, verifying this using \eqref{eq:checkhoptimal} would be a nontrivial task.\\

\section{Locally Coordinated Strategies} \label{sec:distrsoln}

So far we assumed that a central coordinating authority exists that can directly control the robots. However, we are interested in developing a locally coordinated winning strategy\index{Winning strategy!Locally coordinated}. Several questions must be answered: How to adapt the globally coordinated winning strategy to a locally coordinated one? How is authority negotiated\index{Negotiation} between the robots, i.e.\ how do the robots coordinate? Where is the information about the state of the edges and vertices of the graph stored and how is it accessed or observed by sensors? This section is dedicated to answering these questions, and a complete specification is presented in \sec{implmovt}.\\





\subsection{Negotiations} \label{sec:negotiations}

Given a graph $G$ and its cop number $c(G)$, we would like to find $c(G)$ strategies, one for each robot, that together implement a winning strategy\index{Winning strategy} for $G$. No central authority is used, and so the robots must somehow cooperate\index{Cooperation} in order to eventually find the targets. The required communication protocol has already been presented in 
\sec{communication}.\\

Coordinating their movements on the graph restricts the authority each individual robot has. For example, a guarding robot may not move from a vertex $v$ while another robot is engaged in sliding down the out-edges of $v$. This requires the robots to be able to negotiate for authority over each other, so that one robot can, at least temporarily, control the other.\\

Instead of implementing sophisticated negotiation algorithms in the specification, we fix the structure of authority. Thus, each robot has a {\bf master}\index{Master} and a {\bf slave}\index{Slave}, resulting in a circular structure of authority. Indeed this is necessary when recontamination\index{Recontamination} occurs and the robot that was the first guarding robot in the GCS is told to move by its master.\\

This results in symmetrical specifications of the robots, where each robot provides the same services to its master and may expect the same services from its slave. We therefore develop only one local specification\index{Local specification} $\v$ and merely change the names of the variables specific to a robot. In practice, $\v$ is synthesized once and the architecture\index{Architecture} ensures that the communication channels are set up correctly, see \sec{commstructure}.\\

Thus, negotiation is decided at specification time and no longer has to be taken account in the specification. Communication is then merely used to transfer the commands from masters to slaves. The data, i.e.\ the state of the graph, can directly be obtained from the global storage discussed next.\\

\subsection{Global Storage} \label{sec:globstore}

Our implementation of the locally coordinated strategy for the cops and robbers game does not require the robots to have knowledge about the entire state of the graph. Instead, the robots are only provided with the information that is immediately required to form a local decision about the next move. This is achieved by storing the entire state of the graph in a data structure that can be accessed by all robots, but only provides local information to each robot.\\

Keeping the state outside of the searchers is current practice in SAR and is not a new idea. For example, the United Nations INSARAG marking system defines a set of shapes and letter used to mark a building's status to aid the SAR operations.\footnote{INSARAG Guidelines and Methodology, United Nations Office for the Coordination of Humanitarian Affairs, Jan 2011, \url{http://ochanet.unocha.org/p/Documents/INSARAG\%20Guidelines\%202011-Latest.pdf}, accessed 03. Nov 2011} Similar techniques could be implemented for the SAR robots, but we only consider the abstraction to the graph state introduced above.\\

In order to formally state a specification of a distributed solution, we must define how to access the global state. A robot will only ever have access to the state of the vertex it is currently on. We thus include the system variable $\cst_j \in \{\ccl, \cpc, \ccr, \csv\}$ in the system variables of robot $R_j$, indicating the current state of the vertex the robot is on. From the point of view of the robot this acts like a sensor that reads the relevant aspect of the environment.\\

The variable $\cst_j$ takes the value $\csv$ if $R_j$ is on a start vertex and (S1) is being executed. In (S1), $R_j$ is placed on a contaminated vertex $v$, for which $\vstate(v) = \vco$. However, $R_j$ may only choose a start vertex\index{Start vertex} according to the heuristics $\mathcal{H}$. Whether a contaminated vertex is a start vertex is indicated to the robot by the global storage. Thus, if $\vstate(v) = \vco$, but $v$ is not a start vertex, then $\cst_j = \ccr$ from the point of view of the robot $R_j$ if it is on $v$. However, if $v$ is a starting vertex and $R_j$ is on $v$, then $\cst_j = \csv$. Otherwise, $\cst_j$ takes the values $\ccl$, $\cpc$ and $\ccr$ if $\vstate(v) = \vcl$, $\vstate(v) = \vpc$ and $\vstate(v) = \vcr$ respectively. This will later be used in the specification of the controllers in \sec{robspec}.\\

In an implementation, it has to be ensured that the robot indeed gets the current state of the vertex on which it is. This requires in particular that the state of the graph is updated appropriately. Initially, all edges of the graph are contaminated. The state of the vertices only changes with the state of the edges to which they are connected, and the state of the edges changes depending on the movement of the robots on the graph. Thus, the global state of the graph changes only when a robot slides down an edge in steps (S2.2) or (S3.2.1) or the graph is recontaminated\index{Recontamination}. In the latter case the procedure in \tab{recontaminate} is executed.\\

Keeping the state of the graph outside the robots has several advantages: The robots do not need to communicate the local information they have to each other, reducing the need for system and environment variables for communication. Furthermore, the state space of the robots decreases significantly by only querying the immediately relevant local information of the graph. Note that the state of a graph with $n = |V_G|$ vertices is uniquely defined by the states of its edges, giving rise to $\mathcal{O}(2^{n^2})$ configurations in dense graphs.\\

\subsection{Multi-Stage Specification} \label{sec:mstage}

The local specifications must ensure that all robots together execute a sequence of movements that satisfies a GCS, i.e.\ a sequence of clearing moves. Thus a robot may be in one of four modes corresponding to the possible movements necessary to implement a clearing move. The robot $R_j$ with index $j$ therefore has a system variable $M_j \in \{\mcs, \mgu, \mpc, \mcl\}$ keeping track of the {\bf mode}\index{Mode} of the robot. If $M_j = \mcs$, the robot is searching  for a (contaminated) starting vertex in (S1). If $M_j = \mgu$, the robot is guarding a vertex in (S1). If $M_j = \mpc$, the robot is searching a partially cleared vertex in (S3). And if $M_j = \mcl$, the robot is sliding down the out-edges of a vertex in (S2.2) or (S3.2.1).\\

All robots except one are initialized to guard some arbitrary vertex in the graph, i.e.\ they are in mode $\Mgu$. The remaining robot is initialized on some arbitrary vertex but is searching for a starting vertex, i.e.\ it is in mode $\Mcs$. For reference, a summary of the environment and system variables of the robots is given in \tab{inp:robot} and \tab{out:robot} respectively.\\

\subsubsection{Searching for Start Vertices: $\Mcs$}

A robot $R_j$ in mode $M_j = \mcs$ searches for the contaminated vertex that is to be the next start vertex in the sequence of clearing moves $\cseq$ that constitutes a GCS for the graph $G$. The robot searches all vertices, querying their states until a start vertex is found, indicated by $\cst_j = \csv$. This search is done in a circular fashion as outlined in \sec{statsearch}. Using a circular search\index{Circular search}, the robot knows when all vertices have been visited and no contaminated vertex is in the graph. In this case, the graph is known to be cleared already.\\

If a contaminated vertex $v$ is found that matches the six criteria in the heuristics $\mathcal{H}$, it is selected as start vertex. Consequently, the robot enters guard mode $M_j = \mgu$ and stays at $v$ until called by its master to move away. On entering guard mode, the robot's slave, with index $j'$, is instructed to also move to $v$ and start sliding down the out-edges of $v$ to clear them. Thus the slave is \emph{forced} to enter the search for a partially cleared vertex with $M_{j'} = \mpc$.\\

\subsubsection{Guarding: $\Mgu$}

A robot $R_j$ in mode $M_j = \mgu$ is guarding the vertex $v$ on which it is currently positioned. Note that $v$ must necessarily be a start vertex. $R_j$ thus stays at $v$ until instructed by its master $R_{j''}$ to move away. The only way the master $R_{j''}$ can force robot $R_j$ to leave $v$, is to be guarding a vertex $w$ itself and to then instruct robot $R_j$ to enter the clearing mode $M_j = \mcl$ and clear all out-edges of the vertex $w$ guarded by the master.\\

\subsubsection{Searching for Partially Cleared Vertices: $\Mpc$}

This mode is very similar to searching for start vertices. However the difference is that the robot $R_j$ reacts to finding a partially cleared vertex $v$ by directly entering the clearing mode $M_j = \mcl$. A partially cleared vertex does not need to be guarded, and thus in this case the robot that finds such a vertex does not need to instruct its slave to help clearing the out-edges of $v$.\\

\subsubsection{Clearing: $\Mcl$}

In the clearing mode $M_j = \mcl$, robot $R_j$ is using sliding moves to clear the out-edges of a partially cleared or guarded vertex $v_i$. The vertex $v_i$ is stored in the robot's system variable $\store_j = i$, which is set on entering the clearing mode. The out-edges of $v$ are cleared one at a time, which is specified in the guarantees of the robot's specification. Since the order of edges to be cleared by sliding is arbitrary in the definition of a clearing move (see the discussion of sources of nondeterminism below the definition), it is perfectly admissible to fix the order at specification time.\\

To keep track of which out-edges of a vertex have already been cleared, the robot keeps a system variable $c_j$, which is a counter that is initialized to zero when the clearing mode is entered, in addition to setting $\store_j$ to the index $i$ of the vertex $v_i$. Whenever the robot slides down an out-edge of $v_i$, $c_j$ is increased by one. This is guaranteed by properties similar to the following:
\begin{equation*}
	\bigwedge_{i \in [0, n)} \bigwedge_{\gamma \in [0, \bar{\gamma}_i)} \always(\store_j = i \wedge c_j = \gamma \wedge X_i^j \rightarrow \next(X_{w_\gamma^i}^j \wedge c_j = \gamma + 1)),
\end{equation*}
where $\bar{\gamma}_i$ is the number of out-edges of the vertex $i$ and $w_\gamma^i$ is the (unique) vertex such that there is an edge $e$ from $i$ to $w_\gamma^i$ and $e$ is the $\gamma^\mathrm{th}$ out-edge to be cleared by the robot.\\

After clearing an edge, the robot has to return to the stored vertex in $\store_j$ again, which is guaranteed by properties similar to
\begin{equation*}
	\always(\bigvee_{i \in [0, n)} (\store_j = i \wedge \bigvee_{\gamma \in [0, \bar{\gamma}_i)}(c_j = \gamma + 1 \wedge X_{w_\gamma^i}^j)) \rightarrow \eventually \bigvee_{i \in [0, n)} (\store_j = i \wedge X_i^j)).
\end{equation*}

The full specifications that follow the ideas developed in this section are given in \sec{implmovt}.\\

\chapter{Implementations} \label{sec:impl}

In this chapter, we present two implementations of controllers for SAR robots, one for stationary target search in an USAR application, and the other for moving target search in a WiSAR application. In the stationary target search, a central authority, called the ``allocator'' is responsible for dispatching the robots to the vertices in the graph where a target needs to be rescued. In the moving target search, a distributed solution is adopted, where the authority structure among the robots is already negotiated in advance, cf.\ \sec{negotiations}.\\

\section{SAR of Stationary Target} \label{sec:statSAR}

In USAR\index{USAR}, the targets are mostly considered stationary. We develop specifications to build controllers for a team of robots that is managed by a central allocator. The goal is to find all targets in the affected area and rescue them. A robot that is at the location of a target may {\bf engage} with it by performing a rescue task like providing medical support. In this scenario, we will consider a target to be rescued if $\bar{\alpha} = 2$ robots have engaged with it for a sufficient amount of time.\\

The allocator is responsible for gathering information about the targets' status and location, and coordinating the action of the robots by dispatching them to the location of a target. If a robot is not dispatched, it may move around freely. The topology is represented as a graph, which is explained in \sec{maprepresentation}. It is optimistically assumed that the entire graph is known to the robots at synthesis time and that it does not change during the SAR task.\\


When the allocator dispatches a robot to a vertex, it merely asserts the index of a target's vertex on a transmission variable to the respective robot. The robot is then expected to move to this vertex on its own account, so the allocator does not need to know the topology. The allocator knows about a target's needs by receiving a message, or a ``flag'' associated with the location of the target. Targets will be assumed to be stationary since an injured or trapped person will not move.\\

Communication between the allocator and the robots is assumed to be layered on top of some reliable wireless protocol. Delivery of data is guaranteed eventually in finite time, messages are not corrupted, no reordering takes place, no messages get lost --- comparable to a data-layer protocol. Thus the communication is based on the four-phase handshake protocol developed in \sec{4PHP}.\\


\subsection{Architecture} \label{sec:statarch} \index{Architecture}

We synthesize the controllers for a set of robots and the central allocator. The asynchronous composition of the reactive systems resulting from the synthesis must satisfy the global specification of rescuing each flagged target, see~\sec{globsearchspec}.\\

In order to be able to simulate the reactive systems, it is necessary to close them as explained in~\sec{gametheorappr}. This is done by also synthesizing a reactive system for each vertex or ``cell'' in the topology. These reactive systems control the flags according to the environment assumptions of the allocator, which gathers the information about the targets. The complete architecture is shown in \fig{amm2} and the individual parts are explained below.\\


\subsubsection{Allocator and Queues}

At the heart of the architecture is the allocator, denoted by $\AMM$. It is connected to the robots, $\{R_0, R_1, \ldots, R_{m-1} \}$. The cells are also included in the figure to show how the closed-loop system would be controlled. They are shown as FDS's $\{C_0, C_1, \ldots, C_{n-1}\}$.\\

The allocator is informed about whether a target needs to be rescued by the boolean flags $f_0, f_1, \ldots, f_{n-1}$, each associated with one cell. Similarly, the allocator knows about which robots are ready by receiving the boolean variables $r_0, r_1, \ldots, r_{m-1}$, each associated with one robot. If a robot is ready and a vertex has been flagged, the allocator must dispatch a robot to the vertex that has been flagged for the longest period of time.\\

If there are $n$ vertices and $m$ robots, the allocator would need at least $2^{n+m}$ states to account for the $n$ flags and $m$ ready signals of the robots. However, in order to reduce the allocator's state space, two FIFO\index{FIFO} queues $\FIFOf$ and $\FIFOr$ are implemented as separate components that maintain which target and robot has been waiting for the longest period of time respectively. Thus the environment variables of $\FIFOf$ are the flags $f_0, f_1, \ldots, f_{n-1}$ and the environment variables of $\FIFOr$ are the ready signals $r_0, r_1, \ldots, r_{m-1}$.\\

The allocator receives a flag-index $F \in [\e, n)$ from $\FIFOf$, and a robot-index $R \in [\e, m)$ from $\FIFOr$ indicating which robot may be dispatched to which cell. If $\FIFOf$ holds $f_i$ at its head, then it {\bf offers} $F = i$ to the allocator, and symmetrically if $\FIFOr$ holds $r_j$ at its head, then it offers $R = j$. Moreover, the allocator may request the next flag index from $\FIFOf$ by sending a dequeue signal $\Df$, which is acknowledged by $\DfA$. The queue $\FIFOr$ does not have such a dequeuing capability, since this is not required in the specification.\\

The allocator can dispatch robots by setting the dispatch variable $D \in [\e, m)$ to the value in $F$. This indicates to the robot which is currently selected by $R$ that it has to go to the vertex in $D$. Internally, $\AMM$ maintains a counter variable $c \in [0,\bar{\alpha}]$ of the number of robots that have already been dispatched to the cell with index $F$. If possible, $\bar{\alpha} = 2$ robots will be dispatched to a cell, but never more, because $\bar{\alpha}$ is the number of robots that are required to clear the flag.\\

We do not provide specifications for the queues since they are not synthesized in the way explained above. With $n$ inputs a FIFO FDS would have $\mathcal{O}(2^nn!)$ states, making synthesis futile even for modest $n$. This number results from the $\mathcal{O}(n!)$ configurations the internal store of the queue can be in (corresponding to the system variables $\Y$) and the $\mathcal{O}(2^n)$ configurations the boolean inputs can be in (corresponding to the environment variables $\X$). Hence, we consider FIFO's as standard components and use a Python implementation for simulation.\\

\subsubsection{Robots}

For stationary target search, the number of robots required is only as high as the number of robots required to rescue a target, i.e.\ $\bar{\alpha}$. However, we may include an arbitrary number of robots in the architecture to complete the rescue operation more effectively. For example, with $\bar{\alpha} = 2$, if there are two targets and four robots, the robots can be dispatched in teams of two and rescue both targets at the same time. We synthesize one controller that is used on all robots and simply rename the variables with indices to indicate which variable belongs to a particular robot.\\

A robot $R_j$ maintains its own position on the topology both in $\cellID_j \in [\e, n)$ and in the boolean variables $X^j_0, X^j_1, \ldots, X^j_{n-1}$. As explained above, the two representations are merely introduced for notational convenience in the specification. Another variable, $\store_j \in [\e, n)$, stores the destination cell if the robot has been dispatched.\\

The robot $R_j$ knows that it is dispatched from the dispatch signal $D$ sent by $\AMM$, if the associated robot index $R$ from $\FIFOr$ satisfies $R = j$, indicating that $\AMM$ sends $D$ to $R_j$. However, the robot is not interested in whether any other robot is dispatched or not. Therefore, in order to decrease the state space of the robot, auxiliary environment variables $d_0, d_1, \ldots, d_{m-1}$ over the domain $[\e, n)$ are introduced, one for each robot. Robot $j$ receives $d_j = i$ from the allocator if it is dispatched to vertex $v_i$. On reaching the vertex $v_i$, the robot checks its boolean sensor signal $\flag_j$ whether there is still a target on this vertex. The environment variable $\flag_j$ is true iff there is a target on the vertex the robot is currently on. If a target is detected at the vertex to which the robot is dispatched (recognized by $\cellID_j = \store_j \wedge \flag_j = \mathrm{True}$), the robot asserts a boolean engage signal $e_j$ to indicate to the allocator that it has successfully reached the target's vertex.

\newcommand{\connE}[1]{\textstyle{E_{#1} = \sum_{j}(e_j \wedge X^j_{#1})}}
\newcommand{\connF}[1]{\textstyle{\flag_{#1} = \bigvee_{i} (X^{#1}_i \wedge f_i)}}\
\newcommand{\connD}[1]{\textstyle{d_{#1} = \begin{cases} D, &\text{if }R = {#1}\\ \e, &\text{otherwise}\end{cases}}}

\newcommand{\rinputF}{{f_i} \ar[l]}
\newcommand{\rinputD}{{} \ar[l]_<{D, R,}^<{f_i}}
\newcommand{\cinput}{{} \ar[r]^<{\cellID_j}_<{e_j}}

\newcommand{\cell}[1]{\cinput & *+[F]{\connE{#1}} \ar[r]^(0.7){E_{#1}} & *+++[o][F]{C_{#1}}}
\newcommand{\robotM}[2]{*+++[o][F]{R_{#1}} \ar@<.7pc>[r]^(.24){\cellID_{#1}} \ar[#2]_(.3){r_{#1}} & *+[F]{\begin{array}{l}\connD{#1} \\ \connF{#1}\end{array}} \ar@<.7pc>[l]^(.7){\flag_{#1}} & \rinputD }
\newcommand{\middleportion}{*++++[o][F]{\FIFOf} \ar@<1pc>[r]^{\DfA} \ar[r]^{F} & *++++[o][F]{\AMM} \ar@<1pc>[l]^{\Df} \ar[uu]^(0.3){D} & *++++[o][F]{\FIFOr} \ar[l]^{R}}

\begin{sidewaysfigure}
\begin{displaymath}
    \xymatrix@R=1.2pc@C=1.5pc{
\cell{0} \ar[ddr]^(.3){f_0} & & & & \robotM{0}{ddl} \\
\cell{1} \ar[dr]^(.3){f_1} & & & & \robotM{1}{dl} \\
\cell{2} \ar[r]^(.3){f_2} & \middleportion & \robotM{2}{l} \\
 {\vdots} & {\vdots} & {\vdots} & & & & {\vdots} & {\vdots} & {\vdots} \\
\cell{n} \ar[uur]^(.3){f_n} & & & & \robotM{m}{uul} \save "3,4"."3,6"*++[F--]\frm{} \restore
}
\end{displaymath}
\caption{Allocator based SAR architecture. Implicitly, $i \in [0, n)$ and $j \in [0, m)$. The outgoing arrows showing the system variables $e_j$ of the robots are omitted for a clearer layout. Also, internal system variables such as $c$ and $\store_j$ are not shown.}
\label{fig:amm2}
\end{sidewaysfigure}
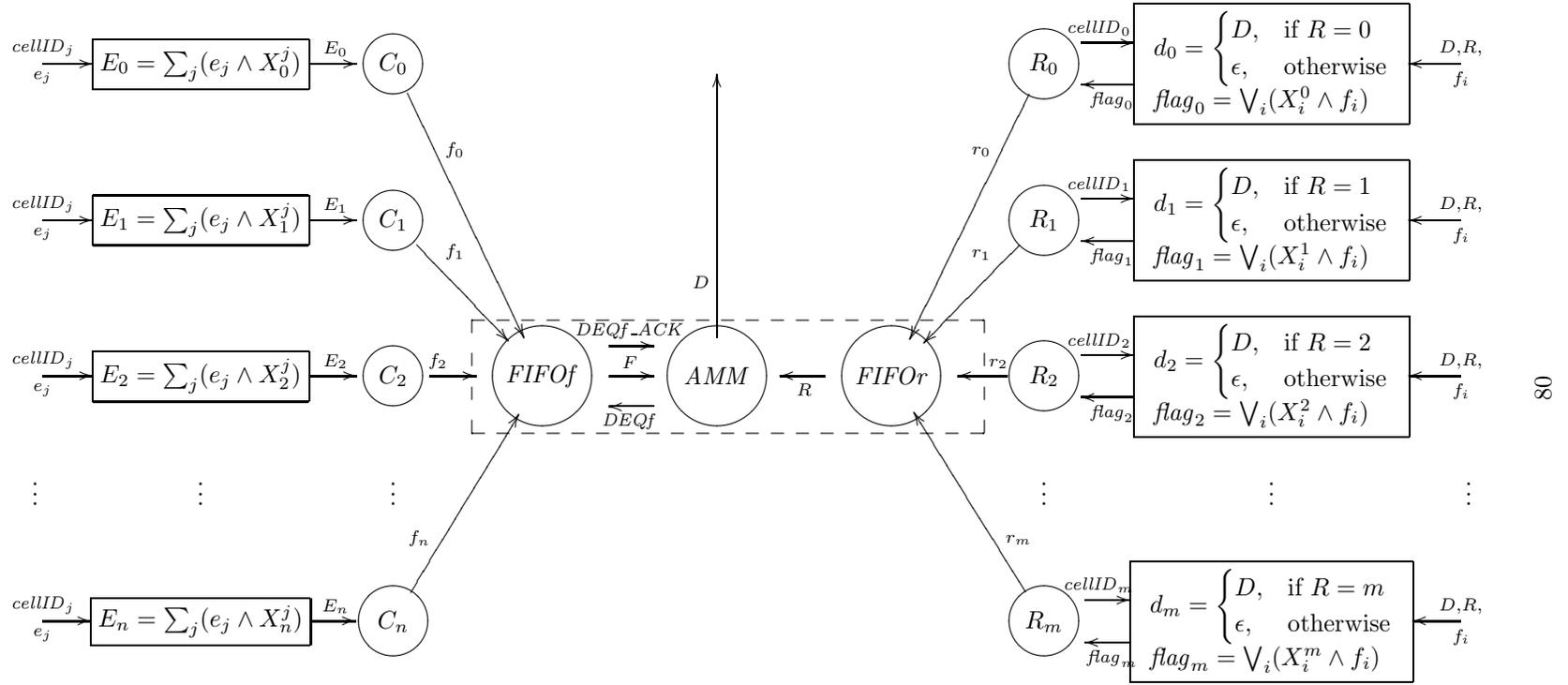

\subsubsection{Connectors} \index{Connector}

The communication\index{Communication} between the individual components in the architecture is shown by the arrows between the individual components. Some environment variables of the robots and cells are functions of the system variables of their peers. For example, the environment variable $E_i$ of cell $C_i$ is a function of the variables $\cellID_j$ and $e_j$ for $j \in [0, m)$. One way to incorporate their relationship into the design would be to have $E_i$ as a system variable and $\cellID_j$ and $e_j$ for $j \in [0, m)$ as environment variables in $C_i$. Then the guarantee
\begin{equation*}
  \always(\connE{i})
\end{equation*}
can be added to the specification of $C_i$. Note that the summation interprets ``$\mathrm{True}$'' as ``$1$'' and ``$\mathrm{False}$'' as ``$0$''. While this would produce the desired result, it introduces many environment variables\index{Environment variables} into the design of the cells. If they cannot be suitably restricted by assumptions in the specification, this will lead to a considerable increase in the number of states in each cell's FDS.\\

To overcome this, these functions are implemented externally to the FDS's in the form of {\bf connectors}. A connector is an invariant proposition of the global state space. It is of the form $\vartheta = \pi(\vartheta_0, \vartheta_1, \vartheta_2, \ldots)$ where $\vartheta$ is an environment variable of an FDS $\M$, called a {\bf connector variable}, $\vartheta_0, \vartheta_1, \vartheta_2, \ldots$ are system variables of peers of $\M$ and $\pi$ is some proposition over $\vartheta_0, \vartheta_1, \vartheta_2, \ldots$ with the additional condition that $\vartheta$ is not a system variable of \emph{any} component in the architecture and $\vartheta$ is an environment variable only of $\M$. Thus, a connector introduces a new variable to the global state variables.\\

When proving properties over the composition of the specifications, these invariants can simply be assumed to hold at all times. In a simulation, these invariants are ensured by calculating the values of a connector variable $\vartheta$ whenever it is accessed by an FDS.\\

In our implementation of stationary target search, there are three types of connectors, shown as rectangles in \fig{amm2}. There is a connector for the engage signal $E_i$ of a cell $C_i$ that indicates how many robots are engaged on vertex $i$. This is expressed by
\begin{equation}
  \connE{i}.
  \label{eq:connE}
\end{equation}
Another connector is for the flag signal $\flag_j$ of a robot $R_j$ that indicates whether the vertex $i = \cellID_j$ the robot is currently on contains a target, i.e.\ $f_i = \mathrm{True}$. This expressed by
\begin{equation}
  \connF{j}.
  \label{eq:connF}
\end{equation}
The third type of connectors is for the dispatch signal $d_j$ of a robot $R_j$. Robot $R_j$ is dispatched to the vertex $i$ iff $d = i$ and $r = j$, i.e.\ if the robot that is offered to the allocator is $R_j$ and the allocator decides to dispatch a robot to the vertex $i$. This is expressed by
\begin{equation}
	\connD{j}.
	\label{eq:connD}
\end{equation}
Note that the connectors have access to the global state, i.e.\ they are not restricted by the position a robot is in. However, the expressions in \eqref{eq:connE}--\eqref{eq:connD} clearly indicate that only the local state is used to calculate the connector variables. By local state we mean that a cell can only use the values of robots on the corresponding vertex and a robot can only use the values of the cell on which it is currently.\\

\subsection{Global Specifications} \index{Global specification}

The composition of all components in \fig{amm2} has to satisfy the global specification for stationary target search discussed in \sec{globsearchspec} in a form that corresponds to the problem setting considered here. A target on vertex $i$ is indicated by the flag $f_i$ being asserted, and the goal of the allocator is the to get $\bar{\alpha} = 2$ robots to arrive at the vertex $i$. This can be expressed analogously to \eqref{eq:globstatgoal} by
\begin{equation*}
	\bigwedge_{i \in [0, n)}\always(f_i \rightarrow \eventually\bigvee_{\substack{j_1 \in [0, m) \\ j_2 \in [0, m) \backslash \{j_1\}}}(X_i^{j_1} \wedge X_i^{j_2})).
\end{equation*}

We now state the specifications of each component so that the asynchronous composition fulfills this global specification. This can be verified for example by using the Feedback Interconnection Refinement Rule in Proposition 1, but here we only note that the simulation yields the desired result.\\

For compacter specifications, define
\begin{align*}
\stay{\pi} &\defines \pi \wedge \next \pi, \\
\rais{\pi} &\defines \neg \pi \wedge \next \pi, \\
\clear{\pi} &\defines \pi \wedge \next \neg \pi,
\end{align*}
for any propositional formula $\pi$.\\

\subsection{Allocator Specification}

\begin{table}
\centering
\begin{tabular}{c|c|l}
\hline
Name & Domain & Description \\
\hline
$R$ & $[\e, m)$ & Indicates a robot that is ready \\
$F$ & $[\e, n)$ & Indicates a vertex that is flagged \\
$\DfA$ & $\mathbb{B}$ & Dequeue acknowledge signal from $\FIFOf$ \\
\hline
\end{tabular}
\caption{Environment variables of the allocator, $\AMM$.}
\label{tab:env:alloc}
\end{table}

\begin{table}
\centering
\begin{tabular}{c|c|l}
\hline
Name & Domain & Description \\
\hline
$\Df$ & $\mathbb{B}$ & Dequeue signal to $\FIFOf$ \\
$c$ & $[0, \bar{\alpha}]$ & Counts how many robots have been dispatched \\
$D$ & $[\e, m)$ & Dispatch signal to robots \\
\hline
\end{tabular}
\caption{System variables of the allocator, $\AMM$.}
\label{tab:sys:alloc}
\end{table}

The state variables of the allocator are shown in tables \tab{env:alloc} and \tab{sys:alloc}. The following specifications are for the FDS $\AMM$ in \fig{amm2}.\\

\subsubsection{Assumptions}

Initially, no robot is ready, so $\FIFOr$ offers no robot and $R = \e$. Moreover, $\FIFOf$ doesn't have a dequeue request to acknowledge, so $\DfA$ is low:
\begin{equation*}
  \ass{A}{1} = (R=\e) \wedge \neg\DfA.
\end{equation*}

$\FIFOr$ keeps track of the robots that are ready, and so, if available, presents a robot to the allocator that can be dispatched. This robot will not change unless the formerly presented robot has been dispatched:
\begin{equation*}
    \ass{A}{2} = \bigwedge_{j \in [0, m)} \always(\clear{R=j} \rightarrow D\neq\e).
\end{equation*}

Once a vertex is flagged, it is not unflagged unless a robot is dispatched to this cell, i.e.\ no target can rescue itself. More precisely, the flag offered by $\FIFOf$ may only change if a robot is dispatched or a dequeue signal is sent to the queue:
\begin{equation*}
    \ass{A}{3} = \bigwedge_{i \in [0, n)} \always(\clear{F=i} \rightarrow \Df \vee D\neq\e).
\end{equation*}

The dequeue acknowledgement stays high as long as the dequeue request stays high:
\begin{equation*}
    \ass{A}{4} = \always(\clear{\DfA} \rightarrow \next\neg\Df).
\end{equation*}

\subsubsection{Guarantees}

Initially, the dequeue signal is low, the counter is zero and no robot is dispatched:
\begin{equation*}
    \guar{A}{1} = \neg\Df \wedge c=0 \wedge D=\e.
\end{equation*}

Once the allocator has dispatched a robot, it monitors the ready signal it receives from $\FIFOr$. As long as this does not change, it does not clear the dispatch signal:
\begin{equation*}
    \guar{A}{2} = \bigwedge_{j \in [0, m)} \bigwedge_{i \in [0, n)} \always(D=i \wedge \stay{R=j} \rightarrow \next(D=i)).
\end{equation*}
However, if $\FIFOr$ presents a different robot to the allocator, this means that the robot to which the dispatch signal has been sent is no longer ready and has successfully reacted to the dispatch request. Thus dispatch is cleared:
\begin{equation*}
    \guar{A}{3} = \bigwedge_{j \in [0, m)} \bigwedge_{i \in [0, n)} \always(D=i \wedge \clear{R=j} \rightarrow \next(D=\e)).
\end{equation*}

If the allocator gets offered a flag $F = i$ from $\FIFOf$ and a robot is offered by $\FIFOr$, it may dispatch a robot to vertex $i$ by asserting $D = i$. In this case the counter $c$ must be increased by one:
\begin{equation*}
    \guar{A}{4} = \bigwedge_{\gamma < \bar{\alpha}} \bigwedge_{i \in [0,m)} \always (F=i \wedge D=\e \wedge c=\gamma \wedge R\neq \e \rightarrow \next(D=i \wedge c=\gamma+1)).
\end{equation*}

The dispatch signal may only be asserted if the counter $c$ is not saturated, some robot is ready and a cell is flagged:
\begin{equation*}
    \guar{A}{5} = \bigwedge_{i \in [0, m)} \always(D=\e \wedge \next(D=i) \rightarrow c<\bar{\alpha} \wedge R\neq\e \wedge F=i).
\end{equation*}
If a robot is dispatched and the dispatch signal is changed, it must be reset:
\begin{equation*}
    \guar{A}{6} = \bigwedge_{i \in [0, m)} \always(\clear{D=i} \rightarrow \next(D=\e)).
\end{equation*}
Moreover, dispatching a robot requires the dispatch signal to be clear:
\begin{equation*}
    \guar{A}{7} = \bigwedge_{i \in [0, m)} \always(\rais{D=i} \rightarrow D=\e).
\end{equation*}

The counter may only increase by one if a robot has been dispatched:
\begin{equation*}
    \guar{A}{8} = \always(\bigvee_{\gamma < \bar{\alpha}} (c=\gamma \wedge \next(c=\gamma+1)) \rightarrow \bigvee_{i \in [0, n)} (D=\e \wedge \next(D=i))).
\end{equation*}
Also, the counter may not decrease (unless it is reset), and it may only go up by one at a time. This requirement is captured by $0 \leq (\bar{\alpha} + \gamma' - \gamma) \mathrel\mathrm{mod} \bar{\alpha} \leq 1$, where $\gamma$ and $\gamma'$ are the values of the counter at the current and next step respectively:
\begin{equation*}
    \guar{A}{9} = \bigwedge_{\substack{\gamma \leq \bar{\alpha}, \gamma' \leq \bar{\alpha} \\ 0 \leq (\bar{\alpha} + \gamma' - \gamma) \mathrel\mathrm{mod} \bar{\alpha} \leq 1}} \always(c=\gamma \wedge \next(c=\gamma') \rightarrow \mathrm{False}).
\end{equation*}
This formulation \emph{forbids} certain changes of the counter $c$. A positive formulation that ensures that $c$ only changes in a permissible way can also be used but results in exactly the same synthesized FDS.\\

If no robot is dispatched, the counter must not change:
\begin{equation*}
    \guar{A}{10} = \bigwedge_{\gamma \leq \bar{\alpha}} \always(c=\gamma \wedge \stay{D=\e} \rightarrow \next(c=\gamma)).
\end{equation*}
The counter is only reset on a raising edge of the acknowledgement received from FIFOf:
\begin{equation*}
    \guar{A}{11} = \always(c=\bar{\alpha} \wedge c=0 \rightarrow \rais{\DfA}).
\end{equation*}
Conversely, if the counter is saturated and a raising edge of the acknowledgement is observed, then the counter is reset:
\begin{equation*}
    \guar{A}{12} = \always(c=\bar{\alpha} \wedge \rais{\DfA} \rightarrow \next(c=0)).
\end{equation*}

If enough robots have been dispatched to the vertex currently offered by $\FIFOf$, then the counter $c$ is saturated. The allocator must then ask $\FIFOf$ to offer the location of the next target, if available. For this purpose, the dequeue signal $\Df$ is raised if no acknowledgement is currently received:
\begin{equation*}
    \guar{A}{13} = \always(c=\bar{\alpha} \wedge \neg\DfA \wedge \neg\Df \rightarrow \next\Df).
\end{equation*}
The dequeue signal may only go high if the acknowledgement is low:
\begin{equation*}
    \guar{A}{14} = \always(\rais{\Df} \rightarrow \next\neg\DfA).
\end{equation*}
Also, the dequeue signal must stay high until an acknowledgement is received from FIFOf:
\begin{equation*}
    \guar{A}{15} = \always(\clear{\Df} \rightarrow \DfA).
\end{equation*}

\subsubsection{GR[1] Specification and Communication} \index{Communication} \index{Specification!GR[1]}

The above formulae are composed into the specification of the allocator:
\begin{equation*}
    \v_A = \bigwedge_{\alpha \in [1, 4]} \ass{A}{\alpha} \rightarrow \bigwedge_{\beta \in [1, 15]} \guar{A}{\beta},
\end{equation*}
which can be translated into a GR[1] specification.\\

The communication with $\FIFOf$ over the request signal $\Df$ and the acknowledgement signal $\DfA$ follows the four-phase handshake protocol of \sec{4PHP}. The assumption $\ass{A}{4}$ corresponds to III in $\M_S$, and the guarantees $\guar{A}{14}$ and $\guar{A}{15}$ correspond to VII and VIII respectively. IX is not required for a boolean request signal and XI is indirectly asserted by $\guar{A}{4}$, $\ass{A}{3}$ and $\guar{A}{13}$, \emph{given that} there are at least $\bar{\alpha}$ robots offered by $\FIFOr$. The property X is not guaranteed from the specification itself, but is required for the communication protocol to be valid. However, model checking\index{Model checking} the synthesized FDS of the allocator $\M_A \models \v_A$ reveals that also $\M_A \models \always(\DfA \rightarrow \next\Df)$. Thus the allocator satisfies the requirements to be a sender in the four-phase handshake protocol.\\

The allocator communicates with a robot also using the four phase handshake protocol. The request signal that is sent by the allocator is a combination of the index $R$ offered by $\FIFOr$ and the index of the target's vertex in $D$. In the notation of \sec{4PHP}, $(r = i) \Leftrightarrow (D = i \wedge R = j)$ for the request signal to robot $R_j$. The acknowledgement from robot $R_j$ is the signal $r_j$ being cleared. However, $r_j$ is not received as an environment variable by the allocator, so the robot index $R$ offered by $\FIFOr$ is used to extract this information: When $R = j$ changes, then robot $j$ has acknowledged the dispatch request. In the notation of \sec{4PHP}, $a \Leftrightarrow \clear{R = j}$. Thus $\guar{A}{2}$ and $\guar{A}{6}$ correspond to VIII and IX respectively, while X is implied by $\guar{A}{3}$ and $\ass{A}{2}$.\\

\subsection{Robot Specification}

\begin{table}
\centering
\begin{tabular}{c|c|l}
\hline
Name & Domain & Description \\
\hline
$d_j$ & $[\e, n)$ & The vertex the robot is dispatched to \\
$\flag_j$ & $\mathbb{B}$ & Indicates whether the vertex the robot is on is flagged \\
\hline
\end{tabular}
\caption{Environment variables of the robot $R_j$. Note that $d_j$ and $\flag_j$ are both provided by a connector.}
\label{tab:env:robot}
\end{table}

\begin{table}
\centering
\begin{tabular}{c|c|l}
\hline
Name & Domain & Description \\
\hline
$r_j$ & $\mathbb{B}$ & Indicates whether the robot is ready \\
$e_j$ & $\mathbb{B}$ & Indicates whether the robot is engaged \\
$\store_j$ & $[\e, n)$ & Stores the current target vertex \\
$\cellID_j$ & $[\e, n)$ & Stores the current position of the robot \\
$X^j_i, i \in [0, n)$ & $\mathbb{B}$ & Stores the current position of the robot \\
\hline
\end{tabular}
\caption{System variables of the robot $R_j$.}
\label{tab:sys:robot}
\end{table}

The environment and system variables of the robot $R_j$ are given in tables \tab{env:robot} and \tab{sys:robot} respectively. In the following specifications, the sub- and superscripts with the robot's index $j$ are omitted.

\subsubsection{Assumptions}

Initially, it is only known that the robot is not dispatched. This is a consequence of the robot also not being ready when it is started up (see guarantees):
\begin{equation*}
    \ass{R}{1} = (d=\e).
\end{equation*}

The allocator must wait for the robot to become ready before it can be dispatched:
\begin{equation*}
    \ass{R}{2} = \always(\clear{d=\e} \rightarrow r).
\end{equation*}

If the allocator sends a dispatch signal, this must remain the same until the robot acknowledges this by clearing its ready signal. Then FIFOr deletes this robot out of its list of robots that are ready, which in turn is (possibly) received by the allocator which only then may clear the dispatch signal:
\begin{equation*}
    \ass{R}{3} = \bigwedge_{i \in [0, n)} \always(\clear{d=i} \rightarrow \neg r).
\end{equation*}

\subsubsection{Guarantees}

Initially, the robot is not engaged, not ready and has no target stored:
\begin{equation*}
    \guar{R}{1} = \neg E \wedge \neg r \wedge \store=\e.
\end{equation*}

The robot becomes ready when no target is stored and it is not ready. From this time on, it will turn out that the atomic propositions $r = \mathrm{True}$ and $\store=\e$ are equivalent, because the robot is ready exactly if no destination is stored:
\begin{equation*}
    \guar{R}{2} = \always(\neg r \wedge \store\neq\e \rightarrow \next r).
\end{equation*}

If the robot receives a dispatch signal when it is ready, it fills its target storage with the appropriate destination:
\begin{equation*}
    \guar{R}{3} = \bigwedge_{i \in [0, n)} \always(\store=\e \wedge d=i \wedge r \rightarrow \next(\store=i)).
\end{equation*}
Moreover, a target may only be stored if the robot is ready. As a result of this, the robot must clear the ready signal $r$:
\begin{equation*}
    \guar{R}{4} = \bigwedge_{i \in [0, n)} \always(\store=\e \wedge \next(store=i) \rightarrow r \wedge d=i \wedge \neg\next r).
\end{equation*}
The target store may only be cleared to $\e$, and only if the robot is at the target location and no flag is detected. Then also the robot must assert that it is ready:
\begin{equation*}
    \guar{R}{5} = \bigwedge_{i \in [0, n)} \always(\clear{\store=i} \rightarrow \neg\flag \wedge X_i \wedge \next(\store=\e \wedge r)).
\end{equation*}

When a destination is stored then the robot eventually reaches the required cell:
\begin{equation*}
    \guar{R}{6} = \always(\store\neq\e \rightarrow \eventually \bigvee_{i \in [0, n)} (X_i \wedge \store=i)).
\end{equation*}

If the robot arrives at the correct cell and the flag is still up, then it engages:
\begin{equation*}
    \guar{R}{7} = \bigwedge_{i \in [0, n)} \always(\store=i \wedge X_i \wedge \flag \rightarrow e).
\end{equation*}
If the robot arrives at the correct cell but the flag is already cleared, then the store is cleared, the robot goes back to ready and disengages:
\begin{equation*}
    \guar{R}{8} = \bigwedge_{i \in [0, n)} \always(\store=i \wedge X_i \wedge \neg\flag \rightarrow \next(r \wedge \store=\e \wedge \neg e)).
\end{equation*}
The robot may only engage if it is in the correct cell and the flag is still sensed:
\begin{equation*}
    \guar{R}{9} = \always(\rais{e} \rightarrow \next\left(\bigvee_{i \in [0, n)} (X_i \wedge \store=i)\right) \wedge \next\flag).
\end{equation*}
Disengaging requires the target storage to be cleared. Note that this implicitly also includes all that is required to clear the target storage:
\begin{equation*}
    \guar{R}{10} = \always(\clear{e} \rightarrow \bigvee_{i \in [0, n)} \clear{\store=i}).
\end{equation*}

The robot is free to move as long as it is not engaged.
\begin{equation*}
    \guar{R}{11} = \bigwedge_{i \in [0, n)} \always(\stay(e) \rightarrow (X_i \leftrightarrow \next X_i)).
\end{equation*}

The allowable moves of the robot depend on the graph $G$ on which the search is performed. Since $G$ is given at specification time, the restrictions due to the topology are encoded as presented in \sec{maprepresentation}, and the formula $\v_G$ is included in the robot's specification.\\

\subsubsection{GR[1] Specification and Communication} \index{Specification!GR[1]} \index{Communication}

The above formulae are composed into the specification of the robot:
\begin{equation*}
    \v_R = \bigwedge_{\alpha \in [1, 3]} \ass{R}{\alpha} \rightarrow \v_G \wedge \bigwedge_{\beta \in [1, 11]} \guar{R}{\beta},
\end{equation*}
which can be translated into a GR[1] specification, see \sec{GR1}.\\

The robot $R_j$ may receive dispatch signals from the allocator via the dispatch signal $d_j$. It acknowledges such a request by clearing its ready signal $r_j$. The allocator does not directly receive $r_j$ but only the signal $R$ from $\FIFOr$, which must be taken into account when reasoning about the correctness of the communication protocols. Thus the acknowledgement signal is equivalent to $r_j = \mathrm{False}$, i.e.\ when the robot declares that it is no longer ready. Except in the initial state, this is also equivalent to $\store_j \neq \e$. The request signal is equivalent to $d_j \neq \e$. Therefore the guarantee $\guar{R}{4}$ corresponds to XIV, and the combination of $\guar{R}{3}$ and $\guar{R}{5}$ corresponds to XII. However, XV and XIII are not directly guaranteed by the robot but are required in Cases 4 and 5 of the proof of Proposition 3. XIII states that $d_j=\e$ necessarily results in $r_j=\mathrm{True}$. However, since the allocator does not require the liveness assumption I, the proof of Case 5 still works without guaranteeing XIII as follows:\\

To prove Precondition 1. we deduce II--VI from the conjuncts of the closures of $\varphi_S^s$ and $\varphi_R^s$. II and III follow from $\always(a = \mathrm{True})$ arising from the existential quantification of $\bar{a}$. IV--VI similarly follow from $\always(r = \e)$. To prove Precondition 2. we take $\bar{\bar{t}} = \mathrm{False}$ in the closure and the result follows trivially.\\

XV states that $\rais{r_j}$ requires $d_j=\e$. But the proof of Case 4 fails when omitting this guarantee. However, let us consider what would happen if the robot raised $r_j$ when $d_j\neq\e$. The allocator only receives $R$ from $\FIFOr$. If $R = \e$ then $D$ must be $\e$ from $\guar{A}{3}$ and $\guar{A}{5}$, so in this case $d_j\neq\e$ is not possible. Thus at that point $R\neq\e$ and so a robot $R_h$, $h \neq j$ is offered to the allocator in $R$. Thus when $r_j$ is raised, $R$ does not change necessarily, since $\FIFOr$ offers $R=h$ first. However, from the connector proposition for $d_j$, when $R\neq j$ then $d_j=\e$. So also in this case $d_j\neq\e$ is not possible.\\

\subsection{Cell and Queue Specifications}

\begin{table}
\centering
\begin{tabular}{l|c|c|l}
\hline
& Name & Domain & Description \\
\hline
Environment & $E_i$ & $[0, m]$ & The number of robots engaged on this cell \\
\hline
System & $f_i$ & $\mathbb{B}$ & Indicates whether the cell is flagged \\
\hline
\end{tabular}
\caption{Variables of the Cell $C_i$}
\label{tab:cellvars}
\end{table}

\begin{table}
\centering
\begin{tabular}{l|c|c|l}
\hline
& Name & Domain & Description \\
\hline
Environment & $r_j, j \in [0, n)$ & $\mathbb{B}$ & The ready signal received from robot $R_j$ \\
\hline
System & $R$ & $[\e, n)$ & The index of the robot offered to $\AMM$\\
\hline
\end{tabular}
\caption{Variables of $\FIFOr$.}
\label{tab:fiforvars}
\end{table}

\begin{table}
\centering
\begin{tabular}{l|c|c|l}
\hline
& Name & Domain & Description \\
\hline
Environment & $f_i, i \in [0, m)$ & $\mathbb{B}$ & The flag received from cell $C_i$\\
            & $\Df$ & $\mathbb{B}$ & The dequeue request from $\AMM$\\
\hline
System & $F$ & $[\e, n)$ & The index of the vertex offered to $\AMM$\\
       & $\DfA$ & $\mathbb{B}$ & The dequeue acknowledgement \\
\hline
\end{tabular}
\caption{Variables of $\FIFOf$.}
\label{tab:fifofvars}
\end{table}

Each cell $C_i$ in the architecture in \fig{amm2} has just one environment variable $E_i$, indicating the number of robots engaged on the cell, and one system variable $f_i$, the flag of the cell. These variables are shown in \tab{cellvars}. In the specification, we will again drop the subscripts from the variables and write just $f$ and $E$.\\

A cell has to satisfy only two guarantees. The flag can only be cleared if there are at least $\bar{\alpha} = 2$ robots engaged on the cell:
\begin{equation*}
	\guar{C}{1} = \always(\clear{f} \rightarrow E \geq \bar{\alpha}).
\end{equation*}
Moreover, we include a liveness property, ensuring that the cell will always be flagged eventually:
\begin{equation*}
	\guar{C}{2} = \always\eventually f.
\end{equation*}

In order for the specification to be realizable, we need to assume that as long as a cell is flagged, the number of robots that are engaged on the vertex associated with that cell will never decrease:
\begin{equation*}
	\ass{C}{1} = \bigwedge_{\gamma \in [0, m)}\always(f \wedge E=\gamma \rightarrow \next(E\geq\gamma)).
\end{equation*}

The FIFO\index{FIFO} queues $\FIFOr$ and $\FIFOf$ are not synthesized as FDS's due to the state space explosion problem, cf.\ \sec{statarch}. Instead, a thread is implemented that guarantees certain properties to the allocator $\AMM$. In particular, the environment assumptions of $\AMM$ on the local system variables controlled by $\FIFOr$ and $\FIFOf$ must be satisfied. The conceptual environment and system variables of the two queues are shown in \tab{fiforvars} and \tab{fifofvars} respectively.\\

From $\ass{A}{1}$, the allocator requires that initially ``$\DfA = \mathrm{False}$'', which can easily be implemented in $\FIFOf$. Also, $\ass{A}{4}$ requires the dequeue acknowledgement $\DfA$ to only be lowered if also the dequeue request $\Df$ is false, which is similarly straightforward to implement.\\

There is only one requirement on $\FIFOr$, imposed by $\ass{A}{3}$. The robot index $R$ that is offered to $\AMM$ may hence only be changed if a robot is dispatched. Since $\FIFOr$ keeps track on which robots are ready, and robots may only lower their ready signals $r$ following a dispatch $D$ (cf.\ the connector \eqref{eq:connD}), this requirement is automatically met.\\

Lastly, from $\ass{A}{3}$, the flag offered to the allocator may only be changed if a dequeue signal $\Df$ is received by $\FIFOf$ or a robot is dispatched. $\FIFOf$ therefore only dequeues a flag either if $\Df$ is high, or the flag is lowered. By the cell's guarantee $\guar{C}{1}$ and the connector \eqref{eq:connE}, a flag can only be lowered if at least $\bar{\alpha}$ robots are engaged on the corresponding cell. This again is only the case if $D \neq \e$, i.e.\ a robot is dispatched.\\

\subsection{Synthesis} \index{Synthesis}

The specifications given above were successfully synthesized using the TuLiP\index{TuLiP} front end. \fig{statstates} shows the number of states of the allocator and the robot when varying the size of the topology and the number of robots available.\\

The size of the state space of the robot does not depend on the number of robots in the architecture, but the size of the state space of both the allocator and the robots increases quadratically with the number of vertices in the graph representation of the topology. Also, the size of the state space of the allocator increases slightly with the number of robots.\\

\begin{figure}
\centering
	\psfrag{alloc2}[ll][cc]{\footnotesize{$\AMM$, $m=2$}}
	\psfrag{alloc3}[ll][cc]{\footnotesize{$\AMM$, $m=3$}}
	\psfrag{alloc4}[ll][cc]{\footnotesize{$\AMM$, $m=4$}}
	\psfrag{alloc5}[ll][cc]{\footnotesize{$\AMM$, $m=5$}}
	\psfrag{alloc6}[ll][cc]{\footnotesize{$\AMM$, $m=6$}}
	\psfrag{robot}[ll][cc]{\footnotesize{Robots $R_j$, any $m$}}
	\psfrag{States}[cc][cc]{\footnotesize{States}}
	\psfrag{Vertices}[cc][cc]{\footnotesize{Vertices, $n$}}
	\includegraphics[width=0.8\textwidth]{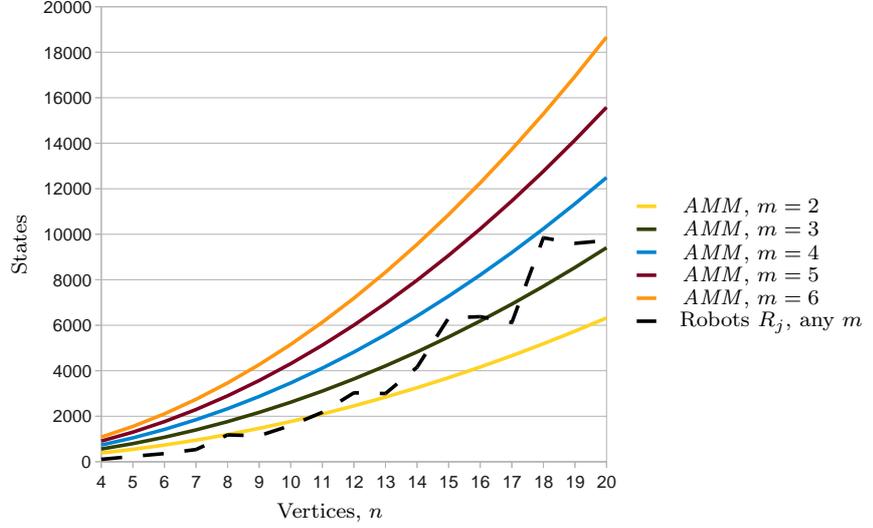}
\caption{States of allocator and robot for the stationary target search. The relation for both allocator $\AMM$ and robots $R_j$ is quadratic in the number of vertices $n$.}
\label{fig:statstates}
\end{figure}

The time for synthesizing the controllers is also polynomial in the size of the topology, as would be expected: The size of the specification increases at most quadratically with $n$, the number of vertices in the graph representing the topology, and the synthesis time is cubic in the size of the specification. By plotting the synthesis time on a logarithmic scale as in \fig{stattime} shows that the complexity must be subexponential. A rough analysis yields polynomial complexity of sixth order in $n$, but this cannot be seen accurately from the available data.\\

The timing measurements have been conducted on a $2.4\unit{GHz}$ Intel Core i7 CPU, with $4\unit{GB}$ of working memory. The available memory is likely to also affect the performance because the JTLV implementation uses the JVM Garbage Collector. Note also that the graph presented in this section was obtained for just one random graph for each value of $n$. Other random graphs give different state space sizes and synthesis times, but the polynomial complexity results still hold.\\

\begin{figure}
\centering
	\psfrag{alloc2}[ll][cc]{\footnotesize{allocator, $m=2$}}
	\psfrag{alloc3}[ll][cc]{\footnotesize{allocator, $m=3$}}
	\psfrag{alloc4}[ll][cc]{\footnotesize{allocator, $m=4$}}
	\psfrag{alloc5}[ll][cc]{\footnotesize{allocator, $m=5$}}
	\psfrag{alloc6}[ll][cc]{\footnotesize{allocator, $m=6$}}
	\psfrag{robot}[ll][cc]{\footnotesize{robots, any $m$}}
	\psfrag{Synthesis Time [s]}[cc][cc]{\footnotesize{Synthesis Time [s]}}
	\psfrag{Vertices}[cc][cc]{\footnotesize{Vertices, $n$}}
	\includegraphics[width=0.8\textwidth]{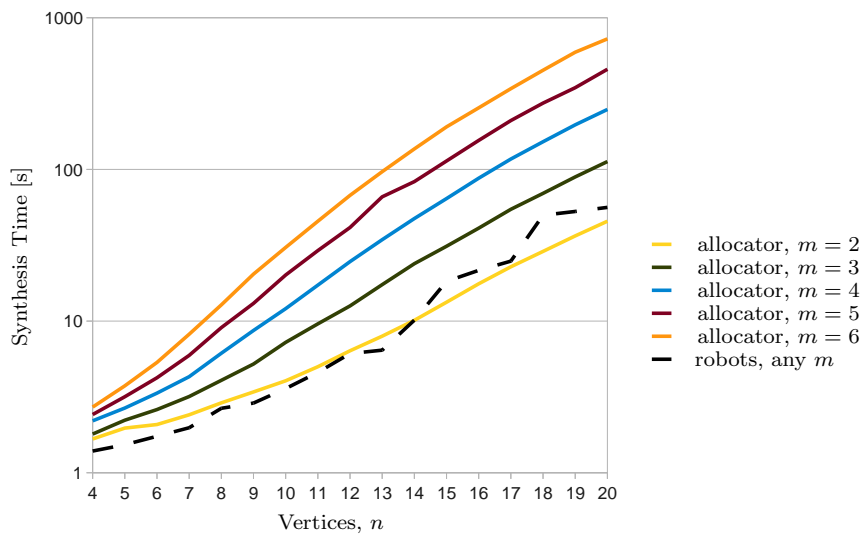}
\caption{Time to synthesize allocator and robot for the stationary target search with $n$ vertices and $m$ robots. Note that the ordinate is logarithmic but the relation is polynomial.}
\label{fig:stattime}
\end{figure}

%
%
%

\subsection{Discussion}

Stationary target search has been successfully simulated for various numbers of robots and vertices. When more than two robots are used, the allocator explicitly requests a new flag to be offered by $\FIFOf$ and dispatches a robot to the corresponding vertex. Thus, with four and more robots several targets can be rescued simultaneously.\\

However, as is clear from \fig{statstates}, the size of the state space grows very fast with the size of the graph. This affects the efficiency of synthesis. The synthesis of the controllers for $20$ vertices and $6$ robots already took about $15\unit{min}$, and much longer times are expected for larger graphs. However, space seems to be even a more pressing issue. Larger problems are susceptible to heap overflow due to unsuccessful JVM garbage collection. For example, synthesis of a controller for $16$ vertices and $5$ robots is not successful when only permitting $1\unit{GB}$ of working memory.\\

The increase of the state space is mostly due to the added environment variables from the communication protocol. The assumptions on these additional environment variables cannot be strengthened sufficiently to reduce the state space further. The communication in this specification already deviates slightly from the four-phase handshake protocol defined in \sec{4PHP} in order to alleviate this problem, as any additional signal like a boolean acknowledgement roughly doubles the state space. Also, beyond a certain graph size, synthesis seems impractical. If this limitation cannot be overcome, then this type of specification seems unsuitable for practical applications.\\

\section{Moving Target Search} \label{sec:implmovt}

In contrast to USAR, the targets in WiSAR\index{WiSAR} are considered to be able to move. That means that the stationary target search methods are no longer sufficient, and we implement the distributed multi stage moving target search strategy developed in \sec{mstage}.\\

There is no central allocator, hence the specification is easier than that of stationary target search. However, as is evident from \sec{movtsearch}, justifying that a given number robots with controllers synthesized from the specification reliably find all moving targets requires more effort than with stationary target search.\\

\subsection{Communication Structure} \label{sec:commstructure} \index{Communication}

As indicated in \sec{negotiations}, the robots are organized in a circular structure of authority and communication. A robot $R_j$ has exactly one slave and exactly one master. \fig{commstructure} shows such a cycle. The state variables of the robots are explained in \tab{inp:robot} and \tab{out:robot}.\\

\newcommand{\dist}{0.2pc}
\newcommand{\ratio}{0.35}
\newcommand{\robot}[1]{*++[o][F-]{R_{#1}}}
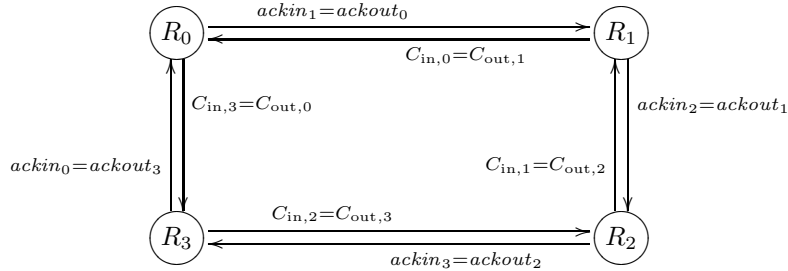
\begin{figure}[h!]
\begin{displaymath}
    \xymatrix@R=2cm@C=5cm{ \robot{0} \ar@<\dist>[r]^(\ratio){\ackin_1 = \ackout_0} \ar@<\dist>[d]^(\ratio){\Cin{,3} = \Cout{,0}} & \robot{1} \ar@<\dist>[d]^(\ratio){\ackin_2 = \ackout_1} \ar@<\dist>[l]^(\ratio){\Cin{,0} = \Cout{,1}}\\ \robot{3} \ar@<\dist>[u]^(\ratio){\ackin_0 = \ackout_3} \ar@<\dist>[r]^(\ratio){\Cin{,2} = \Cout{,3}} & \robot{2} \ar@<\dist>[l]^(\ratio){\ackin_3 = \ackout_2} \ar@<\dist>[u]^(\ratio){\Cin{,1} = \Cout{,2}}}
\end{displaymath}
\caption{Four robots communicating in the moving target search architecture. Note that only the local system and environment variables directly relevant for communication are shown.}
\label{fig:commstructure}
\end{figure}

We use the standard four-phase handshake protocol introduced in \sec{4PHP}. Hence no sophisticated arguments as in the previous section on stationary target search are required to justify the correctness of the communication protocols used. Also, the correctness of the composition of all synthesized components is a simple extension of Proposition 2 in \sec{4PHP}.\\

\subsection{Robot Specifications} \label{sec:robspec}

Similar to the stationary target search, the global specification must state that any target is eventually found on the given topology. Since the search strategies developed in \sec{distrsoln} ensure that the graph is cleared, this global requirement is necessarily satisfied, given that the architecture contains sufficiently many robots. We therefore only state the specifications that are synthesized into robots implementing a winning strategy for the cops. It is then easy to extend this to include sensors to detect targets and react appropriately (e.g.\ by engaging).\\

\begin{table}
\centering
\begin{tabular}{c|c|l}
\hline
Name & Domain & Description \\
\hline
$\Cin{,j}$ & $[\e, n)$ & Vertex the robot is sent to by its master \\
$\ackin_j$ & $\mathbb{B}$ & Acknowledgement received from its slave \\
$\cst_j$ & $\{\ccl, \cpc, \ccr, \csv\}$ & State of the vertex on which the robot is \\
\hline
\end{tabular}
\caption{Environment Variables of Robot $R_j$}
\label{tab:inp:robot}
\end{table}

\begin{table}
\centering
\begin{tabular}{c|c|l}
\hline
Name & Domain & Description \\
\hline
$M_j$ & $\{\mgu, \mcs, \mpc, \mcl\}$ & Mode of the robot \\
$\store_j$ & $[\e, n)$ & Vertex stored for the clearing mode\\
$\cellID_j$ & $[0, n)$ & Current position of the robot \\
$X_i^j, i \in [0, n)$ & $\mathbb{B}$ & True if robot $R_j$ is in position $i$\\
$c_j$ & $[\e, \bar{\gamma}_\mathrm{max}]$ & Counter in clearing mode\\
$\cnt_j$ & $[0, |p|)$ & Counter in search modes\\
$\mcnt_j$ & $[0, |p|)$ & Maximum counter value in search modes\\
$\ackout_j$ & $\mathbb{B}$ & Acknowledgement sent to master \\
$\Cout{,j}$ & $[\e, n)$ & Vertex to send the robot's slave to \\
\hline
\end{tabular}
\caption{System Variables of Robot $R_j$}
\label{tab:out:robot}
\end{table}

In the following specifications, the subscripts and superscripts indicating the index of the robot are omitted for notational convenience.\\

\subsubsection{Guarantees: Topology}

Similar to the stationary target search, the allowable moves of the robot have to be defined in the guarantee part of the specification. The restrictions due to the topology are encoded as presented in \sec{maprepresentation}, and the formula $\v_G$ encoding the topology is included in the robot's specification.\\

Note that the targets are restricted to move on the same topology. However, since they are assumed to be able to move with infinite speed, and the graph is strongly connected, the robot cannot assume any restriction on the movement of the targets.\\

\subsubsection{Guarantees: Guarding Mode}

In the guarding mode $\Mgu$, the robot has to stay at the current vertex unless a command from its master is received via $\Cin{}$:
\begin{equation*}
	\guar{M}{1} = \bigwedge_{i \in [0, n)}\always(\Mgu \wedge \Cin{}=\e \wedge X_i \rightarrow \next(X_i \wedge \Mgu)).
\end{equation*}
However, if a command is received, then the clearing mode $\Mcl$ has to be entered, setting up the store and counter appropriately:
\begin{equation*}
	\guar{M}{2} = \bigwedge_{i \in [0, n)}\always(\Mgu \wedge \Cin{}=i \rightarrow \next(\Mcl \wedge \store=i \wedge c=\e)).
\end{equation*}
If the guarding mode is entered, then the robot's slave is instructed to clear all out-edges of the current vertex $v_i$ by asserting the index $i$ on $\Cout{} = i$. However, if the guarding mode is entered because the counter $c$ is saturated, this means that no vertex was found in $\Mcs$ or $\Mpc$. In this case we know that the graph is cleared and so $\Cout{}$ stays at $\e$.\\
\begin{align*}
	\guar{M}{3} = \bigwedge_{i \in [0, n)}\bigwedge_{\gamma \in [0, |p|)}\always(\Mcs \wedge &\next(\Mgu) \wedge \next(X_i)\,\wedge\\
 &\cnt=\gamma \wedge \mcnt\neq\gamma \rightarrow \next(\Cout{}=i)).
\end{align*}

\subsubsection{Guarantees: Searching Modes}

The searching modes $\Mcs$ and $\Mpc$ can be specified together, as they differ only slightly by the state of the vertex the robot is searching for and the mode that is entered if such a vertex is encountered.\\

First, when entering a searching mode, the counters must be initialized correctly:
\begin{align*}
	\guar{M}{4} = \bigwedge_{i \in [0, n)}\always(&X_i \wedge (\rais{\Mcs} \vee \rais{\Mpc}) \\ &\rightarrow \next(\cnt=\gamma_i \wedge \mcnt=\nextval{\gamma_i} \wedge X_i)),
\end{align*}
where $\gamma_i$ is the index of the first occurrence of the vertex $v_i$ in the path $p$. Note that $\nextval{\gamma} = (\gamma + 1 + |p|) \mathop\mathrm{mod} |p|$ as defined in \sec{globsearchspec} before \eqref{eq:countincrease}.\\

The robot moves along the path $p$ one vertex at a time, making sure that the counter $\cnt$ increases at every transition:
\begin{align*}
	\guar{M}{5} = \bigwedge_{\gamma \in [0, |p|)}\always(&X_{p(\gamma)} \wedge \cnt=\gamma \wedge \mcnt\neq\gamma\,\wedge \\ &((\Mcs \wedge \next(\cst\neq\csv)) \vee (\Mpc \wedge \next(\cst\neq\cpc)))\\ & \rightarrow \next(\cnt=\nextval{\gamma} \wedge X_{p(\nextval{\gamma})})).
\end{align*}

If a start vertex\index{Start vertex} has been found, then enter the guarding mode $\Mgu$:
\begin{equation*}
	\guar{M}{6} = \bigwedge_{i \in [0, n)}\always(\Mcs \wedge \next(\cst=\csv) \wedge X_i \rightarrow \next(\Mgu \wedge c=\e \wedge X_i)).
\end{equation*}
Similarly, if a partially cleared vertex $v$ has been found, then enter the clearing mode $\Mcl$, initialized to clear the out-edges of $v$:
\begin{align*}
	\guar{M}{7} = \bigwedge_{i \in [0, n)}\always(&\Mpc \wedge \next(\cst=\cpc) \wedge X_i \\ &\rightarrow \next(\Mcl \wedge c=\e \wedge \store=i \wedge X_i)).
\end{align*}
Also, if the maximum counter value is reached in $\Mpc$, then there is no partially cleared vertex and the search for a start vertex $\Mcs$ is entered:
\begin{equation*}
	\guar{M}{8} = \bigwedge_{\gamma \in [0, |p|)}\always(\Mpc \wedge \cnt=\gamma \wedge \mcnt=\gamma \wedge \next(\cst\neq\cpc) \rightarrow \next(\Mcs)).
\end{equation*}
Note that the values of the counters are reset to the correct values necessarily by the guarantee $\guar{M}{4}$. Similarly, if the maximum counter value is reached in $\Mcs$, then there is no contaminated (start) vertex and the graph must necessarily be cleared and the robot becomes dormant in guarding mode $\Mgu$. Note that an additional mode can be introduced in this case that allows the robots to move arbitrarily on the graph, but here only the concept of moving target search should be introduced. The corresponding guarantee is:
\begin{equation*}
	\guar{M}{9} = \bigwedge_{\gamma \in [0, |p|)}\always(\Mcs \wedge \cnt=\gamma \wedge \mcnt=\gamma \wedge \next(\cst\neq\csv) \rightarrow \next(\Mgu)).
\end{equation*}

Now a few guarantees are required that prevent the counters from changing when they should not (so called frame axioms). The counter of the position in the path $p$, $\cnt$ may only increase until it overflows, corresponding to the path closing in a cycle.
\begin{align*}
	\guar{M}{10} = \always(&\stay{\Mcs} \vee \stay{\Mpc} \\&\rightarrow \bigvee_{\gamma \in [0, |p|)}(\cnt=\gamma \wedge \next(\cnt=\nextval{\gamma}))).
\end{align*}
Similarly, the value stored as maximum for the counter may not change:
\begin{align*}
	\guar{M}{11} = \always(&\stay{\Mcs} \vee \stay{\Mpc} \\&\rightarrow \bigvee_{\gamma \in [0, |p|)}(\mcnt=\gamma \wedge \next(\mcnt=\gamma))).
\end{align*}
Since $\cnt$ and $\mcnt$ are only used in $\Mcs$ or $\Mpc$, they are set to zero otherwise:
\begin{equation*}
	\guar{M}{12} = \always(\neg(\Mcs \vee \Mpc) \rightarrow \cnt=0 \wedge \mcnt=0).
\end{equation*}

Two more guarantees are required to prevent the robot from leaving the searching mode if no start vertex or partially cleared vertex was encountered or the counter has not reached its maximum value:
\begin{align*}
	&\guar{M}{13} = \always(\clear{\Mcs} \rightarrow \bigvee_{\gamma \in [0, |p|)}(\cnt=\gamma \wedge \mcnt=\gamma) \wedge \next(\cst=\csv)),\\
	&\guar{M}{14} = \always(\clear{\Mpc} \rightarrow \bigvee_{\gamma \in [0, |p|)}(\cnt=\gamma \wedge \mcnt=\gamma) \wedge \next(\cst=\cpc)).
\end{align*}

\subsubsection{Guarantees: Clearing Mode}

The clearing mode is the most involved mode, but also the mode that is responsible for actual progress in clearing edges: no edges can be cleared by a robot that is not in mode $\Mcl$.\\

First, on entering clearing mode, the index of a vertex is written to $\store$ so that the robot knows for which vertex the out-edges must be cleared. This can only happen when the robot is guarding and called by its master to move, or when the robot has found a partially cleared cell that it is now about to clear:
\begin{align*}
	\guar{M}{15} = \bigwedge_{i \in [0, n)}\always(&\store=\e \wedge \next(\store=i) \rightarrow \\ &(\Cin{}=i \wedge \Mgu) \vee (\Mpc \wedge \next(\cst=\cpc) \wedge X_i)).
\end{align*}
Also, $\store$ may only change from $\e$ if also the counter $c$ is reset to $\e$ or alternatively if the robot enters the clearing mode from a search for a partially cleared vertex:
\begin{align*}
	\guar{M}{16} = \always(\clear{\store=\e} \rightarrow &(\next(c=\e)\wedge \Mgu \wedge\next(\Mcl)) \vee \\ &(\Mpc \wedge\next(\cst=\cpc) \wedge\next(\Mcl))).
\end{align*}
Moreover, if $\store\neq\e$, then it may only change to $\e$, i.e.\ it is not permitted to immediately change from sliding down the out-edges of one vertex to sliding down the out-edges of another. Such a change also requires the counter $c$ to be saturated, indicating that all out-edges of the vertex held in $\store$ are successfully cleared:
\begin{equation*}
	\guar{M}{17} = \bigwedge_{i \in [0, n)}\always(\clear{\store=i} \rightarrow \next(\store=\e) \wedge c=\bar{\gamma}_i \wedge \next(c=\e)).
\end{equation*}
Conversely, if the counter $c$ is saturated, then the robot has successfully cleared all out-edges of the vertex held in $\store$ and thus enters the search for a partially cleared vertex:
\begin{equation*}
	\guar{M}{18} = \bigwedge_{i \in [0, n)}\always(\store=i \wedge c=\bar{\gamma}_i \rightarrow \next(\store=\e \wedge c=\e \wedge \Mpc)).
\end{equation*}

The counter $c$ is always diligently set to $\e$ when the clearing mode is entered (see $\guar{M}{26}$), representing that the robot first has to move to the vertex held in $\store$. This latter requirement is expressed using the following property:
\begin{equation}
	\always(\store\neq\e\wedge c=\e \rightarrow \eventually \bigvee_{i \in [0, n)}(\store=i\wedge X_i)).
	\label{eq:gohome}
\end{equation}
Note that we write $\always(\pi \rightarrow \eventually\bigvee_{i \in [0,n)}(\store=i\wedge X_i))$ rather than the more intuitively obvious $\bigwedge_{i \in [0,n)}\always(\pi \wedge \store=i \rightarrow \eventually X_i)$. This is because the latter introduces $n$ response formulae\index{Response formula}, each of which requires a trigger\index{Trigger} variable to be added to the system variables, which would multiply the potential number of states by $2^n$. Using the former contracted response property only requires one trigger and thus the synthesized FDS has fewer states.\\

While the robot moves to the vertex held in $\store$, the counter $c$ must remain at $\e$:
\begin{equation*}
	\guar{M}{19} = \bigwedge_{i \in [0, n)}\always(\store=i \wedge c=\e \wedge \neg X_i \rightarrow \next(c=\e)).
\end{equation*}	
On reaching the vertex, the robot sets the counter to zero. It also stays in the vertex so that the state of the robot is consistent:
\begin{equation*}
	\guar{M}{20} = \bigwedge_{i \in [0, n)}\always(\store=i \wedge c=\e \wedge X_i \rightarrow \next(c=0 \wedge X_i)).
\end{equation*}

Given that the robot is in the vertex held in $\store$ and the counter $c$ has a value not equal to $\e$, an out-edge can be cleared. The corresponding property has already been introduced in \sec{mstage}:
\begin{equation*}
	\guar{M}{21} = \bigwedge_{i \in [0, n)} \bigwedge_{\gamma \in [0, \bar{\gamma}_i)} \always(\store = i \wedge c = \gamma \wedge X_i \rightarrow \next(X_{w_\gamma^i} \wedge c = \gamma + 1)).
\end{equation*}
Recall that $\bar{\gamma}_i$ is the number of out-edges of the vertex $i$ and $w_{\gamma^i}$ is the (unique) vertex such that there is an edge $e$ from $i$ to $w_{\gamma^i}$ and $e$ is the $\gamma^\mathrm{th}$ edge to be cleared by the robot. After clearing the edge, the robot must move back to the vertex held in $\store$, which is expressed by the guarantee
\begin{equation}
	\always\left(\bigvee_{i \in [0, n)} \store = i \wedge \bigvee_{\gamma \in [0, \bar{\gamma}_i)}(c = \gamma + 1 \wedge X_{w_\gamma^i})) \rightarrow \eventually \bigvee_{i \in [0, n)} (\store = i \wedge X_i)\right).
	\label{eq:goback}
\end{equation}
This again is a response property, which would require a trigger to be introduced. Instead of having both response properties \eqref{eq:gohome} and \eqref{eq:goback}, we combine them into one guarantee:
\begin{align*}
	\guar{M}{22} = \always((&\store\neq\e\wedge c=\e) \vee \bigvee_{i \in [0, n)} (\store = i \wedge \bigvee_{\gamma \in [0, \bar{\gamma}_i)}(c = \gamma + 1 \wedge X_{w_\gamma^i})) \rightarrow \\ &\eventually \bigvee_{i \in [0, n)} (\store = i \wedge X_i)).
\end{align*}
This is of course only possible since the right hand side of the implication $\rightarrow$ is the same in both \eqref{eq:gohome} and \eqref{eq:goback}.\\

Again, the counter $c$ must be restricted. It may only increase by one if an edge is cleared:
\begin{equation*}
	\guar{M}{23} = \bigwedge_{\gamma \in [0, \bar{\alpha})}\always\left(c=\gamma \wedge \next(c=\gamma+1) \rightarrow \bigvee_{i \in [0, \bar{\gamma}_i)}(\store=i \wedge X_i \wedge \next X_{w_\gamma^i})\right).
\end{equation*}
Also, the counter may not decrease unless it is reset. Moreover, it may only increase by one:
\begin{equation*}
	\guar{M}{24} = \bigwedge_{i \in [0, n)}\always(\store=i \rightarrow \bigvee_{\substack{\gamma_1 \in [\e, \bar{\gamma}_i] \\ \gamma_2 \in [\e, \bar{\gamma}_i]}}\namefont{increase}(c, \gamma_1, \gamma_2, i)),
\end{equation*}
where $\namefont{increase}(c, \gamma_1, \gamma_2, i)$ is a formula that only holds if the counter changes by an admissible amount, which is defined as follows:
\begin{equation*}
	\namefont{increase}(c, \gamma_1, \gamma_2, i) =
	\left\{
	\begin{array}{ll}
		c=\gamma_1 \wedge \next(c=\gamma_2) & \mbox{if } \gamma_2=\gamma_1+1 \\&\mbox{or } \gamma_1=\bar{\gamma}_i \wedge \gamma_2=\e \\&\mbox{or } \gamma_1=\gamma_2 \\
		\mathrm{False} & \mbox{otherwise}
	\end{array}
	\right.
\end{equation*}
The counter may only be reset to $\e$ if all out-edges have been cleared:
\begin{equation*}
	\guar{M}{25} = \always(\rais{c=\e} \rightarrow \bigvee_{i \in [0, n)}(\store=i \wedge c=\bar{\gamma}_i)).
\end{equation*}

If the robot is not in the clearing mode, then the counter $c$ and $\store$ are not used, and they are therefore set to $\e$:
\begin{equation*}
	\guar{M}{26} = \always(\Mncl \rightarrow c=\e \wedge \store=\e).
\end{equation*}
Finally, the robot has to remain in the clearing mode as long as there are still out-edges of the vertex held in $\store$ to be cleared.
\begin{equation*}
	\guar{M}{27} = \always(\Mcl \wedge \neg\bigwedge_{i \in [0, n)}(\store\neq i \vee c=\bar{\gamma}_i) \rightarrow \next(\Mcl)).
\end{equation*}

\subsubsection{Communication}

The robots communicate with each other using the four-phase handshake protocol introduced in \sec{4PHP}. Each robot communicates with its slave and its master as receiver and sender respectively, see \sec{negotiations}. The sender specifications, i.e.\ for the communication of the robot as master, consist of the following guarantees:
\begin{eqnarray*}
	&\mathrm{VII} &\guar{M}{28} = \always(\Mgu \wedge \clear{\Cout{}=\e} \rightarrow \neg\ackin), \\
	&\mathrm{VIII} &\guar{M}{29} = \always(\Mgu \wedge \rais{\Cout{}=\e} \rightarrow \ackin),\\
	&\mathrm{X} &\guar{M}{30} = \always(\Mgu \wedge \ackin \rightarrow \next(\Cout{}=\e)),
\end{eqnarray*}
and of the following assumptions:
\begin{eqnarray*}
	&\mathrm{II} &\ass{M}{1} = \always(\rais{\ackin} \rightarrow \Cout{}\neq\e), \\
	&\mathrm{III} &\ass{M}{2} = \always(\clear{\ackin} \rightarrow \Cout{}=\e).
\end{eqnarray*}
Note however, that the assumption $\always(\Cout{}=\e \rightarrow \eventually\neg\ackin)$ corresponding to I is not included, since then the specification would not be synthesizable. This is because the rest of the specification implicitly includes $\always(\pi(\ackin) \rightarrow \eventually\pi'(\Cout{}))$ for some propositional formulae $\pi$ and $\pi'$. However, leaving out an assumption does not render the protocol incorrect (while leaving out a guarantee would). The guarantee IX is not necessary, since the request signal $\Cout{}$ is boolean and a specification $\always(\Cout{} \wedge \next\neg\Cout{} \rightarrow \next\neg\Cout{})$ is redundant. The conjunct $\Mgu$ in the antecedents of the guarantees is necessary to ensure realizability since $\Cout{}$ must be permitted to be asserted when guarding mode is entered (cf.\ guarantee $\guar{M}{3}$).\\

The receiver specifications, i.e.\ for the communication of the robot as a slave consist of the following guarantees:
\begin{eqnarray*}
	&\mathrm{XII} &\guar{M}{31} = \always(\Cin{}\neq\e \rightarrow \next\ackout),\\
	&\mathrm{XIII} &\guar{M}{32} = \always(\Cin{}=\e \rightarrow \next\neg\ackout),\\
	&\mathrm{XIV} &\guar{M}{33} = \always(\rais{\ackout} \rightarrow \Cin{}\neq\e),\\
	&\mathrm{XV} &\guar{M}{34} = \always(\clear{\ackout} \rightarrow \Cin{}=\e),
\end{eqnarray*}
and of the following assumptions:
\begin{eqnarray*}
	&\mathrm{VI} &\ass{M}{3} = \always(\rais{\Cin{}=\e} \rightarrow \ackout),\\
	&\mathrm{VII} &\ass{M}{4} = \always(\clear{\Cin{}=\e} \rightarrow \neg\ackout).
\end{eqnarray*}
Note again that the assumption $\always(\ackout \rightarrow \eventually(\Cin{}=\e))$ corresponding to V is not included due to problems with synthesizability. But since all guarantees are included, the communication protocol works as proved above in Proposition 2 in \sec{4PHP}.\\

\subsubsection{Assumptions}

While the specification above is perfectly realizable, due to the large number of possible valuations of the environment variables ($|\Cin{}| \times |\ackin| \times |\cst| = 8n$) it is beneficial to make assumptions on the environment to reduce the number of transitions that have to be included in the synthesized FDS (see \sec{compsingle}).\\

Since the state of a vertex is only required in the searching modes $\Mcs$ and $\Mpc$, we can artificially set $\cst$ to some arbitrary value in these modes. Then the robot can make the following two assumptions on the environment:
\begin{align*}
	&\ass{M}{5} = \always(\Mncs \wedge \Mnpc \rightarrow \next(\cst=\ccr)),\\
	&\ass{M}{6} = \always(\stay{\Mncs \wedge \Mnpc} \rightarrow \next(\cst=\ccr)).
\end{align*}
Here we chose to set $\cst= \ccr$. Note that this has to be guaranteed by the global storage of the graph state, cf. \sec{globstore}.\\

\subsection{Synthesis}
	
The full specification of the robots results is the following
\begin{equation*}
	\v_M = \bigwedge_{1 \leq \alpha \leq 6} \ass{M}{\alpha} \rightarrow v_G \wedge \bigwedge_{1 \leq \beta \leq346} \guar{M}{\beta},
\end{equation*}
which can be translated into a GR[1] specification and hence synthesized with TLV\index{TLV}.\\

The robot specifications given above were successfully synthesized using the TuLiP\index{TuLiP} front end. For varying numbers of vertices in the topology, nine\footnote{For $n = 11$ only four samples were calculated.} random graphs were generated each in order to investigate the state space size and synthesis time,  plotted in \fig{movstates} and \fig{movtimes}. The resulting mean and standard deviation are presented by bold and dashed lines respectively.\\

As with the controllers for stationary target search, the size of the state space depends quadratically on the size of the graph $n$. The synthesis time for the moving target search controllers is polynomial as well, since the curve on the logarithmic plot in \fig{movtimes} is sublinear. However, for a given $n$ the state space size of the controllers exceeds those in stationary target search. The same is true for the time they take to synthesize (cf. \fig{statstates} and \fig{stattime}). \\

\begin{figure}
\centering
	\psfrag{avg_states}[ll][cc]{\footnotesize{Robots $R_j$}}
	\psfrag{States}[cc][cc]{\footnotesize{States}}
	\psfrag{Vertices}[cc][cc]{\footnotesize{Vertices, $n$}}
	\includegraphics[width=0.8\textwidth]{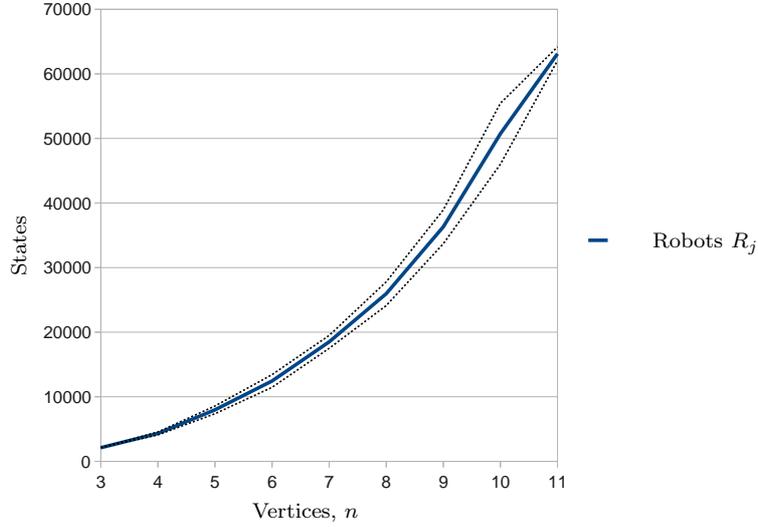}
\caption{States of the robots for the moving target search. The mean and the standard deviation of nine samples are shown in bold and dashed lines respectively.}
\label{fig:movstates}
\end{figure}

\begin{figure}
\centering
	\psfrag{avg_time}[ll][cc]{\footnotesize{Robots $R_j$}}
	\psfrag{Synthesis Time [s]}[cc][cc]{\footnotesize{Synthesis Time [s]}}
	\psfrag{Vertices}[cc][cc]{\footnotesize{Vertices, $n$}}
	\includegraphics[width=0.8\textwidth]{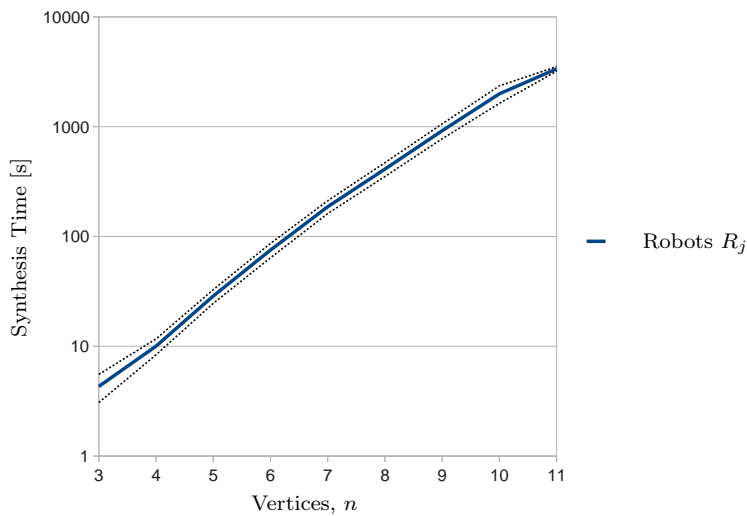}
\caption{Time to synthesize allocator and robot for the moving target search with $n$ vertices. The mean and the standard deviation of nine samples are shown in bold and dashed lines respectively. Note that the ordinate is logarithmic.}
\label{fig:movtimes}
\end{figure}

\subsection{Simulation}

We show a simulation of the moving target search on a small graph $G$ with five vertices. The cop number is $c(G) = 3$, and so we use three robots. The specification is synthesized first and then the resulting FDS is executed in three different threads, corresponding to three robots on the graph  with controllers $\M_{\Cops_0}$, $\M_{\Cops_1}$ and $\M_{\Cops_2}$. The communication between the robots is realized using a shared data structure holding the valuations of the global state variables $V$ of the composite FDS $\M_\Cops = \M_{\Cops_0} \comp \M_{\Cops_1} \comp \M_{\Cops_2}$.\\

As explained in \sec{mstage}, all robots are initialized to $\Mgu$, except one, which is initialized to $\Mcs$. \fig{s0} shows the initial configuration on the graph. Solid edges are contaminated while dashed edges are cleared. A vertex is black, red, green and dashed if it is contaminated, critical, partially cleared or cleared, respectively.\\

The execution of the asynchronous composition of $\M_{\Cops_0}$, $\M_{\Cops_1}$ and $\M_{\Cops_2}$ results in a sequence of states $s_0s_1s_2\ldots \models \M_{\Cops}$. Significant points in this sequence are shown in \fig{movsim}. It can be seen that the graph is successfully cleared with the given number of robots. The specifications do not state how the target is detected, in which case the moving target search would have ended with the target being found (if one exists).\\

\newcommand{\fsize}{\tiny}
\newcommand{\rlabel}[3]{\psfrag{#1}[cc][cc]{\fsize{$R_{#2}\!\!:$#3}}}
\newcommand{\subst}{
	\psfrag{(0)}[cc][cc]{\fsize{$v_0$}}
	\psfrag{(1)}[cc][cc]{\fsize{$v_1$}}
	\psfrag{(2)}[cc][cc]{\fsize{$v_2$}}
	\psfrag{(3)}[cc][cc]{\fsize{$v_3$}}
	\psfrag{(4)}[cc][cc]{\fsize{$v_4$}}
	\rlabel{R0:-1}{0}{$\mgu$}
	\rlabel{R1:-1}{1}{$\mgu$}
	\rlabel{R2:-1}{2}{$\mgu$}
	\rlabel{R0:0}{0}{$\mcs$}
	\rlabel{R1:0}{1}{$\mcs$}
	\rlabel{R2:0}{2}{$\mcs$}
	\rlabel{R0:1}{0}{$\mcl$}
	\rlabel{R1:1}{1}{$\mcl$}
	\rlabel{R2:1}{2}{$\mcl$}
	\rlabel{R0:2}{0}{$\mpc$}
	\rlabel{R1:2}{1}{$\mpc$}
	\rlabel{R2:2}{2}{$\mpc$}
}

\newcommand{\fwidth}{0.45\textwidth}
\newcommand{\statefig}[3]{
	\subfigure[$s_{#1}$: #3]{
	\subst
	\includegraphics[width=\fwidth]{state0#2.eps}
	\label{fig:s#1}
	}
}

\begin{figure}[p!]
\centering
	\statefig{0}{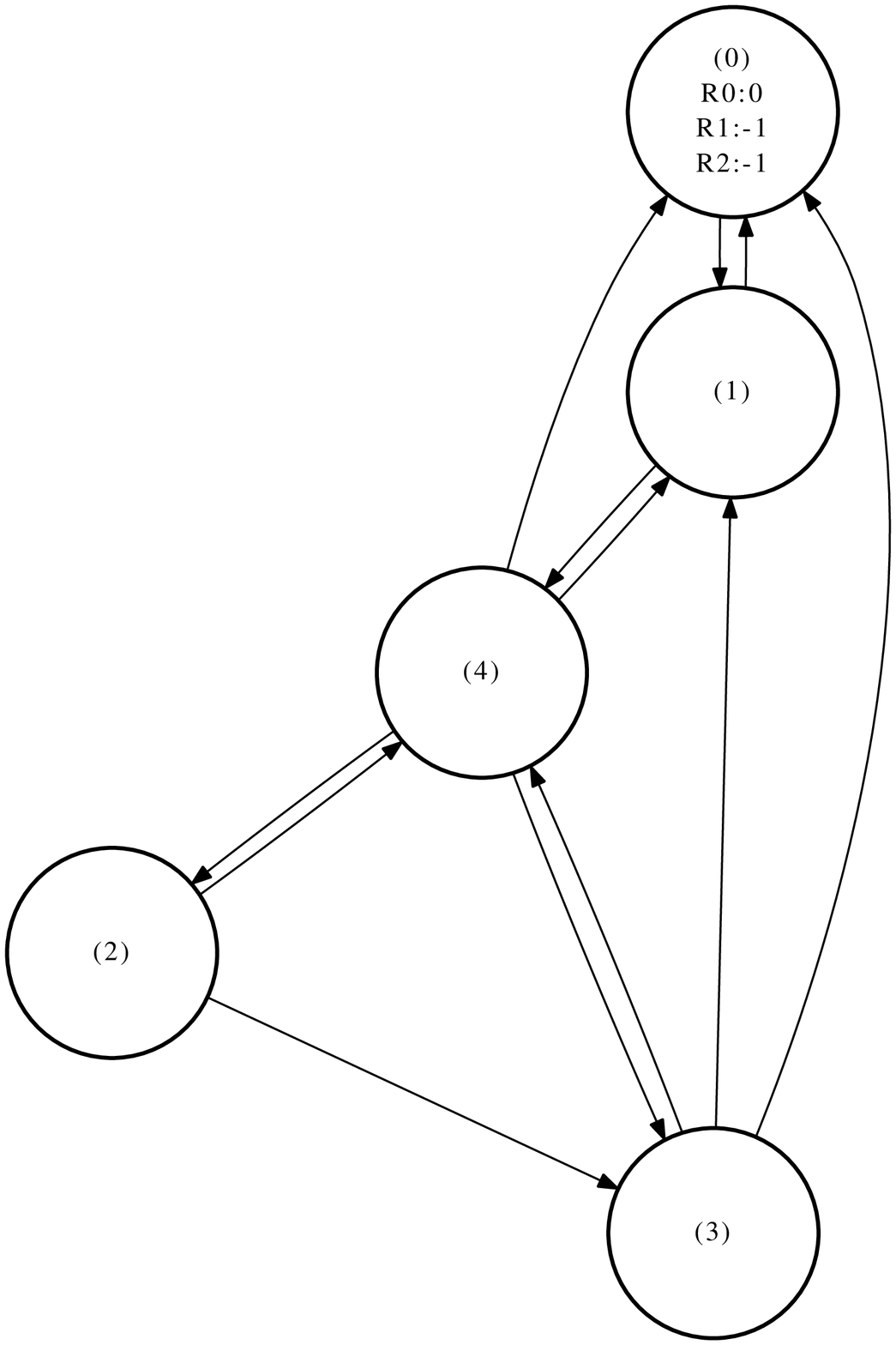}{Initial configuration.}
	\statefig{1}{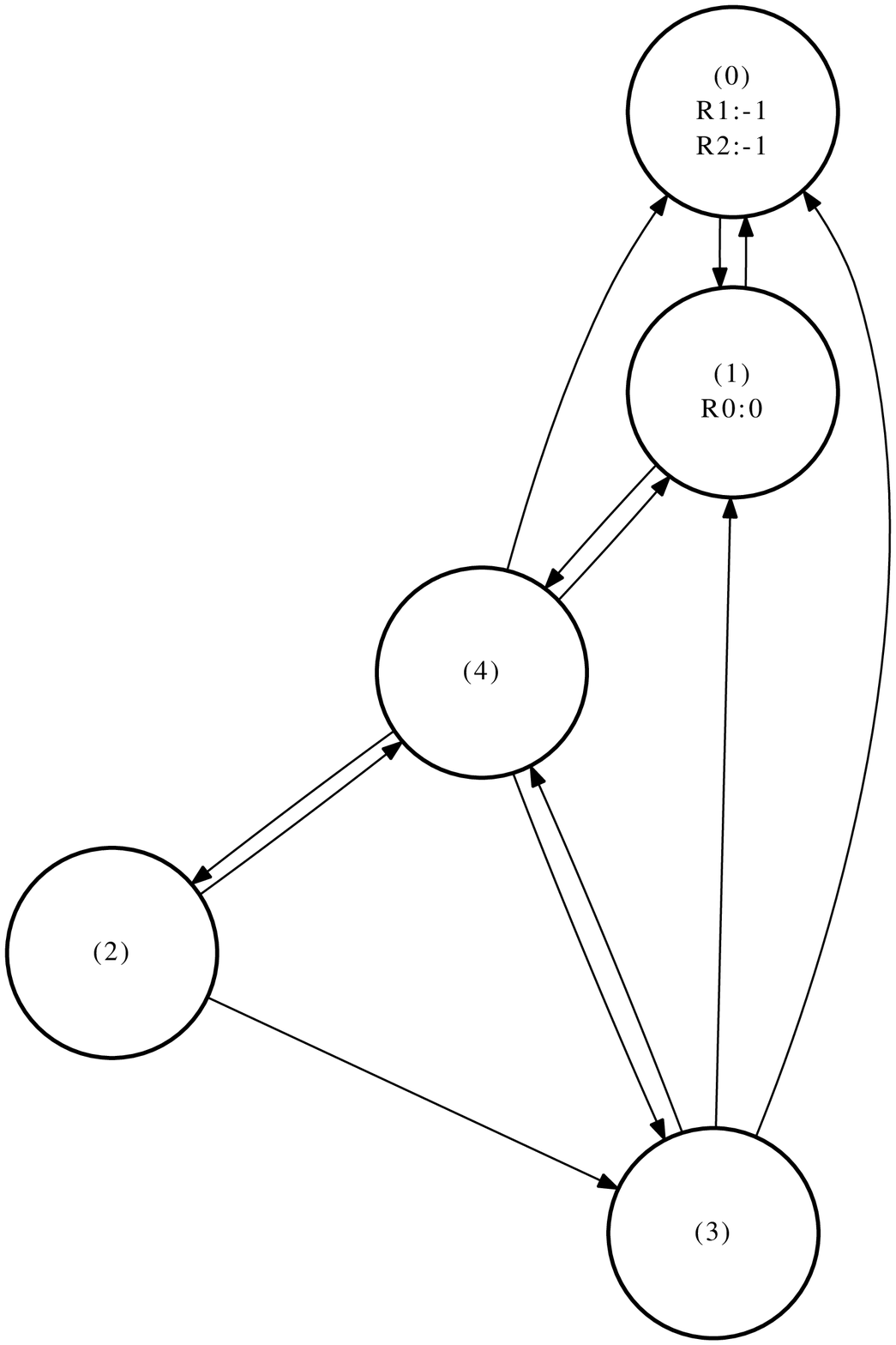}{Robot $R_0$ starts moving.}
	\statefig{6}{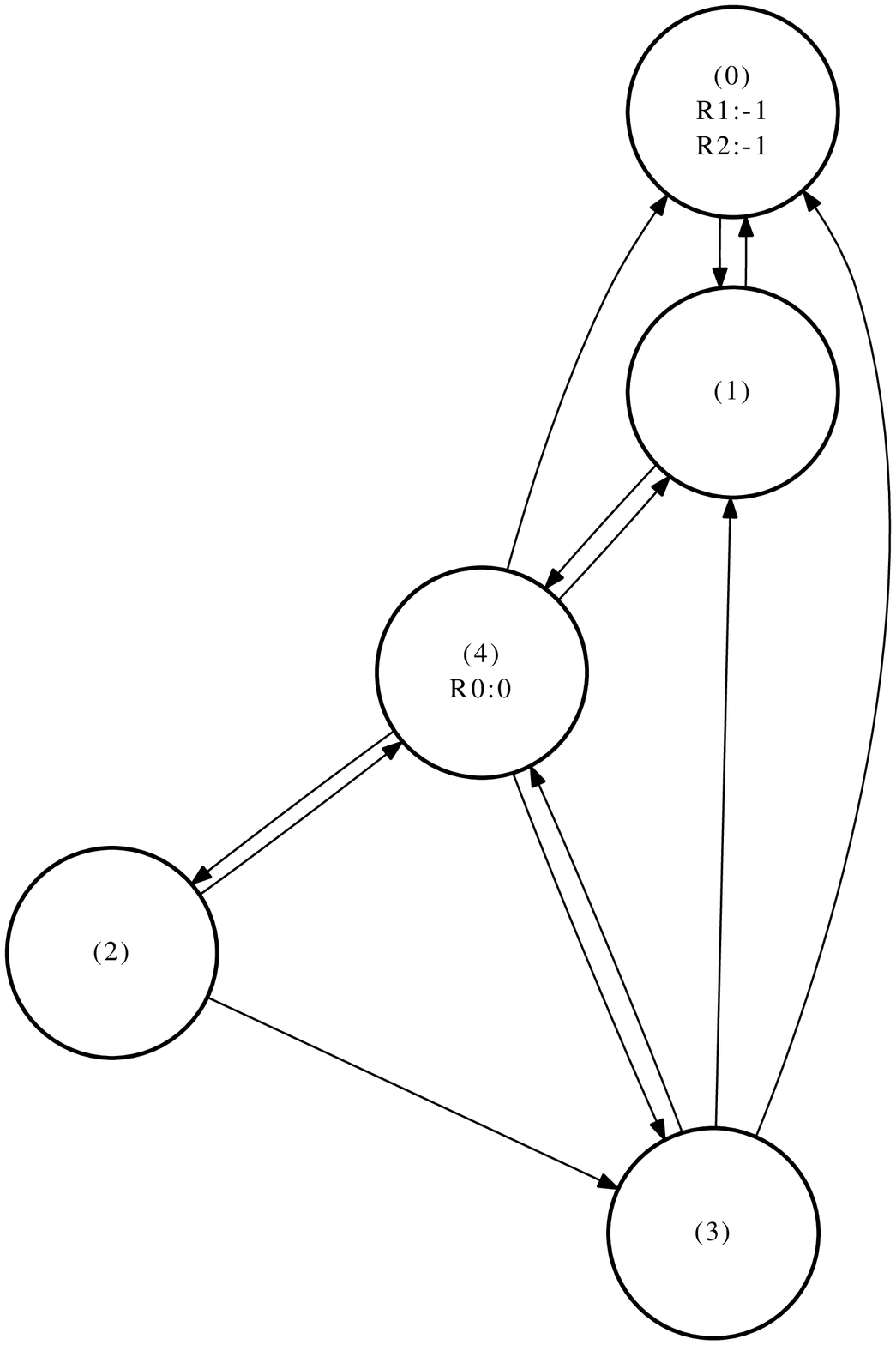}{Robot $R_0$ moves onto the starting vertex $v_4$.}
	\statefig{9}{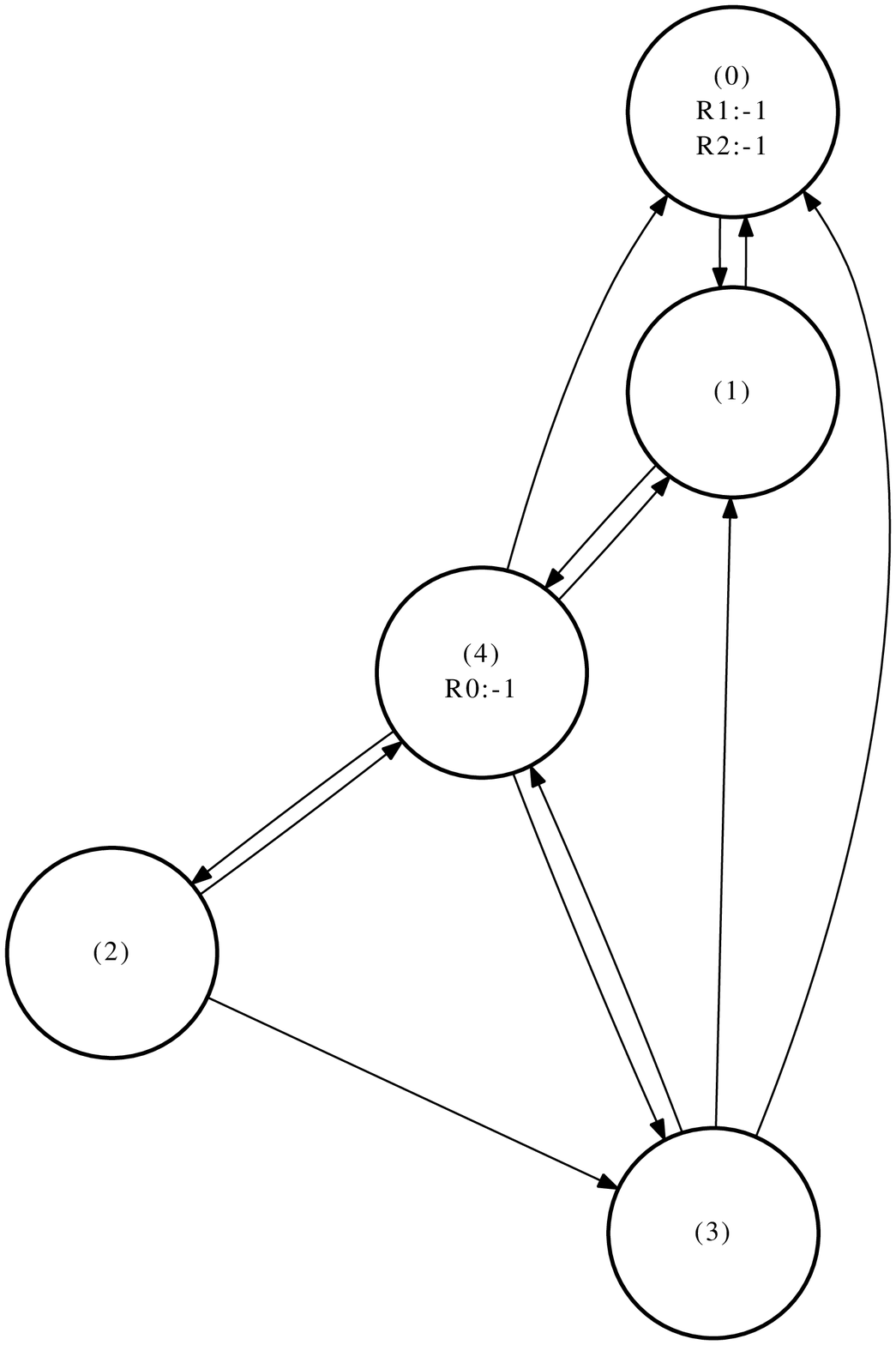}{Robot $R_0$ realizes that $v_4$ is a starting vertex and goes to $\Mgu$, calling its slave $R_1$.}
	\statefig{24}{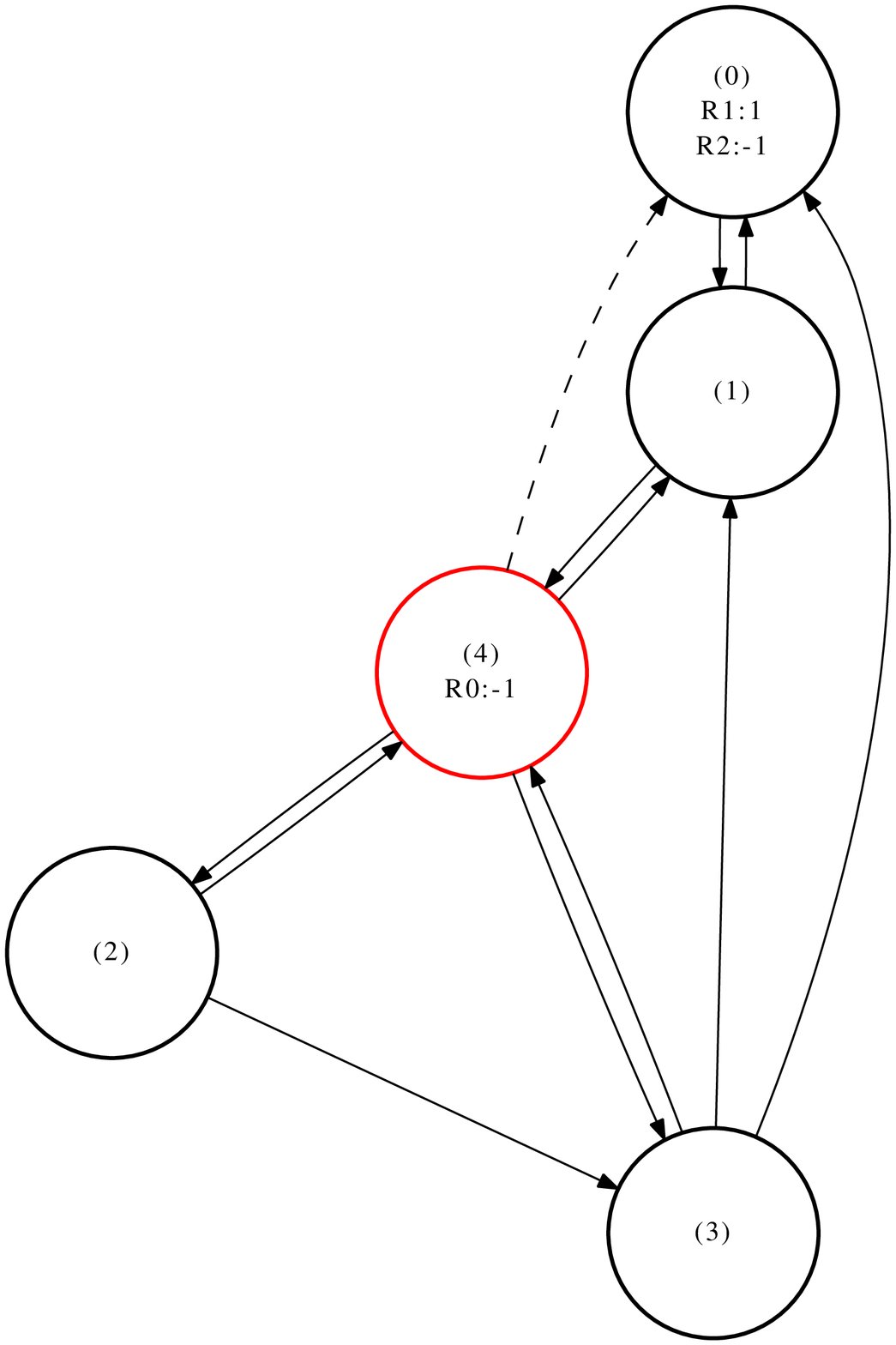}{Robot $R_1$ goes to $\Mcl$ and arrives at $v_4$.}
	\statefig{26}{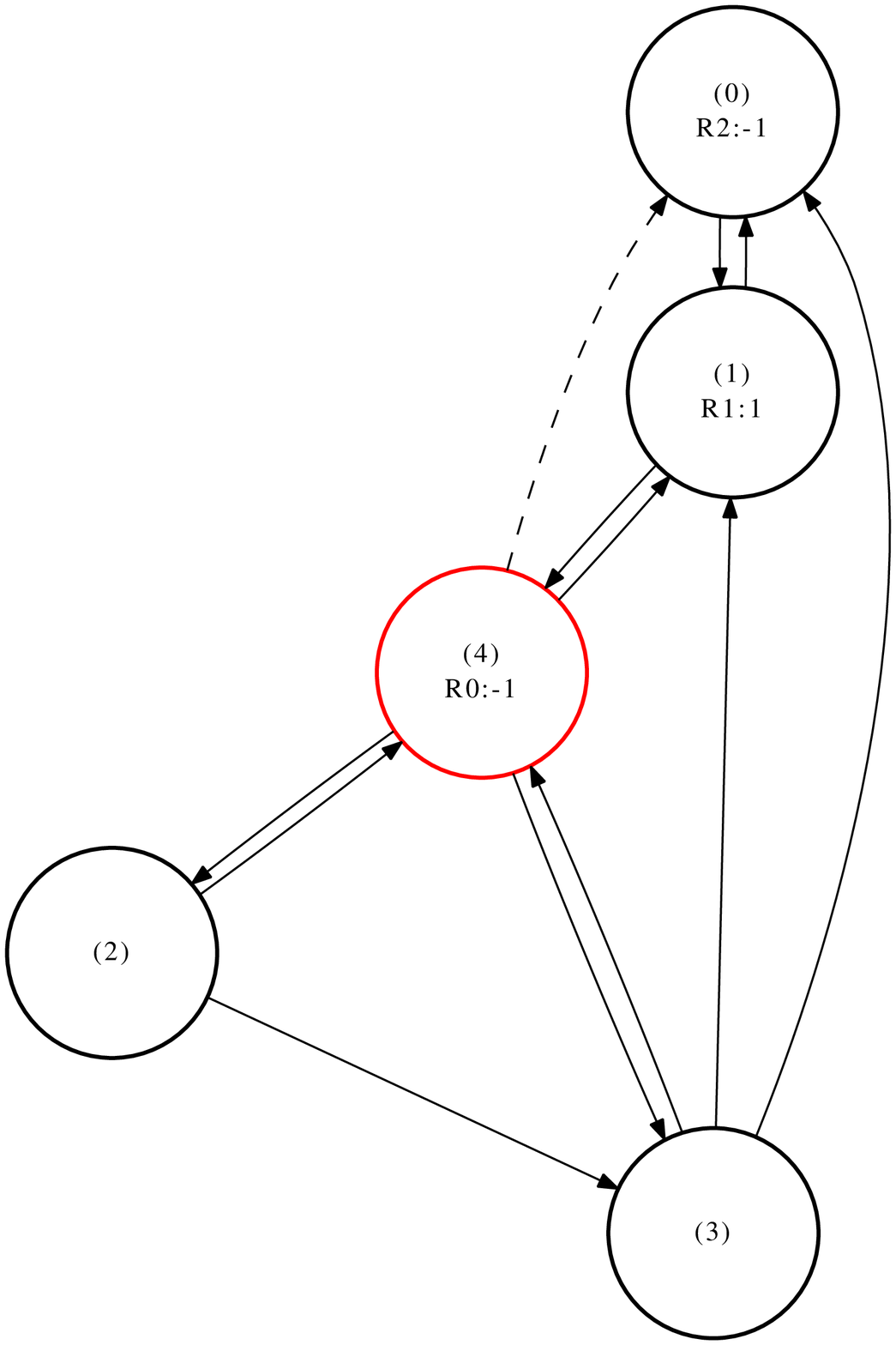}{Robot $R_1$ clears the edge from $v_4$ to $v_0$. $v_4$ becomes critical.}
	\caption{Simulation result of moving target search on a graph with five vertices and three robots.}
		\label{fig:movsim}
\end{figure}

\begin{figure}[p!]
\centering
	\setcounter{subfigure}{6}
	\statefig{47}{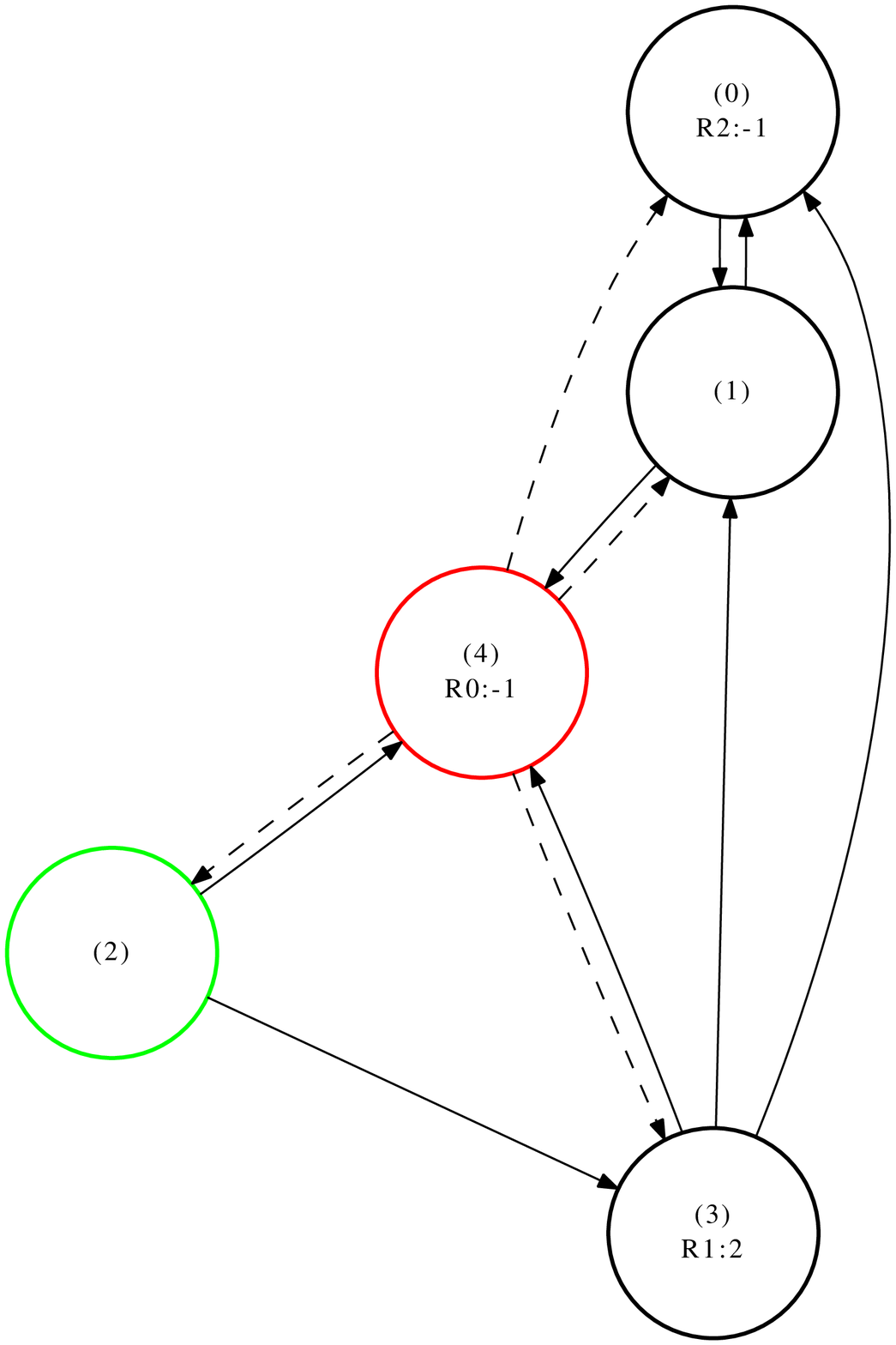}{All out-edges of $v_4$ are cleared.}
	\statefig{50}{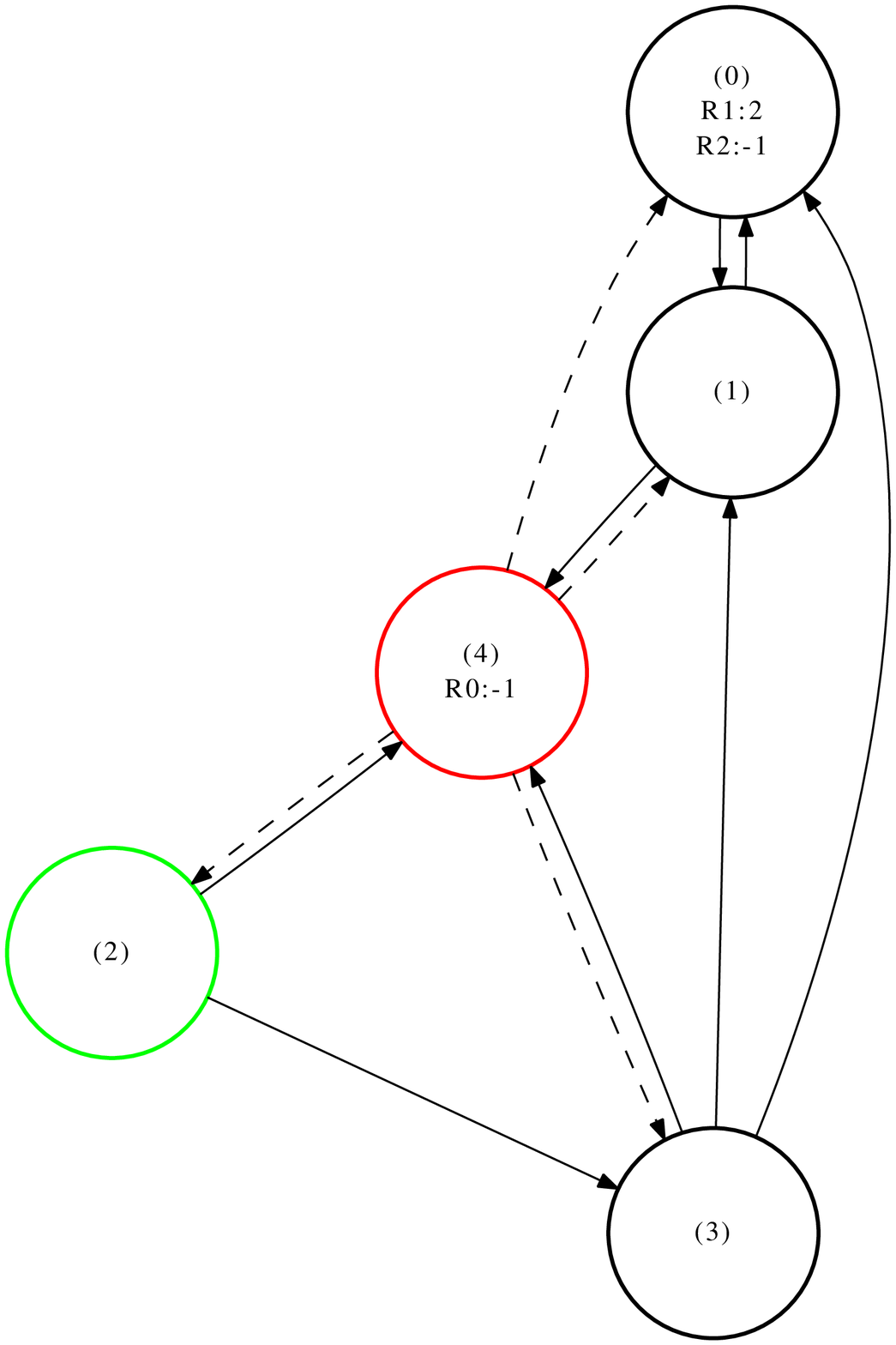}{$R_1$ goes to $\Mpc$ to search for a partially cleared vertex.}
	\statefig{63}{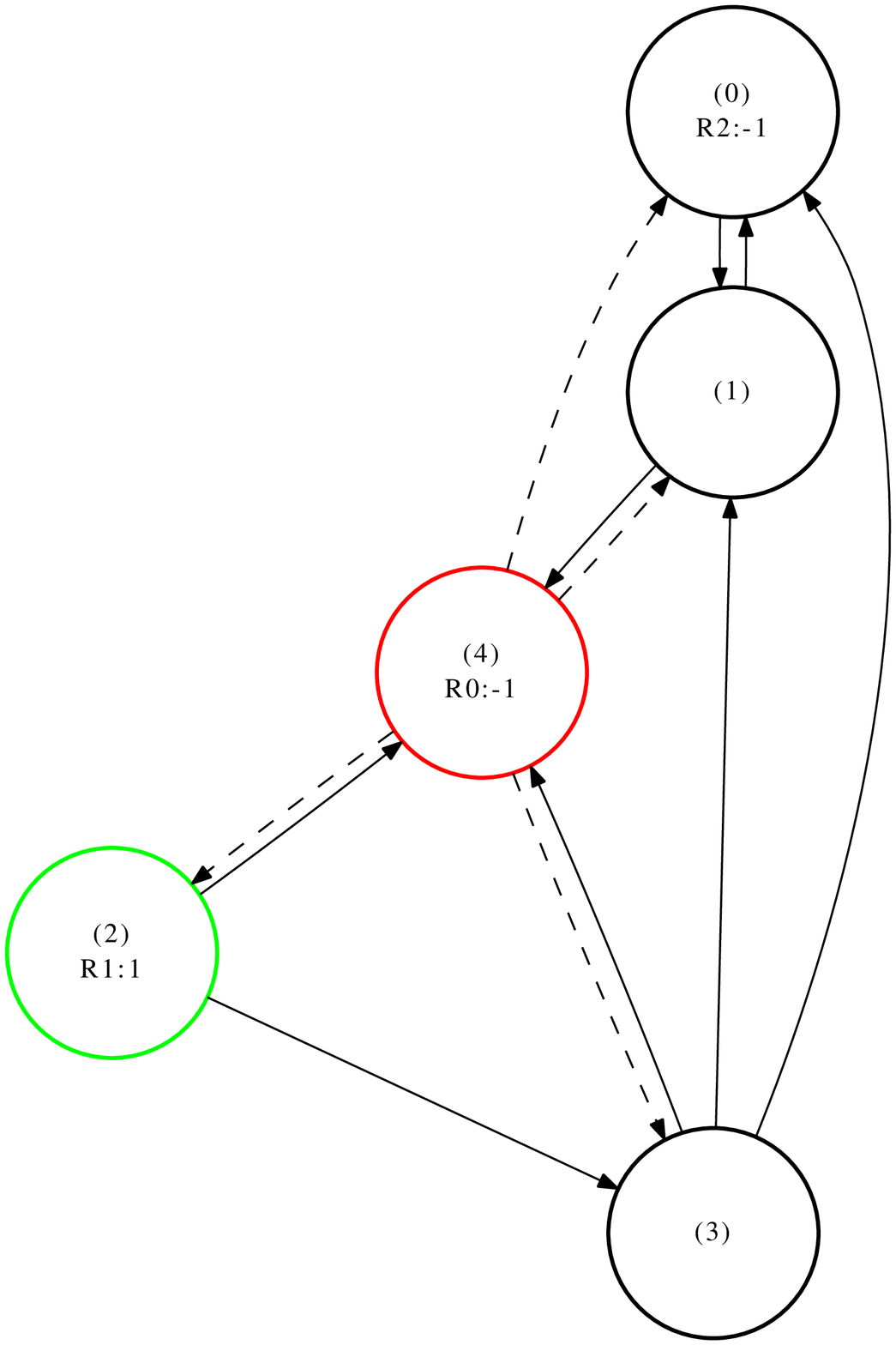}{$R_1$ found $v_2$, a partially cleared vertex.}
	\statefig{80}{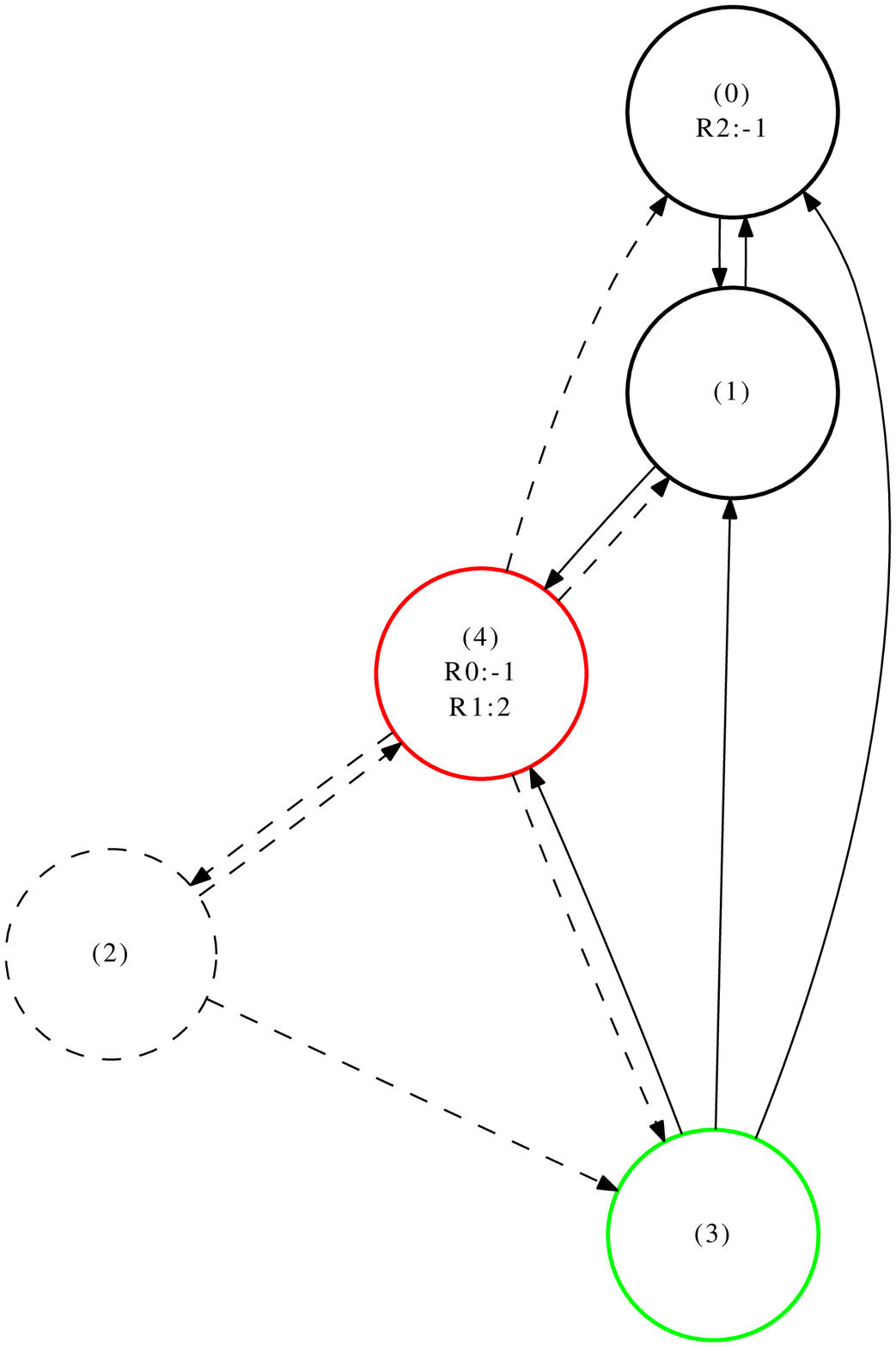}{All out-edges of $v_2$ are cleared. $v_2$ itself is now cleared and $v_3$ is partially cleared. $R_1$ continues looking for partially cleared vertices.}
	\statefig{118}{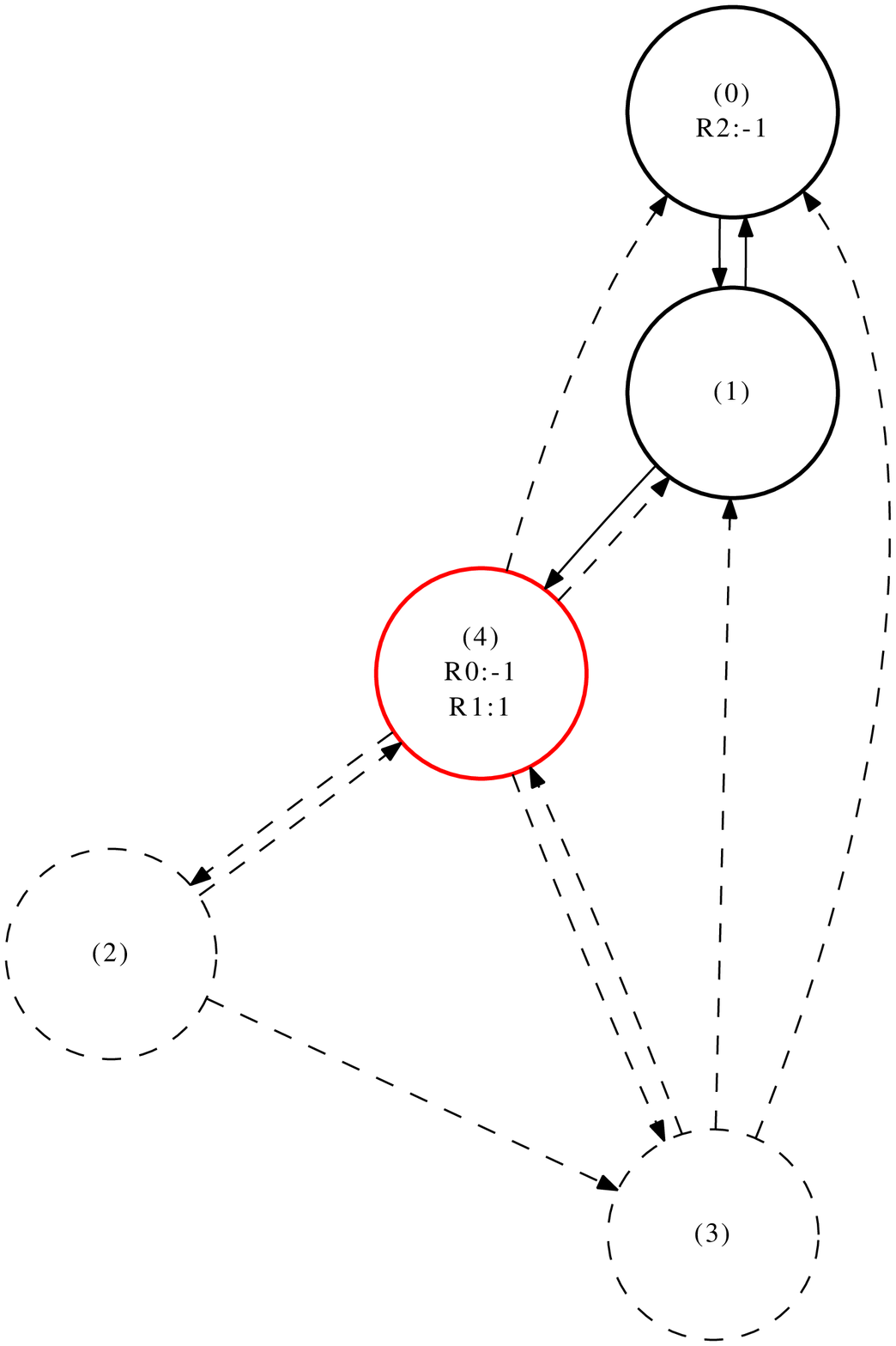}{$R_1$ has finished clearing all out-edges of all partially cleared vertices.}
	\statefig{121}{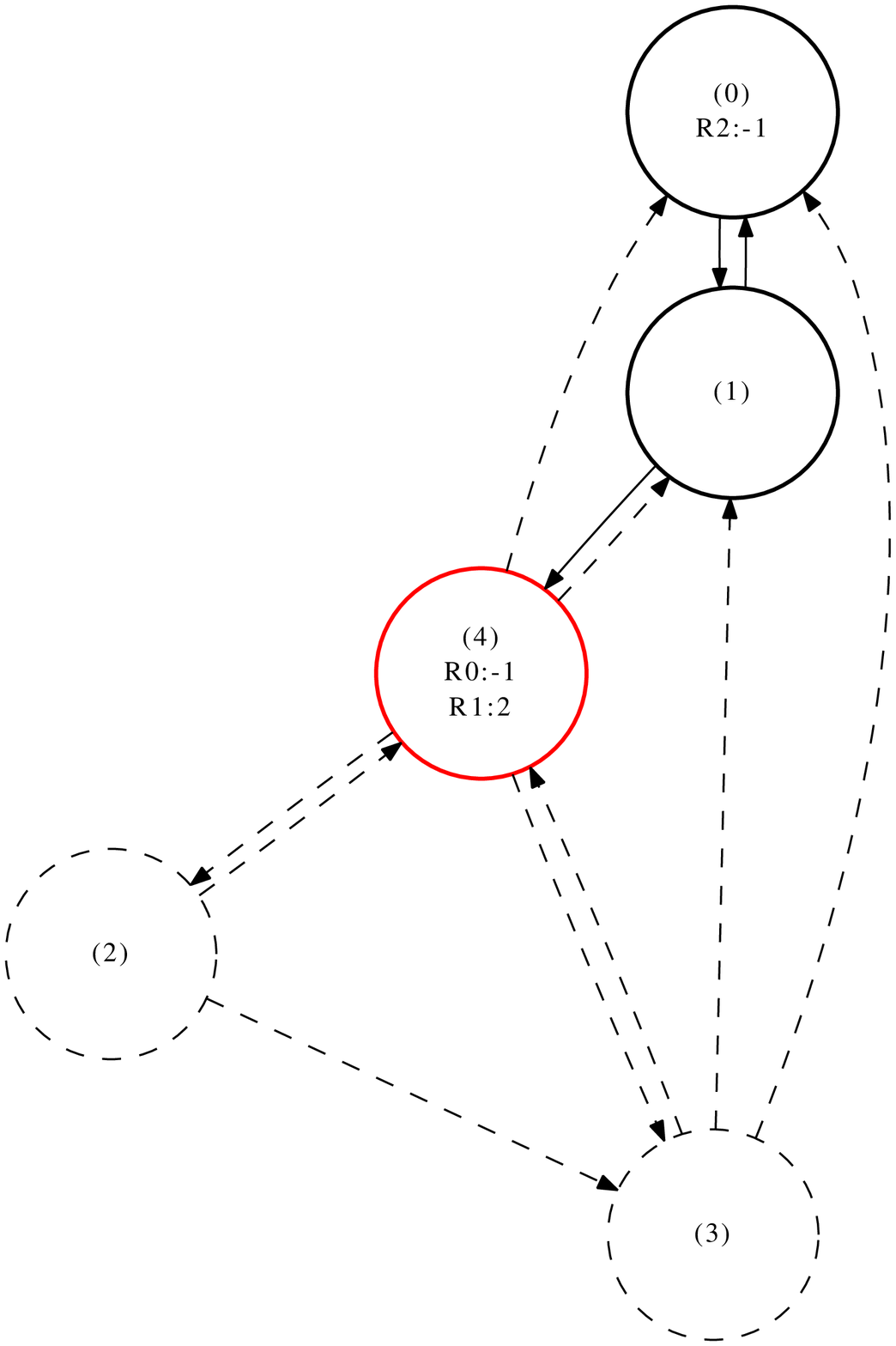}{$R_1$ tries looking for other partially cleared vertices, but will not find any.}
	\caption*{{\bf Figure 5.7:} Simulation result of moving target search on a graph with five vertices and three robots.}
\end{figure}

\begin{figure}[p!]
\centering
	\setcounter{subfigure}{12}
	\statefig{136}{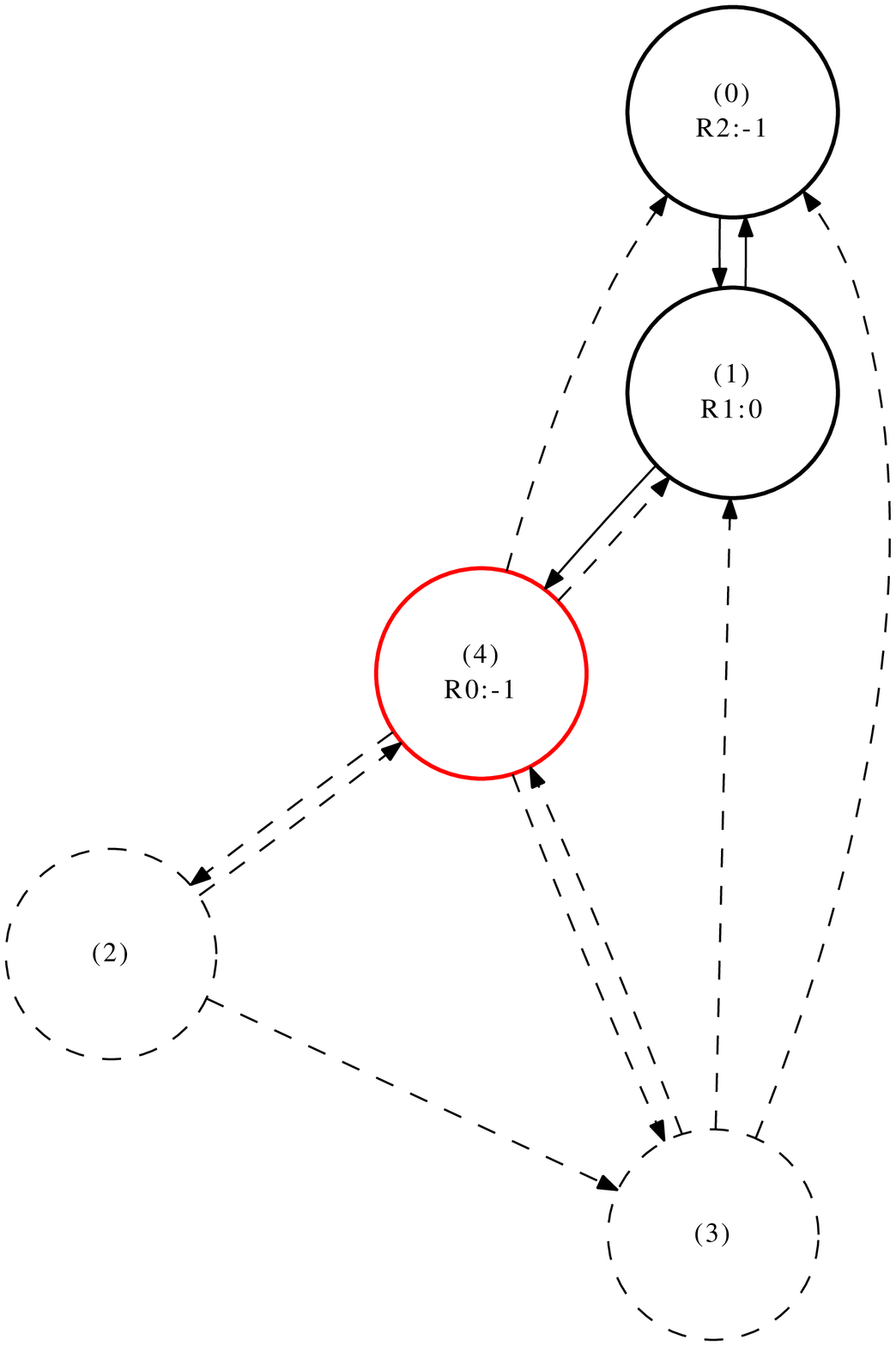}{$R_1$ discovered no partially cleared vertex and thus goes to $\Mcs$ to search for the next starting vertex.}
	\statefig{139}{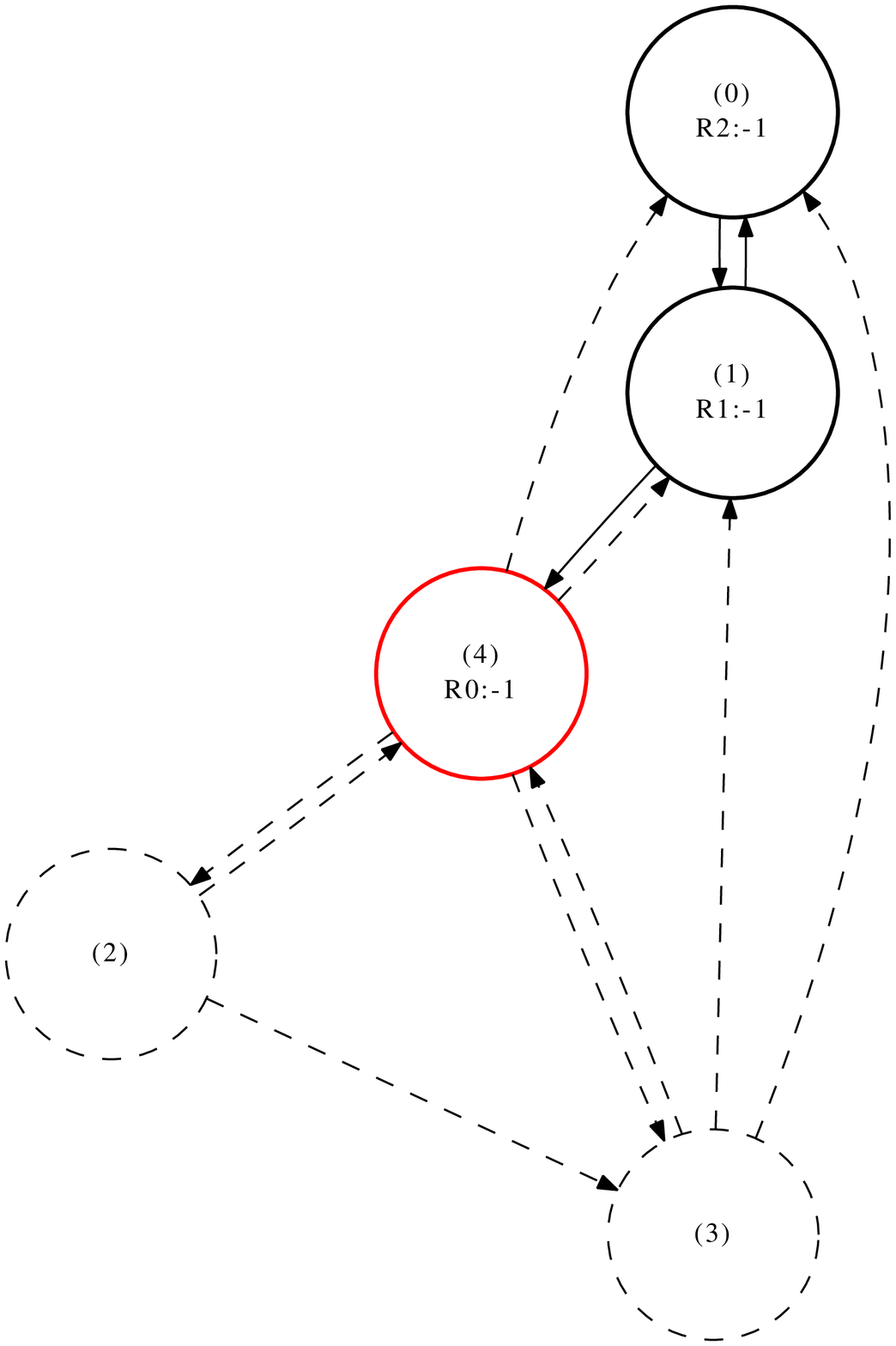}{Coincidentally, the vertex $R_1$ is on already is the starting vertex, so $R_1$ enters guarding mode $\Mgu$ without moving.}
	\statefig{143}{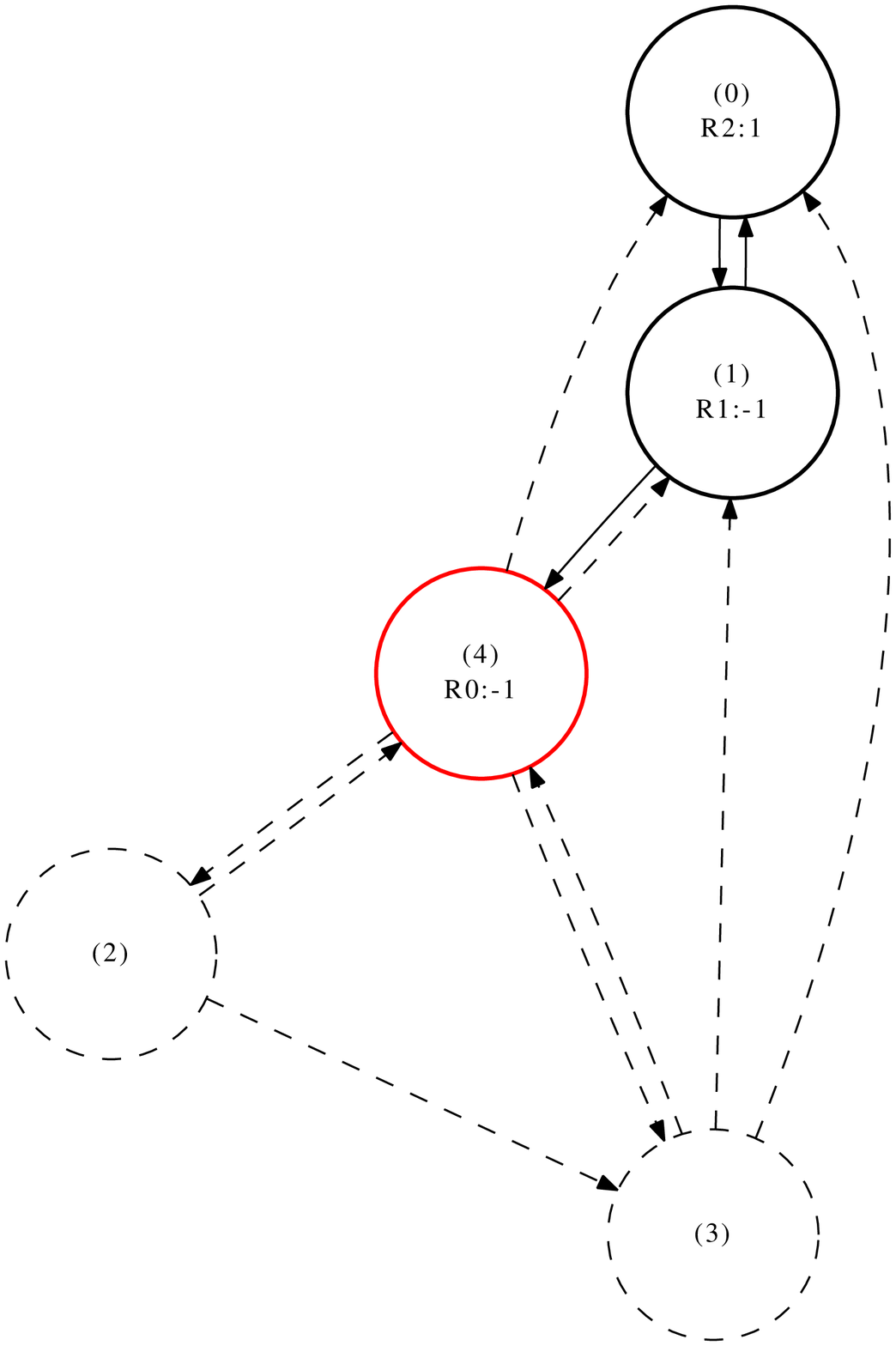}{$R_2$ is the slave of $R_1$ and enters $\Mcl$ to clear the out-edges of $v_1$}
	\statefig{159}{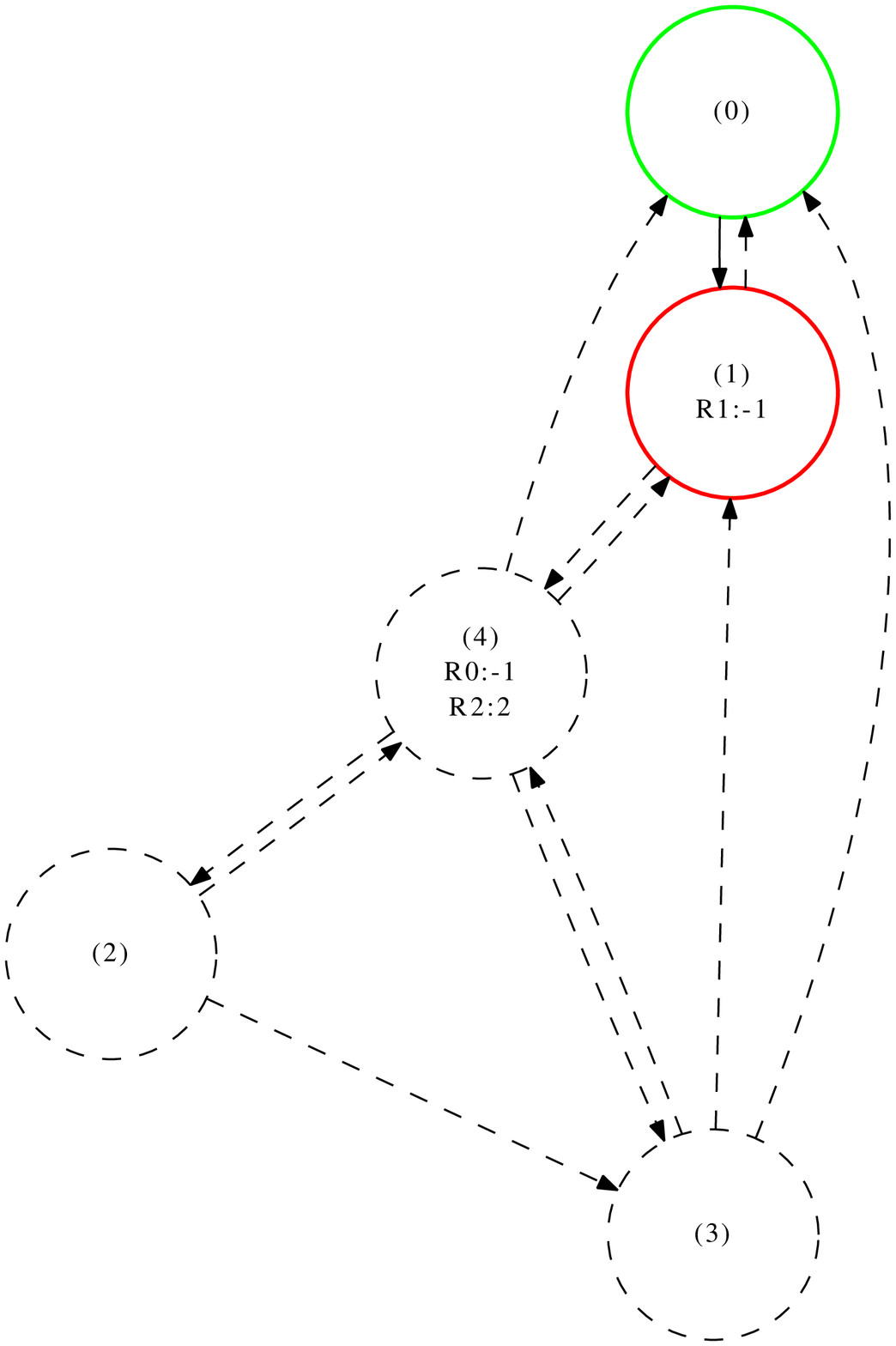}{All out-edges of $v_1$ are cleared, and the graph is already cleared, but the robots continue moving until all edges are cleared.}
	\statefig{183}{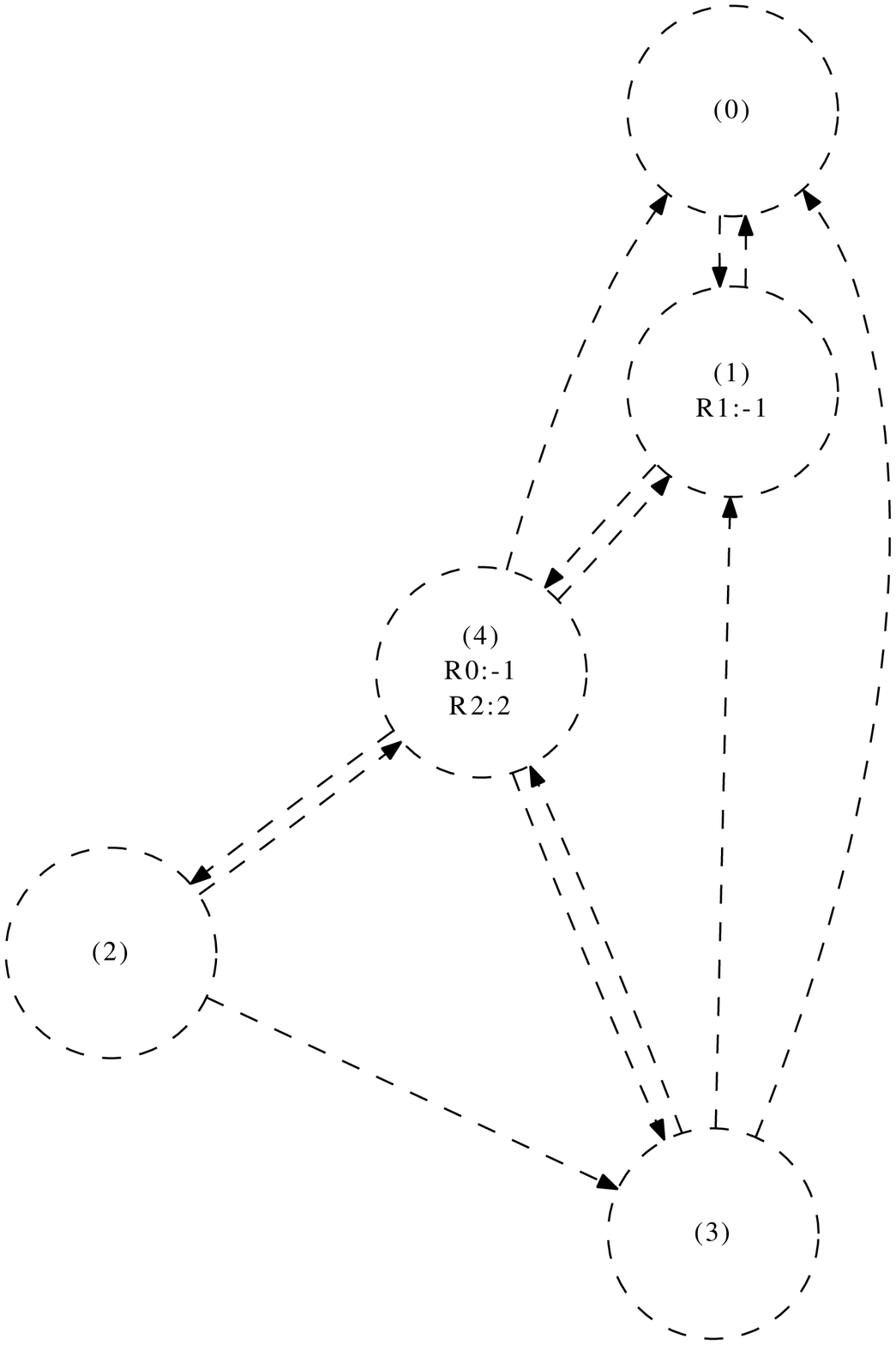}{All edges are cleared after $R_2$ has cleared $v_0$ which was partially cleared in $s_{159}$. $R_2$ is still in $\Mpc$ after clearing $v_0$.}
	\statefig{213}{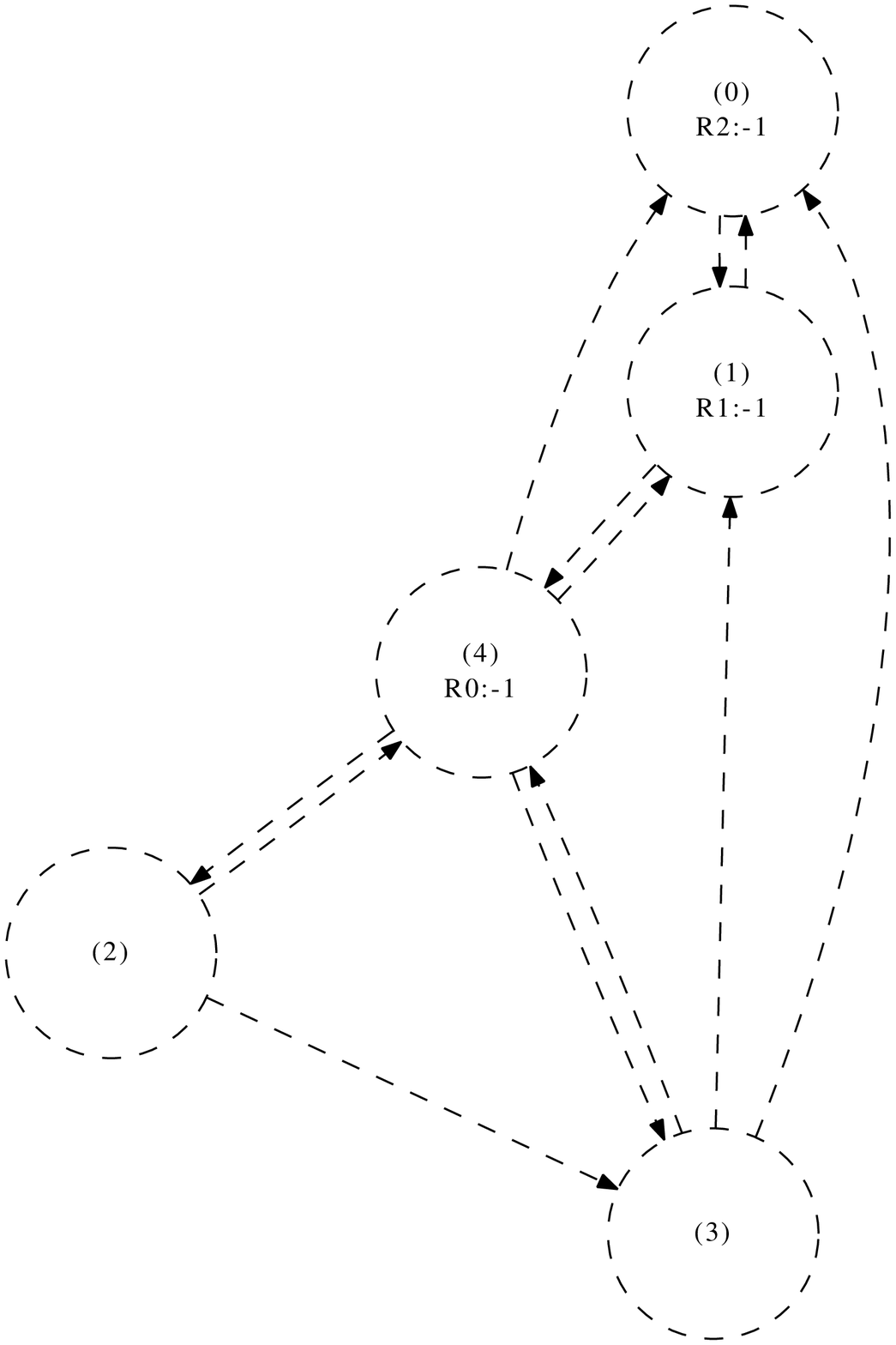}{After $R_2$ has unsuccessfully searched for partially cleared vertices and another starting vertex, it also enters $\Mgu$.}
	\caption*{{\bf Figure 5.7:} Simulation result of moving target search on a graph with five vertices and three robots.}
\end{figure}
%
%
%
%
%
%
%
%
%
%
%
%
%
%
%
%

\subsection{Discussion}

Again, as in the case of the stationary target search, the size of the state space increases quadratically with the number of vertices in the graph. However, the number of robots does not influence the size of the state space, since each robot has exactly one master and one slave and no further negotiation is required. Still, due to the required additional variables for communication, the synthesized FDS's are even larger than those in stationary target search.\\

In the searching modes, the circular search introduced in \sec{statsearch} is used. This results in a rather inefficient moving target search implementation. However, the robots only have local information, so when global properties such as ``for all partially cleared vertices'' need to be implemented in such a distributed way, a tradeoff between information and efficiency is introduced. Note also that by only using $V_G^i$, the heuristics $\mathcal{H}$ uses local information to decide on the start vertex.\\

When a robot is guarding a vertex, it cannot move. In this specification of moving target search, there is always only one robot not in guarding mode unless the graph is already cleared. This might seem inefficient, as only one robot can move. However, the guarding robots are playing an important part in the search; if they move, recontamination\index{Recontamination} could revert the progress of previous sliding moves.\\

One reason why the moving target search strategy can be implemented in such a way is because the global storage (cf. \sec{globstore}) allows the robots to not retain any historical knowledge of the graph state. Also, the heuristics $\mathcal{H}$ is implemented in the global storage, so that the start vertex is offered to a robot $R_j$ simply by letting $\cst_j = \csv$. In a real-world implementation this has to be taken into account.\\

\chapter{Conclusion}

%
%

In this thesis, a framework for specifying and reasoning about communicating and cooperating SAR robots is introduced, and the synthesis of controllers from their specifications is demonstrated. We further define and verify a communication protocol, allowing reliable asynchronous transmission of data between the robots.\\

The methods are demonstrated by successfully synthesizing and simulating controllers for robots performing centrally coordinated SAR of stationary targets, and for robots performing distributed search of moving targets. This demonstrates that it is possible to automatically generate discrete controllers for coordinated SAR operations that are mathematically proven to achieve their high-level goals.

\section{Observations}

The controllers cannot ensure bounds on the time it takes to eventually find a target. This is limitation inherent in the fragment of the specification language that we use and the synthesis method. However, fast searching is paramount, as with increasing time the chances of successful rescue dwindle, which is exacerbated in WiSAR, where it is usually necessary to increase the search radius with time.\\

With respect to the expected time to find a target, there is is a tradeoff between the accuracy and speed of effective search. These two factors are also called {\bf detection} and {\bf coverage}~\cite{WiSAR}. With greater accuracy the search takes longer and the probability of a SAR operation being successful decreases. In typical SAR operations, exhaustive search is only resorted to if other more effective search strategies have failed to yield useful information.\\

The discrete controllers are synthesized to satisfy their specifications with mathematically proven certainty. However, this requirement might be too rigorous for uncertain environments, compromising robustness in the case of unexpected scenarios or failures. For example, a change in the topology would render a controller unusable and cause the robot to stop operating. Also, it might be sufficient to guarantee that targets are found only with very high probability, or almost surely, while ensuring a faster overall operation.\\

With this type of controllers, the tradeoff between detection and coverage is completely ignored, and the search is always exhaustive. The accuracy of detection depends on the sensors being able to reliably detect targets. No such sensor exists, and sensors with high false-alarm rates lead to misallocating rescue robots that could be used otherwise to improve coverage.\\

\section{Viability of Approach}

Two key themes limiting the versatility of the proposed methods throughout this thesis are the computational complexity of the synthesis methods and the size of the state space. It may be objected that all information required for the controller is already encoded in the specification, and that the synthesis is wasteful both in terms of time and space. A program in a procedural language rather than an FDS could be used instead with the same effect but with lower space requirements. However, for complex specifications, program validation can likewise be a lengthy task.\\

In particular for the search strategies, algorithms have been developed that have proven to yield the same results as the tediously specified FDS's. It might not be viable to translate an existing algorithm into LTL and then synthesizing it back into an executable FDS. However, development of rule-based controllers or incrementally subjecting algorithms to more requirements can be handled straightforwardly with LTL specifications (as long as the specification remains realizable), e.g.\ introducing a new traffic rule when controlling an autonomous vehicle for urban traffic.\\

\section{Future Work}

Several improvements and extensions that are relevant for the specification and synthesis of controllers from LTL can be made. In SAR operations, it is often required that there are agents designated to particular tasks, e.g.\ robots specialized for searching large areas, searching on a small scale and extracting victims, or providing medical relief. This heterogeneity can be exploited for more efficient SAR strategies~\cite{WiSAR}. Moreover, a distribution of the resources can be included, such as partitioning the topology among several teams of robots.\\

The currently implemented search strategies are effective, but inefficient. This can be partially alleviated by the above improvements. Another approach would be to allow more robots than the cop number and implement fast search strategies that require each edge to only be traversed once by a robot~\cite{FastSearch}. Also, efficient search methods that rely on probabilistic information on the environment can be synthesized from probabilistic temporal logic specifications~\cite{PCTLSynthesis, TACAS11}.\\

The controller specifications are small compared to the FDS's resulting from synthesis with TLV. It would be beneficial to exploit the compactness of a specification and perform synthesis on-the-fly instead of generating a static FDS. Such dynamic approaches are already used for model checking, e.g.\ by synthesizing deterministic monitors~\cite{Dynamic}. Dynamic synthesis is still in its infancy, but receding horizon, on-the-fly and compositional methods are already being developed~\cite{Acacia, RecedingControl}.\\

Finally, Chung et al.\ \cite{SearchBots} state that
\begin{quote}
``due to poor reliability and the number of possible failures, few multi-robot systems have been tested on the scale necessary to demonstrate pursuit-evasion in complex environments.''
\end{quote}
This thesis shows that theoretically the synthesis of controllers for SAR robots is viable, but neither the technology of robots is yet advanced enough, nor is controller synthesis explored to the extent that large-scale complex SAR scenarii can be implemented based on the methods that we present. The small existing sample implementations do not yet allow reliable communication between the robots~\cite{Waldo}. Given the promises in robot-assisted SAR and the complexity reduction through synthesis, it would be a rewarding endeavor to test the presented methods on hardware.\\


\cleardoublepage
\printindex

\cleardoublepage
\bibliographystyle{plain}
\small
\bibliography{references}

\end{document}